\documentclass[11pt]{article}
\pdfoutput=1 
\usepackage{./jheppub}

\usepackage[T1]{fontenc}

\usepackage{graphicx}
\usepackage{amsfonts,amsmath,amssymb}
\usepackage{verbatim}
\usepackage{array}
\usepackage{mathrsfs}
\usepackage{mathrsfs}
\usepackage{yfonts}
\usepackage{dsfont}
\usepackage{bbm}
\usepackage{colonequals}
\usepackage{amscd}

\usepackage{subcaption}

\usepackage{wrapfig}

\usepackage{suffix,mathtools,cancel,bbm}

\usepackage{multirow}

\usepackage{enumerate}

	\usepackage{tikz, tikz-3dplot, pgfplots}
	\usepackage{tkz-graph}
	\usetikzlibrary[positioning,patterns] 

\newcommand{\CM}[1]{{\bf [CM: #1]}} 

\newcommand{\bn}{\begin{enumerate}}
\newcommand{\en}{\end{enumerate}}

\def\CC{{\cal C}}

\def\CM{{\cal M}}
\def\CN{{\cal N}}
\def\CO{{\cal O}}

\def\CS{{\cal S}}

\def\z{\zeta}

\def\det{{\rm det}}

\newcommand{\beq}{\begin{equation}}
\newcommand{\eeq}{\end{equation}}

\newcommand\nn{\nonumber}

\newcommand{\cA}{\mathcal{A}}
\newcommand{\cB}{\mathcal{B}}
\newcommand{\cC}{\mathcal{C}}
\newcommand{\cD}{\mathcal{D}}

\newcommand{\cF}{\mathcal{F}}
\newcommand{\cG}{\mathcal{G}}
\newcommand{\cH}{\mathcal{H}}
\newcommand{\cI}{\mathcal{I}}
\newcommand{\cJ}{\mathcal{J}}
\newcommand{\cK}{\mathcal{K}}

\newcommand{\cM}{\mathcal{M}}
\newcommand{\cN}{\mathcal{N}}
\newcommand{\cO}{\mathcal{O}}

\newcommand{\cS}{\mathcal{S}}

\newcommand{\cV}{\mathcal{V}}
\newcommand{\cW}{\mathcal{W}}

\numberwithin{equation}{section}

\def\NO#1#2{{\rm NO}(#1,#2)}

\def\bea{\begin{eqnarray}}
\def\eea{\end{eqnarray}}

\DeclarePairedDelimiterX\MeijerM[3]{\lparen}{\rparen}%
{\begin{smallmatrix}#1 \\ #2\end{smallmatrix}\delimsize\vert\,#3}

\newcommand\MeijerG[8][]{%
  G^{\,#2,#3}_{#4,#5}\MeijerM[#1]{#6}{#7}{#8}}

\WithSuffix\newcommand\MeijerG*[7]{%
  G^{\,#1,#2}_{#3,#4}\MeijerM*{#5}{#6}{#7}}

\def\tr{\mathop{\mathrm{tr}}\nolimits}

\def\cH{\mathcal{H}}

\def\cN{\mathcal{N}}

\def \beg#1{\begin{#1}} 

\def \bea{\beg{eqnarray}}
\def \eea{\end{eqnarray}}
\def \ee{\end{equation}}

\def \restr#1#2{{\left.\kern-\nulldelimiterspace#1\vphantom{\big|}\right|_{#2}}}

\def \nn{\nonumber}

\def \CC{\mathcal{C}}

\def\sign{ \text{sign} }

\newcommand{\ba}[1]{\begin{align} #1 \end{align} }

\newcommand{\bs}[1]{\begin{split} #1 \end{split} }

\usepackage{colortbl}
\definecolor{mygray}{gray}{0.93}

\title{\boldmath M5-brane Sources, Holography, and Argyres-Douglas Theories
}

\author[a]{Ibrahima Bah,}
\author[b]{Federico Bonetti,}
\author[c]{Ruben Minasian,}
\author[d]{and Emily Nardoni} 
\affiliation[a]{Department of Physics and Astronomy, Johns Hopkins University, 3400 North Charles Street, Baltimore, MD 21218, USA}
\affiliation[b]{Mathematical Institute, University of Oxford, Woodstock Road, Oxford, OX2 6GG, UK}
\affiliation[c]{Institut de Physique Th\'{e}orique, Universit\'{e} Paris Saclay, CNRS, CEA, F-91191, Gif-sur-Yvette, France}
\affiliation[d]{Mani L. Bhaumik Institute for Theoretical Physics, Department of Physics and Astronomy, University of California, Los Angeles,  CA 90095, USA}

\emailAdd{iboubah@jhu.edu, federico.bonetti@maths.ox.ac.uk, ruben.minasian@ipht.fr, enardoni@ucla.edu}

\abstract{
We initiate a study of the holographic duals of a class of four-dimensional $\mathcal{N}=2$ superconformal field theories that are engineered by wrapping M5-branes on a sphere with an irregular puncture. 
These notably include the strongly-coupled field theories of Argyres-Douglas type. 
Our  solutions are obtained in 7d gauged   supergravity, 
where they take the form of a warped product of $AdS_5$ 
and a ``half-spindle.''
The irregular puncture is modeled by a localized M5-brane source in the internal space of the gravity duals.
Our solutions feature a realization of supersymmetry that is distinct from the usual topological twist, as well as an interesting St{\"u}ckelberg mechanism involving the gauge field associated to a generator of the isometry algebra of the internal space.
We check the proposed duality by computing the holographic central charge, the flavor symmetry central charge, and the dimensions of various supersymmetric probe M2-branes, and matching these with the dual Argyres-Douglas field theories. 
Furthermore, we compute the large-$N$ 't Hooft anomalies of the field theories using anomaly inflow methods in M-theory, and find perfect agreement with the proposed duality.
}

\usepackage{etoolbox}
\appto\appendix{\addtocontents{toc}{\protect\setcounter{tocdepth}{1}}}

\appto\listoffigures{\addtocontents{lof}{\protect\setcounter{tocdepth}{1}}}
\appto\listoftables{\addtocontents{lot}{\protect\setcounter{tocdepth}{1}}}

\begin{document} 

\setcounter{tocdepth}{2}

\maketitle
\flushbottom



\section{Introduction} \label{sec:introduction}



The Argyres-Douglas (AD) field theories have particular significance among four-dimensional $\mathcal{N}=2$ superconformal field theories (SCFTs). As in the original such theory discovered by Argyres and Douglas in \cite{Argyres:1995jj}, many of these SCFTs appear at special singular points on the moduli space of $\mathcal{N}=2$ gauge theories where mutually non-local dyons simultaneously become massless  \cite{Argyres:1995xn,Eguchi:1996vu}. Such phenomena cannot be captured by a Lagrangian in the traditional sense, and thus these theories are intrinsically strongly coupled. Nonetheless, the existence of an interacting superconformal fixed point has been convincingly argued from both field theoretic and string theoretic perspectives. 
 
Several features of the Argyres-Douglas theories set them apart. 
A dramatic example is that they possess relevant chiral operators in their spectrum with fractional scaling dimensions. 
Among all unitary interacting $\mathcal{N}=2$ SCFTs, the theory with smallest $c$-central charge is the original AD theory with one relevant chiral ring generator of dimension $\frac{6}{5}$, and in this sense the ``minimal'' $\CN=2$ SCFT is of Argyres-Douglas type \cite{Liendo:2015ofa}.

Argyres-Douglas SCFTs also appear in the low-energy limit of various string theory configurations via geometric engineering. One such realization involves compactifying the 6d $\mathcal{N}=(2,0)$ SCFTs of $\mathfrak{g}=$ADE type on a punctured sphere, which for $\mathfrak{g}=A_{N-1}$ corresponds to $N$ M5-branes wrapped on the sphere. An infinite class of four-dimensional conformal field theories with varying amounts of supersymmetry can be obtained by compactifying the (2,0) theories on a punctured Riemann surface, while employing a topological twist to preserve supersymmetry in four dimensions, beginning with the $\mathcal{N}=2$ constructions in \cite{Witten:1997sc,Gaiotto:2009we,Gaiotto:2009hg}. A large subset of Argyres-Douglas theories can be thus obtained in the very special case that the Riemann surface is a sphere with a puncture of {\it irregular}, rather than regular, type \cite{Bonelli:2011aa,Xie:2012hs,Wang:2015mra} (also see \cite{Gaiotto:2009hg,Witten:2007td}). While the holographic duals of a large class of 4d $\mathcal{N}=2$ SCFTs in geometric engineering with regular punctures are known \cite{Gaiotto:2009gz}, until now the gravity duals of 4d field theories from irregular punctures have remained mysterious. 

The realization of Argyres-Douglas theories via M5-branes wrapped on spheres with irregular punctures offers the prospect of studying their properties in holography.  In this paper we present the first gravity duals of AD theories in M-theory, and provide a new perspective on both the geometry of the irregular puncture and the curious field theoretic properties of these SCFTs.  An important motivation to this work has been the recent work on branes wrapping spindle geometries\footnote{~These geometries also appear in the study of holographic duals of $\mathcal{N}=1$ class $\mathcal{S}$ theories, where they determined the structure of probe branes and aspects of the moduli space of the dual field theories \cite{Bah:2013wda}.}  \cite{Ferrero:2020laf,Ferrero:2020twa} as a novel way of preserving supersymmetry beyond the paradigm of the topological twist in supergravity  \cite{Maldacena:2000mw}.  Our construction will provide yet another way of preserving supersymmetry, by wrapping branes on a disk with a nontrivial $U(1)$ holonomy at the boundary.  Our setup can be thought of as M5-branes wrapping a ``half-spindle''.

%

A distinctive feature of our 11d solutions 
is the presence of localized
M5-brane sources in the internal space.
These appear as singularities in the low-energy supergravity
approximation, but correspond to well-defined objects in
the full M-theory. As demonstrated in several examples
\cite{Brandhuber:1999np, Apruzzi:2013yva, Gaiotto:2014lca, Apruzzi:2015wna, DHoker:2016ujz, DHoker:2017mds, Bah:2017wxp, Bah:2018lyv}, brane sources are useful
ingredients in holography. In particular, they provide
an avenue to realizing arbitrary flavor symmetries.
In our solutions the M5-brane source is instrumental,
and is in fact dual to the irregular puncture on the sphere.
This novel connection between irregular punctures and sources in supergravity paves the way to further investigations and generalizations to other brane constructions. 

Another peculiar property of our  solutions
is related to the interplay between the isometry algebra
of the internal space and the algebra of global zero-form
symmetries in the SCFT. In particular,
we identify a $U(1)$ isometry  generator that is \emph{not}
mapped to a generator of a continuous $U(1)$ global zero-form
symmetry of the dual field theory.
Indeed, the would-be massless $U(1)$ gauge field
associated to this isometry generator is actually
massive in the $AdS_5$ low-energy effective action,
by virtue of a novel St\"uckelberg mechanism
involving an axion field originating from the
expansion of the M-theory 3-form.
The interplay between the background $G_4$-flux
supporting the holographic solution and the isometry group
of the internal space can be elegantly described
in the language of equivariant cohomology.
Our physical analysis in terms of a St\"uckelberg mechanism
detects an obstruction 
to finding a closed equivariant
completion of $G_4$---and provides a recipe to overcome it.

The rest of this paper is organized as follows.
In section \ref{sec:spindlesolutions} we present a new class of $AdS_5$ solutions in 11d supergravity that preserve $\CN=2$ superconformal symmetry. We first describe the solutions in 7d $U(1)^2$ gauged supergravity, and then give their uplift on $S^4$ to eleven dimensions. The solutions feature an M5-brane source, and their flux configuration is encoded by three positive integers.  We compute the holographic central charge, as well as the charges of various supersymmetric probe M2-branes wrapping two-cycles in the internal space. 

In section \ref{sec:anomaly} we use the machinery of anomaly inflow to extract the global symmetries and 't Hooft anomalies of the SCFTs dual to the aforementioned supergravity solutions. We verify that the central charge thus computed is compatible with the holographic central charge, and additionally compute the flavor central charge. An important ingredient in the matching of the global symmetries is a St{\"u}ckelberg mechanism, in which one $U(1)$ generator of the isometry algebra of the internal space is spontaneously broken.   

In section \ref{sec:fieldtheory} we describe the proposed 4d $\CN=2$ field theories dual to our supergravity solutions, and perform tests of the holographic duality. The field theories are of Argyres-Douglas type, and arise from $N$ M5-branes wrapped on a sphere with one irregular puncture and one regular puncture. We test the duality by matching the $\CN=2$ R-symmetry generators, the large-$N$ central charge, the flavor central charge associated to the regular puncture, the rank of the flavor symmetry, and the field theory operators dual to M2-brane probes.

Finally, several appendices elaborate on derivations and ideas used in the main text. Appendix \ref{sec:appsugra} provides a full derivation of the 7d gauged supergravity solutions.  Appendix \ref{sec:appLLM} casts the uplifted 11d solutions into canonical $\mathcal{N}=2$ form. Appendix \ref{sec:appanomalies} collects useful formulae on the 't Hooft anomalies of 4d $\CN=2$ SCFTs. Appendices \ref{sec:AD} and \ref{sec:lagrangian} serve as select reviews of the literature on Argyres-Douglas theories: 
appendix \ref{sec:AD} gives an overview of the landscape of four-dimensional field theories of Argyres-Douglas type,  while appendix \ref{sec:lagrangian} reviews  the dual quiver Lagrangian description found in \cite{Agarwal:2017roi,Benvenuti:2017bpg} of a subclass of the AD theories dual to our supergravity solutions. 

A brief summary of some  results of the supergravity solutions and checks of the proposed duality
were first reported in 
 \cite{Bah:2021mzw}.

\section{Supergravity Solutions} \label{sec:spindlesolutions}

This section is devoted to a discussion of a new class of 11d supergravity 
$AdS_5$ solutions.
They are first obtained in 7d gauged supergravity and then uplifted to 
eleven dimensions.

\subsection{Solutions in 7d Supergravity} \label{sec_7dsols}

The reduction of 11d supergravity on $S^4$ yields the 7d $\cN = 4$ $SO(5)$ gauged supergravity of 
\cite{Pernici:1984xx}. In this work we consider a further truncation to the Cartan subgroup $U(1)^2$
of $SO(5)$. We follow the notation and conventions of \cite{Liu:1999ai}. The bosonic field content of the $U(1)^2$ truncated model
consists of the 7d metric $g_{\mu\nu}$, two real scalars $\lambda_1$, $\lambda_2$,
two $U(1)$ gauge fields $A^{(1)}_\mu$, $A^{(2)}_{\mu}$, and a real 3-form potential $C_{\mu\nu\rho}$.
(The indices $\mu$, $\nu$, \dots are curved 7d spacetime indices.)
The equations of motion and BPS equations for this supergravity model
are recorded in appendix \ref{app_EOMs}.
The mass scale of the  model is denoted $m$.
In our conventions, the $AdS_7$ vacuum solution  
has radius $L_{ AdS_7} = 2/m$.
The gauge coupling of the model is denoted $g$, and supersymmetry relates it to $m$ 
as $g = 2 \, m$.

As derived in appendix \ref{sec:appsugra}, the following bosonic field configurations
preserve 4d $\cN = 2$ superconformal symmetry and solve all 
 equations of motion.
The 7d metric is given by
\begin{align} \label{seven_metric}
m ^2 \, ds_7^2  & = \frac{2\,B \, w^{3/5}}{ \sqrt{\kappa \, (1-w^2)}} \, \bigg[
ds^2(AdS_5)+ds^2(\Sigma)
\bigg] \ , \nn \\ 
ds^2(\Sigma) &= \frac{dw^2}{2 \, w \, h(w) \, [ \kappa \, (1-w^2)]^{3/2}}
+ \frac{\cC^2 \, h(w) \,dz^2}{B} \ .
\end{align}
Here $ds^2(AdS_5)$ is the unit-radius metric in $AdS_5$,
$w$ is an interval coordinate whose range is discussed below, $z$ is an angular coordinate, $B$ is a positive constant, $\cC$ is a real constant, $\kappa \in \{ 1,-1\}$ is a sign,
and the function $h(w)$ is given by
\beq \label{h_def}
h(w) = B - 2 \, w \, \sqrt{\kappa \, (1-w^2)} \ .
\eeq
The scalar fields $\lambda_1$, $\lambda_2$ depend on the coordinate $w$ only
and are given as
\beq \label{scalar_fields}
\lambda_1 = \frac 35 \, \log w \ , \qquad
\lambda_2 = - \frac 25 \, \log w \ . 
\eeq
The gauge field $A^{(1)}$ 
has field strength $F^{(1)} = dA^{(1)}$ given by
\beq \label{gauge_1_expr}
F^{(1)} = -2 \, m^{-1} \, \cC \, w \, dw \wedge dz \ ,
\eeq
while the other gauge field $A^{(2)}$ and the 3-form potential $C$ are set to zero.
We observe that the angular coordinate $z$ enters the 7d metric and the field strength
$F^{(1)}$ always in the combination $\cC \, dz$. Without loss of generality we can then
assign periodicity $2\pi$ to  the coordinate $z$.


\begin{figure}
\centering
\includegraphics[width = 7 cm]{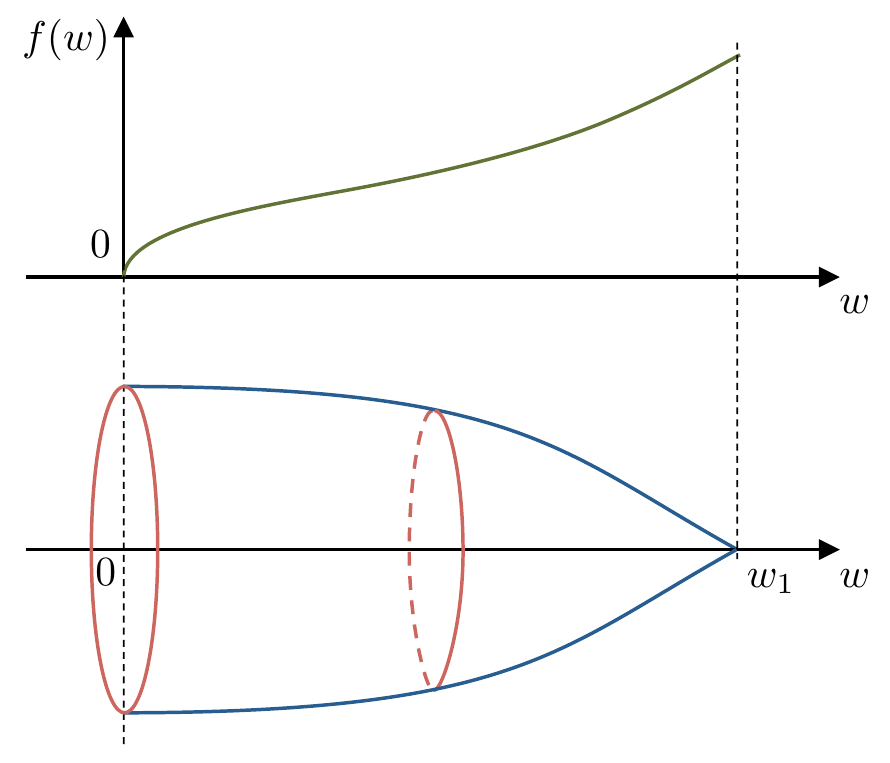}
\caption{
A schematic depiction of the internal geometry for the choice
of parameters and range of the $w$ coordinate specified in  \eqref{y_range_1}.  
The  $z$ circle is fibered  over the $w$ interval to yield $\Sigma$ with  metric
$ds^2(\Sigma)$ in 
\eqref{seven_metric}. $\Sigma$ has the topology of a disk with a $\mathbb Z_\ell$ orbifold
singularity at the center.
We also depict the qualitative behavior of the $AdS_5$ warp function $f(w) = 2\,B \, w^{3/5}/\sqrt{1-w^2}$.
}
\label{Case_I_fig}
\end{figure}

The range of the coordinate $w$ is constrained by requiring that $\lambda_1$, $\lambda_2$ be real
and the 7d metric positive-definite. Depending on the parameters $\kappa$ and $B$
there are various cases,   listed in appendix \ref{app_others}. The case of main interest
for this paper is
\beq \label{y_range_1}
\kappa = 1 \ , \qquad 0 < B < 1 \ , \qquad 0 < w < w_1 :=  \tfrac 12 \, \Big( \sqrt{1+B} -  \sqrt{1-B}\Big) \ .
\eeq
Let us describe the behavior of the metric near the two endpoints $w = w_1$
and $w = 0$ in turn.

In the vicinity of $w = w_1$, the $AdS_5$ warp factor is smooth,
and the $z$ circle shrinks. (The function $h$ has a simple zero at $w = w_1$.)
By tuning the constant parameter $\cC$ we can ensure that $z$ shrinks smoothly.
More generally, if we  impose
\beq \label{orbifold_condition}
|\mathcal C| =\frac{1}{\ell \, \sqrt{1-B^2}} \ , \qquad \ell = 1,2,3,\dots  \  , 
\eeq
the shrinking of the $z$ circle gives an  
 orbifold point $\mathbb R^2/\mathbb Z_\ell$. For more details,
see appendix~\ref{app_others}. 
Near $w = 0$ the $AdS_5$ warp factor vanishes and
the 7d metric
becomes conformal to the direct product of $AdS_5$ and a cylinder.
This can be seen setting $w = r^2$ and observing that
\beq
m^2 \, ds^2_7 \approx 2 \, B\, r^{6/5} \, \Big[  ds^2(AdS_5) + 2 \, B^{-1} \, dr^2 +  \cC^2\, dz^2  \Big] \ , \qquad
r \rightarrow 0^+ \ .
\eeq
The locus $r= 0$ is a curvature singularity of the total 7d metric.
Figure~\ref{Case_I_fig} gives a schematic depiction of $\Sigma$ and the $AdS_5$
warp factor.

The space $\Sigma$, equipped with the metric $ds^2(\Sigma)$ as in \eqref{seven_metric},
has the topology of a disk, with the origin at $w = w_1$ and the boundary at $w = 0$.
Indeed, the $z$ circle does not shrink at $w =0$ in the metric $ds^2(\Sigma)$.
As observed above, we have 
a $\mathbb Z_\ell$ orbifold singularity at the origin  of the disk $\Sigma$.
There exists  a gauge choice  
such that $A^{(1)}$ is well-defined  near $w = w_1$,
\beq \label{nice_gauge}
A^{(1)} = - m^{-1} \, \cC \, (w^2 - w_1^2) \, dz \ .
\eeq
Notice that we have fixed the ambiguity in $A^{(1)}$ by a shift
by a constant times $dz$ by requiring that the prefactor of $dz$ vanishes
at $w = w_1$. In this gauge, $A^{(1)}$ is globally defined on the disk $\Sigma$.
In appendix \ref{app_regularity} we verify that the Killing spinor on $\Sigma$ is also well-defined
near $w = w_1$, and is therefore globally defined on the disk $\Sigma$.

A brief digression about the normalization of $A^{(1)}$ is necessary.
To find the natural normalization, we observe that $A^{(1)}$
is identified with the $ab=12$ component of the field strength $F^{ab}$
of the full $SO(5)$ gauged supergravity model (the indices $a$, $b = 1,\dots,5$
are vector indices of $SO(5)$).
In the conventions of this paper---see also \cite{Pernici:1984xx,Liu:1999ai}---the expression for
$F^{ab}$ is $F^{ab} = dA^{ab} + g \, A^{ac} \wedge A_c{}^b$,
where $g = 2m$ is the gauge coupling constant of the supergravity theory.
It is natural to rescale $A^{ab}$ to eliminate the factor $g$ between the
linear and quadratic terms in the field strength:
we set
\beq
A^{ab} = \frac{1}{2m} \, \mathsf A^{ab} \ ,
\eeq 
so that the field strength of $\mathsf A^{ab}$ is $\mathsf F^{ab} = d\mathsf A^{ab}
+ \mathsf A^{ac}  \wedge \mathsf A_c{}^b$.
The rescaling of $A^{ab}$ induces an analogous rescaling of $A^{(1)}$,
\beq
A^{(1)} = \frac{1}{2m} \, \mathsf A^{(1)} \ .
\eeq
Since in the gauge \eqref{nice_gauge}  $\mathsf A^{(1)}$ is globally defined on the disk
$\Sigma$,  the flux of the field strength $\mathsf F^{(1)} = d \mathsf A^{(1)}$
through $\Sigma$ equals minus the holonomy of $\mathsf A^{(1)}$ along the boundary
at $w = 0$,
\begin{align} \label{good_flux}
{\rm hol}_{\partial \Sigma}(\mathsf A^{(1)}) : = \oint_{w=0} \frac{\mathsf A^{(1)}}{2\pi}
= - \int_\Sigma \frac{\mathsf F^{(1)}}{2\pi} =  2 \, \cC \, w_1^2
= \cC \, (1 - \sqrt{1-B^2}) \ .
\end{align}
We assign positive orientation to $dw \wedge dz$, with $w$ increasing from $0$ to $w_1$.

The parameters $B$, $\cC$ in the 7d solution
can be expressed in terms of the integer $\ell$ and the holonomy
${\rm hol}_{\partial \Sigma}(\mathsf A^{(1)})$, 
\beq
\cC = {\rm hol}_{\partial \Sigma}(\mathsf A^{(1)}) + \frac 1 \ell \ , \qquad 
\sqrt{1-B^2} = \frac{1}{1 + \ell \, {\rm hol}_{\partial \Sigma}(\mathsf A^{(1)})} \ .
\eeq
We have anticipated that $\cC$ is positive, which will be
verified when we perform the uplift to eleven dimensions in section \ref{sec_geometry}.
We think of $\ell$ and ${\rm hol}_{\partial \Sigma}(\mathsf A^{(1)})$
as the geometric and gauge-theoretic input data that specify the solution.
At this stage ${\rm hol}_{\partial \Sigma}(\mathsf A^{(1)})$ is an arbitrary real quantity.
We will see that, in the uplifted solutions, it is identified with the ratio
of two integer $G_4$-flux quanta.

Let us compute the Euler characteristic $\chi(\Sigma)$ of $\Sigma$ 
from the line element $ds^2(\Sigma)$ in \eqref{seven_metric} using the Gauss-Bonnet theorem,
following similar computations in \cite{Ferrero:2020laf,Ferrero:2020twa}.
A potential contribution originates from the boundary of $\Sigma$. One verifies,
however, that the boundary at $w=0$ is a geodetic in the metric $ds^2(\Sigma)$, and thus has vanishing geodetic curvature. As a result, the only contribution to $\chi(\Sigma)$
originates from integrating the Ricci scalar of $ds^2(\Sigma)$ against the volume form
of the metric $ds^2(\Sigma)$,
\beq
\chi(\Sigma) = \frac{1}{4\pi} \, \int_\Sigma R_\Sigma \, {\rm vol}_\Sigma = 
\frac{\sqrt 2 \, \sqrt w_1 \, \cC \, (1-2w_1^2) \, (1-w_1^2)^{1/4}}{\sqrt B}
=
\cC \, \sqrt{1-B^2} = \frac 1 \ell \ .
\eeq
This is the expected result for  a disk in $\mathbb R^2/\mathbb Z_\ell$
centered at the origin.\footnote{~This can be verified by equipping the disk with the flat metric of $\mathbb R^2/\mathbb Z_\ell$: in this case, the only non-zero contribution to the Euler characteristic comes from the geodetic curvature of the boundary of the disk.}

Our 7d solutions can be compared to the spindle
solutions of \cite{Ferrero:2020laf,Ferrero:2020twa,Hosseini:2021fge,Boido:2021szx,Ferrero:2021wvk}. 
As in those references, the 2d space $\Sigma$ is not equipped with a constant curvature metric.
The gauge field $A^{(1)}$ does not cancel the spin connection on $\Sigma$,
and the Killing spinor has a non-trivial profile in the $w$ direction
(its explicit expression is recorded in \eqref{final_spinor}).
These features signal that supersymmetry is realized in a way that deviates
from the standard topological twist paradigm.
In contrast to  \cite{Ferrero:2020laf,Ferrero:2020twa,Hosseini:2021fge,Boido:2021szx,Ferrero:2021wvk}, however, our internal space $\Sigma$ has the topology of a disk,
with a non-trivial holonomy of the gauge field $A^{(1)}$ along its boundary.
This is qualitatively different from the spindle geometries.
Our $\Sigma$ may be intuitively thought of as a ``half spindle''
and leads to a 
 new way of realizing supersymmetry.

\subsection{Uplift to Eleven Dimensions} \label{sec:11dsols}

The uplift on $S^4$ of solutions 
to the 7d $U(1)^2$ gauged supergravity model considered above
has been analyzed in \cite{Cvetic:1999xp}.
To perform the uplift, we find it convenient to 
 make use of the  formulae in \cite{Nastase:1999kf}. 
It is useful to keep in mind that 
the authors of  \cite{Nastase:1999kf} set implicitly $m=1$; it is straightforward to restore
factors of $m$ in their expressions.
The 11d metric is given as
\beq \label{uplift_the_metric}
ds_{11}^2 = (T_{ab} \, Y^a \, Y^b)^{1/3} \, ds^2_7
+ m^{-2} \,  (T_{ab} \, Y^a \, Y^b)^{-2/3}   \, (T^{-1})_{ab} \, DY^a \, DY^b \ .
\eeq
The indices $a,b=1,\dots,5$ are $SO(5)$ indices
and are raised/lowered with $\delta$.
The quantities $Y^a$ are constrained coordinates on $S^4$, satisfying
$Y^a \, Y_a = 1$.
The symmetric, unimodular matrix $T_{ab}$ is constructed with the 
scalar fields $\lambda_1$, $\lambda_2$ as
\beq
T_{ab}  = {\rm diag}( e^{2\lambda_1} , e^{2\lambda_1} , e^{2\lambda_2 } , e^{2\lambda_2} , e^{-4 \lambda_1 - 4 \lambda_2}) \ .
\eeq
The 1-form $DY^a$ is defined as 
\beq
DY^a =  dY^a + g \, A^{ab} \, Y_b \ , 
\eeq
(recall that $g = 2 \, m$)
where $A^{[ab]}$ is an $SO(5)$ connection with legs on 7d spacetime.
Its only non-zero components are
\beq
A^{12} = A^{(1)} \ , \qquad A^{34}  = A^{(2)} \ .
\eeq
The expression for $G_4$ is
\begin{align} \label{uplift_the_flux}
G_4   =  \frac{1}{8 \, m^3} \, \epsilon_{abcde} \, \bigg[ 
&\frac 43 \, DY^a \, DY^b \, DY^c \, DY^d \, Y^e
- \frac 13 \, DY^a \, DY^b \, DY^c \, DY^d \, \widetilde Y^e
\nn \\
& + 2 \, g \, F^{ab} \, DY^c \, DY^d \, \widetilde Y^e
+ g^2 \, F^{ab} \, F^{cd} \, Y^e \bigg] + dC_3 \ ,
\end{align}
where $C_3$ is the 3-form potential of the 7d supergravity model.
We have suppressed wedge products and we have used the quantities
\beq
F^{ab} = dA^{ab} + g \, A^{ac} \, A_c{}^b  \ , \qquad
\widetilde Y^a := \frac{T^{ab} \, Y_b}{T_{cd} \,Y^c \, Y^d} \ .
\eeq

We parametrize the constrained coordinates $Y^a$ as
\beq
Y^1 = \sqrt{1-\mu^2} \, \cos \phi \ , \qquad
Y^2 = \sqrt{1-\mu^2} \, \sin \phi  \ , \qquad
Y^{3,4,5} = \mu \, \hat y^{1,2,3} \ ,
\eeq
where the three real coordinates $\hat y^{1,2,3}$ are subject to the constraint
$(\hat y^1)^2 + (\hat y^2)^2 + (\hat y^3)^2 = 1$ and thus parametrize an $S^2 \subset \mathbb R^3$.
The coordinate $\mu$ has range $[0,1]$ and the angular coordinate $\phi$ has periodicity $2\pi$.
Using the 7d line element \eqref{seven_metric},
the 7d scalar fields \eqref{scalar_fields}, and the 7d gauge field \eqref{gauge_1_expr},
 the uplift formula
\eqref{uplift_the_metric} yields
\begin{align} \label{11d_metric}
m^2 \, ds^2_{11} & =\frac{2 \, B\, w^{1/3} \,  \cH(w,\mu)^{1/3}}{\sqrt{  \kappa \, (1-w^2)} }   \, \bigg[
 ds^2(AdS_5) 
+ \frac{dw^2}{2 \, w \, h(w) \, [ \kappa \, (1-w^2)]^{3/2}}
+ \frac{\cC^2 \, h(w) \,dz^2}{B} 
\nn \\
& +  \frac{\sqrt{\kappa \, (1-w^2)}}{2 \, B} \, \bigg(  \frac{d\mu^2}{w\,(1-\mu^2)}  + \frac{(1-\mu^2)\, D\phi^2}{w\, \cH(w,\mu)}\, 
+ \frac{w \, \mu^2 \, ds^2(S^2)}{\cH(w,\mu)}  \bigg)
\bigg] \ .
\end{align}
We have introduced the notation
\beq \label{shorthand}
\cH(w,\mu) =  \mu^2 + w^2 \, (1-\mu^2)     \ .
\eeq
The function $h(w)$ was defined in \eqref{h_def}.
The quantity $ds^2(S^2)$ is the metric on the round unit 2-sphere
parametrized by $\hat y^{1,2,3}$, while the 1-form $D\phi$ is given as\footnote{~The 1-form
$D\phi$ is computed in the gauge $A^{(1)} = - m^{-1} \, \cC \, (w^2 - \tfrac 12)$,
which differs from \eqref{nice_gauge}. As explained in appendix \ref{app_regularity},
in the gauge \eqref{nice_gauge} the 7d Killing spinor $\eta$ depends on $z$ via the phase
factor $e^{\frac{iz}{2\ell}}$. Using the combined transformation
of $A^{(1)}$ and $\eta$ recorded in \eqref{spinor_and_gauge},
one verifies that, in the new gauge $A^{(1)} = - m^{-1} \, \cC \, (w^2 - \tfrac 12)$, the spinor 
$\eta$ is independent of $z$. 
A different choice of gauge is equivalent to a redefinition of $\phi$, $z$ of the   form
$(\phi,z) \mapsto (\phi + c \, z, z)$, where $c$ is a constant.
We also notice that the choice of gauge for 
$A^{(1)}$ does not affect the coefficient of $\partial_z$ in $\partial_\chi$
in equation \eqref{Rsym_and_flavor} below, which is the data that is mapped to the field theory side in \eqref{eq:translate}. 
}
\beq \label{Dphi_def}
D\phi = d\phi + \cC \, (2 \, w^2 -1) \, dz \ .
\eeq
The expression for $G_4$ that follows from \eqref{uplift_the_flux} is
\beq \label{G4flux}
G_4 =- \frac{1}{ m^3} \, {\rm vol}_{S^2} \, d\bigg[
\frac{ \mu^3}{\mu^2 + w^2 \, (1-\mu^2)} \, D\phi
\bigg] \ ,
\eeq
where ${\rm vol}_{S^2}$ is the volume form on the 2-sphere of unit radius.

\subsection{Internal Geometry and Flux Quantization} \label{sec_geometry}

For the rest of this section we specialize to the choice of parameters
and range for $w$ given in \eqref{y_range_1}. The other possibilities discussed in appendix
\ref{app_others} may also be uplifted to eleven dimensions and discussed in a similar fashion.

\subsubsection{Geometry of the Internal Space}

\begin{figure}
\centering
\includegraphics[width = 8.5 cm]{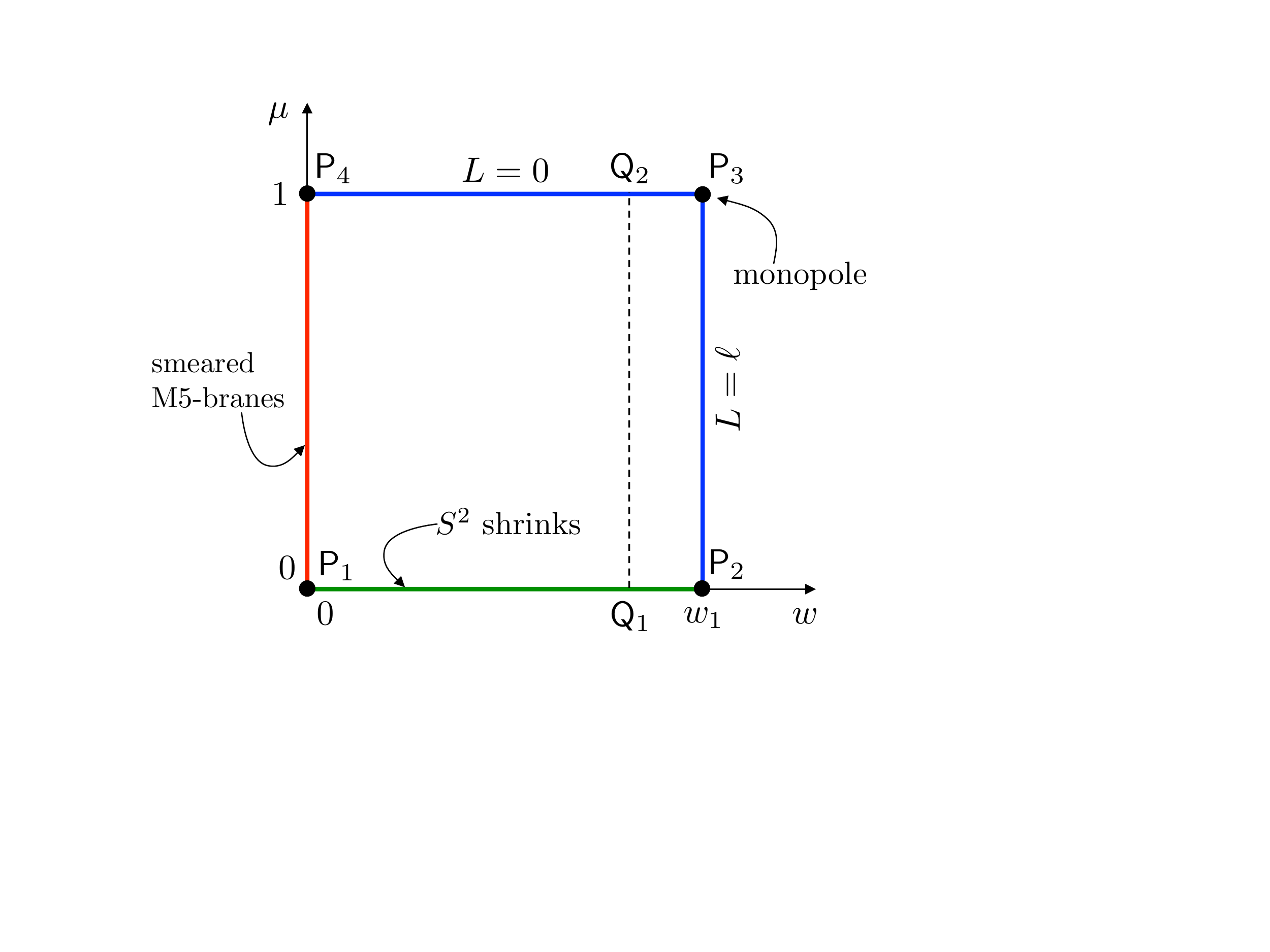}
\caption{
The internal space in the 11d solution is an $S^1_\phi \times S^1_z \times S^2$ fibration
over the rectangle in the $(w,\mu)$ plane delimited by the points $\mathsf P_1$, $\mathsf P_2$,
$\mathsf P_3$, $\mathsf P_4$.
We indicate the constant value of the function $L$ 
in the line element \eqref{puncture_metric} on the sides 
$\mathsf P_2 \mathsf P_3$ and $\mathsf P_3 \mathsf P_4$.
The point $\mathsf P_3$ is the location of a monopole of charge $\ell$ for the $Dz$ fibration
in \eqref{puncture_metric}.
The warp factor vanishes along the side $\mathsf P_1 \mathsf P_4$.
The geometry in this region   is interpreted in terms of  smeared M5-branes.
We also include the segment $\mathsf Q_1 \mathsf Q_2$ which is
used in section \ref{sec:G4flux} in the discussion of $G_4$-flux quantization.
}
\label{rectangle}
\end{figure}

The 6d internal space in the 11d line element \eqref{11d_metric} can be regarded as an
$S^1_\phi \times S^1_z\times S^2$ fibration over the 2d base space $B_2$ parametrized by
$w$ and $\mu$, which is the rectangle $[0,w_1] \times [0,1]$, see Figure \ref{rectangle}.
Let us describe in greater detail the features of the internal geometry
near the following three regions of the boundary of the rectangle  $B_2$:
\begin{itemize}
\item Region I: a neighborhood of the side $\mathsf P_1 \mathsf P_2$ (depicted in green).
\item Region II: a neighborhood of the union
of the sides $\mathsf P_2 \mathsf P_3$ and $\mathsf P_3 \mathsf P_4$  (depicted in blue). 
\item Region III: a neighborhood of the side $\mathsf P_1 \mathsf P_4$   (depicted in red).
\end{itemize}

\paragraph{Geometry of Region I.}
As we approach a point along the $\{ \mu=0\}$ side of the rectangle $B_2$, at generic $w\in (0,w_1)$,
the $S^2$ shrinks smoothly, capping off the internal space.
Both Killing vector fields $\partial_z$ and $\partial_\phi$ have a finite norm
as we approach $\mu = 0$.

\paragraph{Geometry of Region II: Regular Puncture.}
To describe the geometry of this region
we make use of the angular coordinates $\phi$, $z$,
but we break up the 1-form $D\phi$ and
complete instead the $dz$ square.
The resulting line element takes the form
\begin{align} \label{puncture_metric}
ds^2_{11} & = \frac{2\, B \,w^{1/3}\, \cH^{1/3}}{m^2 \, \sqrt{1-w^2}} \, \bigg[ ds^2(AdS_5)
+ \frac{\sqrt{1-w^2} \, \mu^2 \, w \, ds^2(S^2)}{2 \, B \, \cH}
  \\
& + \frac{ dw^2}{2 \, h \, w \, (1-w^2)^{3/2}}
+ \frac{\sqrt{1-w^2} \, d\mu^2}{2 \, B \, w \, (1-\mu^2)}
+ R_{\phi}^2 \, d \phi^2 + R_z^2 \, Dz^2   \bigg] \ , \qquad
Dz = dz - L \, d\phi \ . \nn
\end{align}
The function $h$ is defined in \eqref{h_def},
while $\cH$ is defined in \eqref{shorthand}.
The metric functions $R_{  \phi}^2$,
$R_z^2$ and the function $L$ inside $Dz$ are given as
\begin{align}
R_{  \phi}^2 & =
 \frac{h \, (1-\mu^2) \, \sqrt{1-w^2} }{B \, \Big[ 2 \, h \, w\, \cH 
+ (2w^2 -1)^2 \, (1-\mu^2) \, \sqrt{1-w^2} \Big] } 
\ , \nn \\
R_z^2 & = \frac{2 \, h \, w\, \cH + (2w^2-1)^2 \, (1-\mu^2) \, \sqrt{1-w^2}  }{2 \, B \, w\, \cH} \, \cC^2 \ , \nn \\
L &= \frac{ - (2w^2-1) \, (1-\mu^2) \, \sqrt{1-w^2} }{\cC \, \Big[ 2 \, h \, w\, \cH 
+ (2w^2 -1)^2 \, (1-\mu^2) \, \sqrt{1-w^2} \Big] }  \ .
\end{align}
We are  describing the internal space in terms of $S^2$ and 
the 4d space spanned by $w$, $\mu$, $\phi$, $z$.
The latter is an $S^1_z$ fibration over the 3d space
spanned by $w$, $\mu$, $\phi$.
This description is modeled after \cite{Gaiotto:2009gz}
and the local   geometries 
that describe regular punctures for M5-branes wrapped on a Riemann surface 
\cite{Bah:2018jrv, Bah:2019jts}.
The $Dz$ fibration over 
$w$, $\mu$, $\phi$ is a convenient device to
keep track of the two different linear combinations of the Killing vectors
$\partial_\phi$, $\partial_z$ whose norms go to zero on the two sides
$\{ w =w_1\}$ and $\{ \mu =1 \}$ of the rectangle $B_2$.

We observe that $R_{  \phi}^2$ is the radius squared
of the $\phi$ circle in the 3d base,
and that it goes to zero both along $\mu = 1$ and $w = w_1$.
More precisely, one can verify that
\begin{align}
\mu &= 1- \varrho^2 \ , & \varrho & \rightarrow 0^+ \ , &
\frac{\sqrt{1-w^2} \, d\mu^2}{2 \, B \, w \, (1-\mu^2)}
+ R_{  \phi}^2 \, d  \phi^2
&\approx \frac{\sqrt{1-w^2}}{2 \, B \, w} \, (d\varrho^2 + \varrho^2 \, d  \phi^2) \ ,   \\
w &= w_1- \varrho^2 \ , &  \varrho & \rightarrow 0^+ \ , &
\frac{\sqrt{1-w^2} \, dw^2}{2 \, h \, w \, (1-w^2)^2}
+ R_{  \phi}^2 \, d  \phi^2
&\approx 
\frac{2 \, (1-w_1^2)^{-3/2}}{w_1 \, (-h'(w_1))} \, (d\varrho^2 + \varrho^2 \, d  \phi^2) \ .    \nn
\end{align}
These relations demonstrate that, in the 3d base of the $Dz$ fibration,
the shrinking of the $  \phi$ circle is smooth.
The 3d base space is thus locally $\mathbb R^3$ in the vicinity of the boundary of $B_2$,
with $  \phi$ playing the role of an azimuthal angle in cylindrical coordinates.
The radius squared $R^2_z$ of the $z$ circle,
on the other hand, is only zero at the corner $(w,\mu) = (w_1,1)$.

The function $L(w,\mu)$
is piecewise constant along the sides $\{ w = w_1\}$ and $\{ \mu = 1 \}$
of the rectangle $B_2$. More precisely, one finds
\beq
L(w, 1) = 0 \ , \qquad 
L(w_1, \mu) =    \frac{1}{\cC \, \sqrt{1-B^2}} \ .
\eeq
The jump in $L$ at the corner $(w,\mu) = (w_1,1)$ signals
the presence of a monopole source 
for the $Dz$ fibration.
The monopole charge must be an integer.
We find it convenient to adopt the same orientation
conventions as in the   discussion of the local puncture geometries of \cite{Bah:2019jts}.
In particular, the function $L$ is non-negative and decreasing as we move 
along the axis of the $\mathbb R^3$ fiber (spanned by $w$, $\mu$, $\phi$),
starting from the point where the $S^2$ shrinks (point $\mathsf P_2$ in Figure \ref{rectangle})
and  moving upwards towards $\mathsf P_3$ and then past the monopole
towards $\mathsf P_4$. These considerations imply
that $\cC$ is positive, so that
\beq \label{relation_with_ell}
  \frac{1}{\cC \, \sqrt{1-B^2}} = \ell \ , \qquad \ell = 1,2,3, \dots \ .
\eeq
The integral quantization of the monopole 
charge can also be 
 confirmed by a local analysis of the metric near
the corner $(w,\mu) = (w_1,1)$.
More precisely, 
we  trade $w$, $\mu$ for coordinates $R>0$ and $\theta\in [0,\pi]$ defined via
\beq
\mu = 1 - \frac{w_1^{2/3}}{  2} \, R^2 \, \cos^2 \frac \theta 2  \  , \qquad
w = w_1 -  \frac{  (1-B^2)^{1/2} \, w_1^{5/3} \, (1-w_1^2)^2 }{B^2} \, R^2 \, \sin ^2 \frac \theta 2 \ .
\eeq
In the limit $R \rightarrow 0$, the 11d line element reads
\begin{align}
m^2 \, ds^2_{11} & \approx  4 \, w_1^{4/3} \, ds^2(AdS_5)
+ w_1^{4/3} \, ds^2(S^2) \nn \\
& + dR^2 + R^2 \, \bigg\{  
\frac{d\theta^2 + \sin^2 \theta \, d\phi^2}{4}
+  \cC^2 \, (1-B^2)  \, \bigg[ dz  - \frac{1 +\cos \theta}{2\, \cC \, \sqrt{1-B^2}} \, d\phi \bigg]^2
\bigg\} \ .
\end{align}
If $ \cC  \, \sqrt{1-B^2} =1$, the line element in curly
brackets in the second line is a round $S^3$ presented as a standard
Hopf fibration. 
The Hopf fiber is parametrized by $z$ with period $2\pi$,
while the Hopf base is spanned by $\theta \in [0,\pi]$ and $\phi$ with periodicity $2\pi$.
More generally, we can allow the quantity 
$ \cC  \, \sqrt{1-B^2}$ to be $1/\ell$ for any positive integer $\ell$,
as indicated in \eqref{relation_with_ell}.
The quantity in curly brackets is then the metric on $S^3/\mathbb Z_\ell$.
When the latter is combined with the radial direction $R$,
we obtain the metric on $\mathbb R^4/\mathbb Z_\ell$.
Thus, for $\ell >1$ the geometry in Region II has an orbifold
singularity at the   location of the monopole, and is smooth elsewhere.
We have   demonstrated that the relation 
\eqref{orbifold_condition} 
in the 7d gauged supergravity solution
is reinterpreted in the uplifted 11d solution as the quantization
of a monopole charge.

\paragraph{Geometry of Region III: Smeared M5-branes.}
This region requires special care because the warp factor in front of the $AdS_5$
metric goes to zero as $w$ approaches 0. 
The 11d line element  can be approximated at small $w$ as
\begin{align}
m^2 \, ds^2_{11} & \approx 
w^{1/3} \, \bigg[ 2 \, B   \, \mu^{2/3} \, ds^2(AdS_5) 
+ 2 \, \cC^2 \, B   \, \mu^{2/3} \, dz^2 \bigg]
\nn \\
& + w^{-2/3} \, \bigg[ \mu^{2/3} \, \Big (dw^2   
+    w^2 \, ds^2(S^2)  \Big )
+  \frac{ \mu^{2/3}}{ 1-\mu^2 } \, d\mu^2
+  \mu^{-4/3} \,  (1-\mu^2) \, D\phi^2  \bigg] \ .
\end{align}
This line element is interpreted as originating from smeared M5-brane sources.
More precisely, the M5-branes are:
\begin{itemize}
\item extended along the $AdS_5$ and $z$ directions;
\item localized at the origin of the $\mathbb R^3$ parametrized by $S^2$ and $w$, $ds^2(\mathbb R^3) = dw^2 + w^2 \, ds^2(S^2)$;
\item smeared along the $\mu$ and $\phi$ directions.
\end{itemize}
After smearing, the branes are effectively real codimension-3 objects.
Notice that $w$ is identified with the radial coordinate away from the smeared branes.
The relevant harmonic function for a real codimension-3 problem is $H \propto 1/w$.
As appropriate for an M5-brane solution, we find a prefactor $H^{-1/3}$ in front
of the six directions along which the M5-branes extend,
while we find a factor $H^{2/3}$ in front of the five directions in which the branes
are localized or smeared.

We can confirm the presence of an M5-brane source from
the expression of $G_4$ near $w = 0$,
\begin{align} \label{local_G4}
G_4 =   - \frac{{\rm vol}_{S^2} \wedge d \mu \wedge D\phi }{m^3} + \dots \ , \qquad w \rightarrow 0 \ .
\end{align}
In particular, the integral of the RHS along the $S^2$, $\mu$, $\phi$ directions
is finite as $w \rightarrow 0$. This signals the presence of a source of the 
schematic form
$dG_4 \sim \delta(w) \, dw \wedge {\rm vol}_{S^2} \wedge d \mu \wedge D\phi$.
The total charge of the source is 
computed integrating 
\eqref{local_G4} and is equal to the flux quantum $N$ defined below  in \eqref{N_definition},
which is identified with the number of M5-branes on the stack
wrapping $\Sigma$.

\subsubsection{$G_4$-Flux Quantization} \label{sec:G4flux}

In our conventions for the normalization of $G_4$ in 11d supergravity,
the quantity that has integrally quantized fluxes is $G_4/(2\pi \ell_p)^3$,
where $\ell_p$ is the 11d Planck length.
We find it convenient to define
\beq \label{G4_rescaling}
\overline G_4 =  - \frac{G_4}{(2 \pi \ell_p)^3} \ ,
\eeq
with the sign chosen for future convenience.
The integral of the quantity $\overline G_4$ over any 4-cycle in the internal
space must be an integer.

In the discussion of the non-trivial 4-cycles in the internal geometry
it is convenient to use the presentation \eqref{puncture_metric}
and to make contact with the analysis of \cite{Bah:2018jrv, Bah:2019jts} (see also \cite{Gaiotto:2009gz}).
To this end, let us express $\overline G_4$ in terms of $d\phi$ and $Dz$,
using \eqref{G4flux} and the definition of $Dz$ in \eqref{puncture_metric}.
We find
\beq
\overline G_4 = \frac{{\rm vol}_{S^2}}{4\pi} \wedge d\bigg[ 
Y \, \frac{d\phi}{2\pi} - W \, \frac{Dz}{2\pi}
\bigg] \ ,
\eeq
where the 0-forms $Y$ and $W$ are given as
\beq
Y  = \frac{1}{\pi m^3 \ell_p^3} \, \frac{\Big[ 1 + \cC \, L \, (2w^2-1) \Big] \, \mu^3}{ \mu^2 + w^2 \, (1-\mu^2) }
\ , \qquad
W = \frac{1}{\pi m^3 \ell_p^3} \, \frac{\cC \, (1-2w^2) \, \mu^3}{ \mu^2 + w^2 \, (1-\mu^2) } \ .
\eeq
The function $Y(w,\mu)$
is piecewise constant along the $\mathsf P_2 \mathsf P_3$ and $\mathsf P_3 \mathsf P_4$
segments: $Y(w_1,\mu) = 0$, $Y(w,1) = 1/(\pi m^3 \ell_p^3)$.
These properties of $Y$ are in line with the general analysis of~\cite{Bah:2018jrv, Bah:2019jts}.

A first non-trivial 4-cycle, which we denote $\cC_4$,
is obtained  by considering the segment $\mathsf Q_1 \mathsf Q_2$
(see Figure \ref{rectangle})
and combining it with the $S^2$ and with the circle that shrinks
along the $\mathsf P_3 \mathsf P_4$ segment.
As we have seen above, the latter is 
the $d\phi$ circle in the base of the $Dz$ fibration.
Since along the segment $\mathsf P_3 \mathsf P_4$ we have $L = 0$,
$Dz = dz$ and the shrinking circle is simply $d\phi$.
The 4-cycle $\cC_4$ has the topology of $S^4$ and we identify it
with the $S^4$ fiber on top of a generic point on $\Sigma$
spanned by $w$, $z$.
Having defined $\cC_4$, we can now compute
\beq \label{N_definition}
\int_{\cC_4} \overline G_4 = \int_{\cC_4} 
\frac{{\rm vol}_{S^2}}{4\pi} \wedge d( Y + L\,W) \wedge  
\frac{d\phi}{2\pi}   
= (Y + L\, W) \Big|_{\mathsf Q_1}^{\mathsf Q_2} = \frac{1}{\pi m^3 \ell_p^3} 
=: N \in \mathbb N \ .
\eeq
We have assigned positive orientation to $d\mu \wedge d\phi$.
The positive integer $N$ is identified with the number of M5-branes on the stack wrapping  
$\Sigma$.

A different 4-cycle, denoted $\cB_4$, can be constructed as follows.
Let us consider the segment $\mathsf P_2 \mathsf P_3$
and let us combine it with $S^2$ and the $Dz$ fiber.
We get a 4-cycle because the $S^2$ shrinks as we approach $\mathsf P_2$,
while the radius of $Dz$ goes to zero as we approach the monopole location at $\mathsf P_3$. 
The flux through $\cB_4$ is
\begin{align}
\int_{\cB_4} \overline G_4 &= - \int_{\cB_4} 
\frac{{\rm vol}_{S^2}}{4\pi} \wedge dW \wedge  
\frac{Dz}{2\pi}   
= W  \Big|_{\mathsf P_2}^{\mathsf P_3}    =  \frac{\cC \, (1-2w_1^2)}{\pi m^3 \ell_p^3} 
=   \frac{\cC \, \sqrt{1-B^2}}{\pi m^3 \ell_p^3}  
= \frac N \ell     \ .
\end{align}
We have used 
\eqref{N_definition}, \eqref{relation_with_ell} and we have
chosen the orientation of $\cB_4$
in such a way that $\cB_4 \cong \cC_4$ in the case $\ell = 1$.
For $\ell >1$ the 4-cycles $\cB_4$ and $\cC_4$ are inequivalent.
Flux quantization through $\cB_4$ demonstrates that
$N$ must be a multiple of $\ell$,
\beq
\frac N \ell \in \mathbb N \ .
\eeq

Finally, let us consider the 4-cycle $\cD_4$,
which is the analog of the 4-cycle $\cB_4$
based on the segment $\mathsf P_3 \mathsf P_4$.
More precisely, we combine this segment with the $Dz$ fiber and the $S^2$.
We know that $Dz$ shrinks at $\mathsf P_3$.
The total radius of the $S^2$ in the 11d metric 
goes to zero as we approach $\mathsf P_4$,
because of the vanishing of the warp factor.
The flux through $\cD_4$ is
\begin{align}
\int_{\cD_4} \overline G_4 & = - \int_{\cD_4} 
\frac{{\rm vol}_{S^2}}{4\pi} \wedge dW \wedge  
\frac{Dz}{2\pi}   
= W  \Big|_{\mathsf P_3}^{\mathsf P_4}   
 =    \frac{\cC \,  2w_1^2 }{\pi m^3 \ell_p^3}
 = \frac N \ell \, \frac{1-\sqrt{1-B^2}}{\sqrt{1-B^2}}   =: K  \in \mathbb N \  . 
\end{align}
We observe that the 4-cycle $\cD_4$ leads to a novel
integral flux $K$, which is positive because $0<B<1$.

In summary, 
the topology and flux configuration of the solutions
we are studying are encoded in three positive integers:
$\ell$, $N$, and $K$.
Moreover, $\ell$ divides $N$.
The constant parameters $B$, $\cC$ can we written in terms of $\ell$, $N$, $K$ as
\beq \label{in_terms_of_integers}
\sqrt{1-B^2} = \frac{N}{N + K \, \ell} \ , \qquad
\cC =   \frac{N + K \, \ell}{N \, \ell} \ .
\eeq
Using these identifications, we can revisit the
expression \eqref{good_flux} for the flux of $\mathsf F^{(1)}$ on $\Sigma$,
which is also equal to the monodromy of $\mathsf A^{(1)}$ at $w= 0$
(in the gauge \eqref{nice_gauge} in which $\mathsf A^{(1)}$ is globally defined on the disk $\Sigma$),
\begin{align}  
\oint_{w=0} \frac{\mathsf A^{(1)}}{2\pi} = - \int_\Sigma \frac{\mathsf F^{(1)}}{2\pi} = \frac K N  \ .
\end{align}
As anticipated, this 7d holonomy is identified with the ratio between two integer flux quanta
in eleven dimensions.

\subsubsection{11d Solutions in Canonical $\cN = 2$ Form}
The most general $AdS_5$ solution of 11d supergravity preserving 4d $\cN = 2$ superconformal
symmetry was characterized in Lin-Lunin-Maldacena (LLM) \cite{Lin:2004nb}. The 11d metric
and flux are given as  \cite{Gaiotto:2009gz}
\begin{align} \label{LLM}
ds^2_{11} & =  \frac{e^{2 \widetilde \lambda}}{m^2} \, \bigg[
  ds^2(AdS_5)  + \frac{y^2  \, e^{- 6 \widetilde \lambda}}{4} \, ds^2(S^2)
+ \frac{ D\chi^2}{1 - y \, \partial_y D}
+ \frac{- \partial_yD}{4\,y} \, \Big( dy^2 + e^D \, ((dx^1)^2 + (dx^2)^2 )\Big) \bigg] \ ,  \nn \\
G_4 & = \frac{1}{4\, m^3}\, {\rm vol}_{S^2} \wedge \bigg[
D\chi \wedge d(y^3 \, e^{- 6 \widetilde \lambda})
+ y \, (1 - y^2  \, e^{-6\widetilde \lambda}) \, dv
- \frac 12 \, \partial_y e^D \, dx^1 \wedge dx^2
\bigg] \ .
\end{align}
The line elements on $AdS_5$ and $S^2$ have unit radius.
The warp factor $\widetilde \lambda$ and the function $D$ depend on $y$,
$x^1$, $x^2$ and are related by
\beq \label{lambda_and_D}
e^{- 6\widetilde \lambda} = \frac{- \partial_y D}{y \, (1 - y \, \partial_y D)} \ .
\eeq
The function $D$ satisfies the Toda equation
\beq \label{Toda_equation}
\partial_{x^1}^2 D + \partial_{x^2}^2 D + \partial_y^2 e^D =  0 \ .
\eeq
The coordinate $\chi$ is an angular coordinate with period $2\pi$. The 1-form
$D\chi$ is defined as
\beq \label{v_def}
D\chi = d\chi + v \ , \qquad v =  - \frac 12 \, \Big(
 \partial_{x^1} D \, dx^2 - \partial_{x^2} D \, dx^1 \Big)   \ .
\eeq
The 2-form  ${\rm vol}_{S^2}$ is the   volume form
on a unit-radius round $S^2$. The Killing vector $\partial_\chi$
is dual to the $U(1)_r$ R-symmetry of the 4d $\cN =2$ SCFT,
while the isometries of $S^2$ are mapped to the $SU(2)_R$ R-symmetry.

The 11d solutions presented in section \ref{sec:11dsols}
can be cast into the canonical LLM form \eqref{LLM}.
Let us summarize here the salient feature of this match,
referring the reader to appendix \ref{app_LLM_change} for more details.
It is useful to introduce polar coordinates $r$, $\beta$
on the $x_1$, $x_2$ plane,
\beq \label{LLMpolar}
x^1 = r \, \cos \beta \ , \qquad x^2 = r \, \sin \beta \ .
\eeq
The angular coordinates $\chi$, $\beta$
are related to the angular coordinates $\phi$, $z$ in \eqref{11d_metric}
as
\beq \label{angular_vars}
\chi = \bigg( 1 + \frac 1 \cC \bigg) \,  \phi -  z \ , \qquad
\beta = - \frac 1 \cC \, \phi + z \ .
\eeq
The coordinates $y$ and $r$ are given in terms of $w$ and $\mu$ as
\begin{align} \label{y_and_r}
y = \frac{4\, B \, w\, \mu}{  \sqrt{\kappa\, (1-w^2) }  } \ , \qquad
r = (1-\mu^2)^{-\frac{1}{2\cC} } \, \cG(w) \ ,
\end{align}
where the function $\cG(w)$ is given in \eqref{explicit_G}.
The quantity $D$, expressed in terms of $w$ and $\mu$, is given as
\beq \label{expD}
e^D =
\frac{16\, B  \,  \cC^2 \,  \left(1-\mu ^2\right)^{1 +1 / \cC} \,  \left [ B - 2  \,  w \,  \sqrt{\kappa \,(1  - 
   w^2) }\right] }{     \kappa \,\left(1 - w^2\right)   \, \cG(w)^2 }  \ .
\eeq
Using the expressions of $y$, $r$, $D$ in terms of $w$ and $\mu$
and the properties of the function $\cG(w)$,
one can verify that the Toda equation for $D$ is satisfied.
Finally, we have checked explicitly that the expression \eqref{LLM} for $G_4$ 
matches with \eqref{G4flux}.

The formulae presented above apply to any choice of the sign $\kappa$
and range of $w$. Let us end this section with some remarks that apply to the case of interest
\eqref{y_range_1}.
Combining the relations \eqref{angular_vars}
with 
 \eqref{in_terms_of_integers},
we can write
 \beq \label{Rsym_and_flavor}
\partial_\chi = \partial_\phi  + \frac{N \, \ell }{N +  K \, \ell }   \, \partial_z \ , \qquad
\partial_\beta = \partial_\phi + \bigg[ 
1 + \frac{ N \, \ell }{N +  K \, \ell }
\bigg]  \, \partial_z \ .
\eeq
The $U(1)_r$ superconformal R-symmetry
is given by a non-trivial mixing between the $\partial_z$ isometry direction
on $\Sigma$ and the $\partial_\phi$ isometry of the topological $S^4$ fiber on top
of a generic point on $\Sigma$.
The Killing vector $\partial_\beta$ is naively associated to a $U(1)$ flavor symmetry of the SCFT.
As we will see in section \ref{sec:anomaly}, however, 
this expectation is incorrect.

\subsection{Holographic Central Charge and Supersymmetric Wrapped M2-branes} \label{sec:holo_central}

This subsection is devoted to the analysis of two holographic observables.
We consider the choice of parameters and range of $w$ 
specified in \eqref{y_range_1}.
Firstly, we extract the holographic central charge from the (warped) volume of the
internal space.  
Secondly, we study probe M2-branes wrapping calibrated 2-submanifolds in the internal space.

\subsubsection{Holographic Central Charge}

As already observed in \eqref{LLM},
the 
11d line element is conveniently parametrized as
\beq \label{11d_general_param}
ds^2_{11} =\frac{ e^{2 \widetilde \lambda} }{m^2} \, \Big[ ds^2(AdS_5) + ds^2(M_6) \Big] \ ,
\eeq
where   $AdS_5$ has unit radius  and $\widetilde \lambda$
is the warp factor.
With this notation, the holographic central charge reads \cite{Gauntlett:2006ai}
\beq
c = \frac{1}{2^7   \pi^6 m^9 \ell_p^9} \,\int_{M_6} e^{9 \widetilde \lambda} \,  {\rm vol}_{M_6} \ ,
\eeq
where ${\rm vol}_{M_6}$ is the volume form of the metric $ds^2(M_6)$.
For our solutions, we 
extract $\widetilde \lambda$, $ds^2(M_6)$ by comparing 
\eqref{11d_metric} and \eqref{11d_general_param}, and we
compute
\beq
c = \frac{B^2 \,  \cC }{2 \, \pi^3 \, (m \, \ell_p)^9} \, \int \frac{w \, \mu^2}{(1-w^2)^2} \, dw \wedge d\mu
= \frac{B^2 \,  \cC }{12 \, \pi^3 \, (m \, \ell_p)^9} \, \bigg[  \frac{1}{(1-w^2)} \bigg]_{w_{\rm mix}}^{w_{\rm max} } \ .
\eeq
For the case of interest \eqref{y_range_1}, $w_{\rm min} = 0$ and $w_{\rm max}= w_1$, yielding
a
finite central charge
\beq
c =   \frac{B^2 \,  \cC  \, w_1^2}{12 \, \pi^3 \, (m\, \ell_p)^9 \, (1-w_1^2)} \ .
\eeq
The RHS can be written in terms of $N$, $K$, and $\ell$ by making use of
\eqref{y_range_1}, \eqref{in_terms_of_integers}, 
\beq \label{central_charge}
c = \frac{\ell \, N^2 \, K^2}{12 \, (N +  K\, \ell)} \ .
\eeq
We get a well-defined, finite result even though the 11d solution
has singularities.

\subsubsection{Supersymmetric Wrapped M2-Brane Probes} \label{sec:wrappedM2s}

A probe M2-brane wrapping a calibrated 2d submanifold in the internal space
gives a BPS particle in the external $AdS_5$ spacetime.
Our solutions preserve 4d $\cN = 2$ superconformal symmetry,
but we find it convenient to study the calibration conditions
with reference to a 4d $\cN = 1$   subalgebra.
More precisely, any solution of the form \eqref{LLM}
admits a doublet $\xi^\cI$, $\cI = 1,2$ of Killing spinors on $M_6$,
constructed out of Killing spinors on $S^2$ and suitable
spinors in the 4d space spanned by $\chi$, $y$, $x^1$, $x^2$.
We  select a linear combination $\xi = c_\cI \, \xi^\cI$
of the two Killing spinors and study calibration with respect to $\xi$.
We refer the reader to appendix \ref{app_spinors_and_calibration}
for a more thorough discussion of Killing spinors
for solutions of the form \eqref{LLM}, and their relation
to the most general supersymmetric $AdS_5$ solution of \cite{Gauntlett:2004zh}.

The calibration condition 
for an internal 2d submanifold  $\cC_2$ can be written as \cite{Gauntlett:2006ai}
\beq \label{calibration}
Y' \Big|_{\cC_2} = {\rm vol}_{M_6}( \cC_2) \ ,
\eeq
where ${\rm vol}_{M_6}( \cC_2)$ is the volume form on $\cC_2$
induced by the metric $ds^2(M_6)$ and the 2-form $Y'$
is constructed as a spinor bilinear,
\beq
Y' = \frac 14 \, \overline \xi \, \gamma_{mn} \, \xi \, dy^m \wedge dy^n \ .
\eeq
In the previous expression the indices $m$, $n$ are curved indices on $M_6$,
with local coordinates $dy^m$, and $\gamma_{mn} = \gamma_{[m} \gamma_{n]}$ with
Euclidean gamma matrices $\gamma_m$ in six dimensions.
To write $Y'$ we find it convenient to write the
quantity $ds^2(S^2)$ that enters \eqref{LLM}
as
\beq \label{break_up_S2}
ds^2(S^2) = \frac{d\tau^2}{1- \tau^2} + (1-\tau^2) \, d\varphi^2  \ ,
\eeq
where the coordinate $\tau$ lies in the interval $[-1,1]$
and the angle $\varphi$ has period $2\pi$.
The expression for $Y'$ in terms of the quantities that enter the canonical LLM
form of the solution is
\begin{align}  \label{Yprime}
Y ' & = \frac{1}{4 }  \, y^3 \, e^{-9  \widetilde  \lambda} \, {\rm vol}_{S^2}
 + \frac{1    }{2 } \, y \, e^{-3 \widetilde \lambda} \, (1 - y^2 \, e^{-6\widetilde \lambda}) \, d\tau \wedge D\chi
\nn \\
&  - \frac{  1 }{2  } \, \tau \, e^{-3 \widetilde \lambda }\,  D\chi \wedge dy
 - \frac{ 1  }{4} \,  
\frac{  y \, e^{-9 \widetilde \lambda} \, \tau \, e^D }{1 - y^2 \, e^{-6 \widetilde \lambda}}
  \, dx^1 \wedge dx^2  \ .
\end{align}
We refer the reader to appendix \ref{app_calibration_2form} for the expression of $Y'$ in terms of
the $\mu$, $w$, $z$, $\phi$ coordinates.
The conformal dimension $\Delta$
of the operator associated to the BPS particle originating from a probe M2-brane
on the calibrated subspace $\cC_2$ is given by  \cite{Gauntlett:2006ai}
\beq \label{dimension_formula}
\Delta = \frac{1}{4  \pi^2  m^3 \ell_p^3} \, \int_{\cC_2} \, e^{3 \widetilde \lambda} \,  {\rm vol}_{M_6}( \cC_2)  \ ,
\eeq
where we are still adopting the parametrization \eqref{11d_general_param} of the 11d metric.

We identify two calibrated submanifolds that can support supersymmetric
M2-brane probes. Firstly, we consider the 2-cycle $\cC_2$ in $M_6$ defined
by taking the $S^2$ on top of the point $\mathsf P_3$ in the $(w,\mu)$ plane,
with coordinates $w = w_1$, $\mu = 1$. At this point both the $\phi$ and $z$ circles shrink.
The calibration 2-form $Y'$ restricted on $\cC_2$ is most easily evaluated
by using the expression~\eqref{ugly_Y},
\beq
Y' \Big|_{\cC_2}  = \frac{w_1^{3/2} \, (1-w_1^2)^{3/4}}{\sqrt 2 \, B^{3/2}} \, {\rm vol}_{S^2} = \frac 14 \, {\rm
vol}_{S^2} \ .
\eeq
On the other hand, the metric on $\cC_2$ induced from $ds^2(M_6)$
is readily extracted from \eqref{11d_metric},
\beq
ds^2(\cC_2) = \frac{w_1 \, \sqrt{1-w_1^2}}{2 \, B} \, ds^2(S^2) = \frac 14 \, ds^2(S_2) \ .
\eeq
We see that the calibration condition \eqref{calibration} is satisfied.
Let $\cO_1$ denote the M2-brane operator associated to the calibrated
2-cycle $\cC_2$. The dimension of $\cO_1$ is computed with
\eqref{dimension_formula},
\beq
\Delta(\cO_1) = \frac{N \, K \, \ell}{N + K \, \ell} \ . \label{eq:o1}
\eeq

Another calibrated submanifold $\cB_2$ is realized by considering
 the segment $\mathsf P_3 \mathsf P_4$ (see Figure~\ref{rectangle})
and the combination of $S^1_\phi$ and $S^1_z$ that does \emph{not} shrink
in the interior of $\mathsf P_3 \mathsf P_4$. 
This combination corresponds to the fiber $Dz$  in the presentation \eqref{puncture_metric}.
This submanifold is not a 2-cycle. It rather describes
  an open M2-brane  that ends on the 
 M5-brane source at $w=0$. The M2-brane sits at a point on the $S^2$.
The calibration 2-form $Y'$ on $\cB_2$ can be computed using 
\eqref{ugly_Y} and is given by
\beq \label{Y_for_B2}
Y' \Big|_{\cB_2}  = \frac{\cC \, (1-w^2)^{-3/4}}{\sqrt 2 \, \sqrt B \, \sqrt w} \, \tau_* \, dw \wedge Dz  \ .
\eeq
Here $\tau_*$ is the value of the $\tau$ coordinate on the $S^2$ at which the M2-brane
is located. The induced metric on $\cB_2$ is extracted from \eqref{puncture_metric},
\begin{align} \label{metric_for_B2}
ds^2(\cB_2) = \frac{dw^2}{2 \, w \, h \, (1-w^2)^{3/2}} + \frac{\cC^2 \, h}{B} \, Dz^2 \ .
\end{align}
Comparing \eqref{Y_for_B2} and \eqref{metric_for_B2} we see that the calibration condition
\eqref{calibration} is satisfied, provided that the M2-brane sits at north pole of $S^2$,
$\tau_* = 1$. (We are defining the orientation of the volume form on $\cB_2$
to be $dw \wedge Dz$.)
Since $\cB_2$ describes an open M2-brane,
it corresponds to a collection of operators, which we denote collectively
as $\cO_2^i$. The label $i$ runs over various possible boundary conditions
for the M2-brane ending on the M5-brane source.
All operators $\cO_2^i$ have the same dimension, computed from
\eqref{dimension_formula} to be
\beq \label{dim_O2}
\Delta(\cO_2^i) = K \ .
\eeq

The degeneracy of the operators $\cO_2^i$ (\emph{i.e.}~the range of the label $i$)
can be estimated as follows.
On general grounds, the M2-brane ending on the M5-brane sources can have several boundary components. Thus, for each M5-brane in the source stack, we can decide whether the M2-brane
ends on that M5-brane or not. 
Recalling that the number of   M5-branes at the source is $N$, this gives a total of $2^N$
possibilities. We have to subtract 1, however, because the M2-brane must end on at least
one of the M5-branes. In conclusion, this counting argument gives a degeneracy of $2^N-1$
for the operators $\cO_2^i$.

The charges of the operators $\cO_1$, $\cO_2^i$
under the $U(1)_r \times SU(2)_R$ R-symmetry can be computed on the gravity side
using the methods of \cite{Gauntlett:2006ai}. The derivation is reported in appendix \ref{app_brane_charges}.
The result reads
\beq \label{brane_Rcharges}
r(\cO_1)= 2 \, \Delta(\cO_1) \ , \qquad R(\cO_1) = 0 \ , \qquad
r(\cO_2^i) = 0 \ , \qquad R(\cO_2^i) = \Delta(\cO_2^i) \ .
\eeq
In the previous expressions $R$ denotes the Cartan generator
of $SU(2)_R$, normalized as to have integer eigenvalues.

\section{Symmetries and 't Hooft Anomalies} \label{sec:anomaly}

In this section we analyze the global symmetries and 't Hooft anomalies of   
 the SCFTs dual
to the 11d solutions of section \ref{sec:spindlesolutions}, for the choice of parameters
and range of $w$ specified in \eqref{y_range_1}.
We observe that  the solutions admit a $\mathfrak u(1)^2 \oplus \mathfrak{su}(2)$ isometry algebra, but the 
algebra of continuous (0-form) symmetries of the dual SCFTs
is only 
$\mathfrak u(1) \oplus \mathfrak{su}(2)$, which is identified with the R-symmetry algebra
of 4d $\cN = 2$ superconformal symmetry. This apparent discrepancy
is explained via a St\"uckelberg mechanism.
Moreover, 
we also compute the 't Hooft anomalies for the
R-symmetry and the flavor symmetry associated to the $\mathbb R^4/\mathbb Z_\ell$
orbifold singularity, and we extract the corresponding flavor central charge.

%
%
%

The symmetries and 't Hooft anomalies of a holographic SCFT with a smooth M-theory dual can be
extracted systematically using the 
methods developed in \cite{Bah:2019rgq}, building on
\cite{Freed:1998tg,Harvey:1998bx}.
Let $M_6$ denote the internal space of the holographic solution,
as in the parametrization \eqref{11d_general_param} for the 11d line element.
The analysis makes use of an auxiliary 12-manifold $M_{12}$, realized as a 
fibration of $M_6$ over closed 6-manifold $\cM_6$,
\beq \label{M12_def}
M_6 \hookrightarrow M_{12} \rightarrow \cM_6 \ .
\eeq 
The space $\cM_6$  is interpreted
as external spacetime, Wick-rotated to Euclidean signature, and extended 
from four to six dimensions as appropriate for application of the standard descent
formalism for the anomaly polynomial.
The fibration \eqref{M12_def} of $M_6$ over $\cM_6$ includes non-zero connections
for the isometry algebra of $M_6$. These connections are interpreted as background gauge fields
for the global symmetries of the SCFT. We will see momentarily, however, that this general expectation
requires some refinement for the setups of interest in this paper.

A central role in the analysis of the symmetries and  anomalies
of the SCFT is played by the 4-form $E_4$,
which enjoys the following properties:
\begin{itemize}
\item $E_4$ is a globally defined 4-form on the total space $M_{12}$.
\item $E_4$ is closed.
\item $E_4$ has integral periods on 4-cycles in $M_{12}$.
\item $E_4$ restricted to the $M_6$ fiber over a generic point in $\cM_6$ reproduces
the closed 4-form $G_4/(2\pi \ell_p)^3$ that describes the $G_4$-flux configuration
that supports the $AdS_5$ solution.
\end{itemize}
We interpret $E_4$ as the object that encodes the boundary conditions
for the M-theory 3-form in the vicinity of the stack of M5-branes that supports the SCFT.
In the next section, we describe in detail the construction of $E_4$.
Crucially, we will encounter a difficulty in constructing $E_4$
while incorporating background connections for the full $\mathfrak{u}(1)^2 \oplus \mathfrak{su}(2)$
isometry algebra of $M_6$. The resolution of this difficulty will yield a physical
explanation of 
the absence of a continuous 0-form flavor symmetry in the dual SCFT corresponding to this isometry.

Once the 4-form $E_4$ is constructed, the 6-form anomaly polynomial
of the SCFT, at leading order in the large $N$ limit, is computed as
\beq
I_6^\text{SCFT, large $N$} = \frac 16 \, \int_{M_6} E_4 \wedge E_4 \wedge E_4   \ ,
\eeq
where $E_4 \wedge E_4 \wedge E_4$ is a 12-form on $M_{12}$ and
$\int_{M_6}$ denotes integration along the $M_6$ fibers.

\subsection{Construction of $E_4$}

In this section we construct the 4-form $E_4$.
We find it convenient to proceed in steps.
Firstly, we discuss the inclusion in $E_4$ of background connections for the $\mathfrak u(1)^2$ isometry 
algebra of $M_6$ associated to the Killing vectors $\partial_z$ and $\partial_\phi$. 
We encounter an obstruction in the construction of $E_4$, which is resolved by 
 demonstrating that only one linear combination of the 
$\mathfrak u(1)^2$ isometry generators translates to a continuous symmetry of the SCFT,
with the other combination being spontaneously broken by a St\"uckelberg mechanism.

\subsubsection{Obstruction in the Construction of $E_4$}

Our starting point is the $G_4$-flux background $G_4$,
conveniently rescaled to $\overline G_4$ as in \eqref{G4_rescaling}
to have integral periods. We aim at constructing a local expression for the 4-form $E_4$,
including background connections for the $\mathfrak u(1)^2$ isometry algebra
of $M_6$ associated to the Killing vectors $\partial_z$ and $\partial_\phi$.
(We postpone the discussion of the $\mathfrak{su}(2)$ isometry algebra of the $S^2$.)
It is actually more convenient to use the linear combinations 
$\partial_\chi$, $\partial_\beta$ of $\partial_z$ and $\partial_\phi$ defined in \eqref{Rsym_and_flavor}.
The expression of $\overline G_4$ in terms of $d\chi$, $d\beta$ is
extracted from \eqref{G4flux} and \eqref{angular_vars} and takes the form
\beq \label{better_Gbar}
\overline G_4 = N \, \frac{{\rm vol}_{S^2}}{4\pi} \wedge \bigg[ 
d\alpha_{0\chi} \wedge 
\frac{d\chi}{2\pi}
+ d\alpha_{0\beta} \wedge
\frac{d\beta}{2\pi}  \bigg]  \ ,
\eeq
where $\alpha_{0\chi}$, $\alpha_{0\beta}$ are 0-forms on $M_6$ given by
\beq \label{alpha0_defs}
\alpha_{0\chi} = \frac{2 \, w^2 \,  \mu^3}{\mu^2 + w^2 \, (1-\mu^2)} \ , \qquad
\alpha_{0\beta} = \frac{ \mu^3 \, (2 \, w^2 + 2 \, \cC \, w^2 - \cC)  }{\mu^2 + w^2 \, (1-\mu^2)} \ .
\eeq

The 4-form $E_4$ is globally defined on $M_{12}$ and can be expanded
as a polynomial in the field strengths of the external connections
associated to the Killing vectors $\partial_\chi$, $\partial_\beta$. 
As a result, the na\"ive ansatz for $E_4$ takes the form
\beq \label{naive_ansatz}
E_4 = \overline G_4^{\rm g} + \sum_{I = \chi,\beta} \frac{F^I}{2\pi} \wedge \omega_{2 I}^{\rm g}
+ \sum_{I,J= \chi,\beta}\frac{F^I}{2\pi} \wedge \frac{F^J}{2\pi} \, \sigma_{0IJ} \ .
\eeq
The 2-forms $F^I = dA^I$ are the field strenghts of external $U(1)$ gauge fields $A^I$
associated to the Killing vectors $\partial_\chi$, $\partial_\beta$.
The objects $\omega_{2I}$ are   2-forms on $M_6$,
while $\sigma_{0IJ} = \sigma_{0(IJ)}$ are 0-forms on $M_6$, to be determined.
The superscript `g' stands for ``gauged'' and indicates the operation of
taking a $p$-form on $M_6$ and making the replacements
\beq \label{gauging}
d\chi \rightarrow (d\chi)^{\rm g} := d\chi + A^\chi \  , \qquad 
d\beta \rightarrow (d\beta)^{\rm g} := d\beta + A^\beta \ .
\eeq
This replacement is necessary to promote a globally defined $p$-form
on the fiber $M_6$ to a globally defined $p$-form on the total space $M_{12}$.
Since 0-forms are unaffected by this prescription, we omit the superscript `g' on
$\sigma_{0IJ}$.

We must demand closure of $E_4$. This translates into a set of conditions
on the unspecified forms $\omega_{2I}$, $\sigma_{0IJ}$,
\beq \label{wanna_be_closed}
2\pi \iota_I \overline G_4 + d\omega_{2I} = 0 \ , \qquad
2\pi \iota_{(I} \omega_{2 J)} + d \sigma_{0IJ} = 0  \ .
\eeq
The symbol $\iota_\chi$, $\iota_\beta$ denotes the interior
product of a $p$-form with the vector field $\partial_\chi$, $\partial_\beta$, respectively.
The 4-form $E_4$ must be globally defined on $M_{12}$, which means that $\omega_{2I}$ must be
globally defined on $M_6$. As a result, the first condition in \eqref{wanna_be_closed}
can only be satisfied if the 
 3-forms $2\pi \iota_\chi \overline G_4$ and $2\pi \iota_\beta \overline G_4$
are exact 3-forms on $M_6$.

The 3-forms $2\pi \iota_\chi \overline G_4$ and $2\pi \iota_\beta \overline G_4$ are readily
computed using \eqref{better_Gbar},  
\beq
2\pi \iota_\chi \overline G_4 =- N\,  \frac{{\rm vol}_{S^2}}{4\pi} \wedge d\alpha_{0\chi}  \ , \qquad
2\pi \iota_\beta \overline G_4 = -N\,  \frac{{\rm vol}_{S^2}}{4\pi} \wedge d\alpha_{0\beta} \ .
\eeq
Both these 3-forms  are manifestly closed.
They are also globally defined on $M_6$,
because $\overline G_4$ and the Killing vector fields $\partial_\chi$,
$\partial_\beta$ are globally defined on $M_6$.
The 3-form $2\pi \iota_\chi \overline G_4$ is   exact: we can write
\beq \label{the_exact_one}
2\pi \iota_\chi \overline G_4 =    d\bigg[
- N\, \alpha_{0\chi}\, \frac{{\rm vol}_{S^2}}{4\pi}
\bigg] \ , \qquad
\eeq
and the 2-form inside the total derivative on the RHS is globally defined on $M_6$,
because the 0-form $\alpha_{0\chi}$ goes to zero at the loci
$\{ \mu = 0 \}$, $\{ w = 0\}$ where the $S^2$ shrinks.
A similar manipulation for $2\pi \iota_\beta \overline G_4$ fails,
because  the 0-form $\alpha_{0\beta}$ does not go to zero at $w = 0$.
We can confirm that the 3-form 
$2\pi \iota_\beta \overline G_4$ is closed but not exact by computing its integral over
the
3-cycle $\cC_3$ defined as follow (see Figure \ref{rectangle}).
Consider a path in the $(w,\mu)$ plane
connecting a generic point $\mathsf Q_1$ on the $\mathsf P_1 \mathsf P_2$
segment to the point $\mathsf P_4$.
Combining this path with the $S^2$ we get a 3-cycle,
because the $S^2$ shrinks both at $\mathsf Q_1$ and $\mathsf P_4$. 
 The integral of $2\pi \iota_\beta \overline G_4$
over $\cC_3$ is indeed non-zero, and evaluates to
\beq \label{3cycle_check}
\int _{\cC_3}  2\pi \iota_\beta \overline G_4 = \Big[
- N \, \alpha_{0\beta}
\Big]^{\mathsf P_4}_{\mathsf Q_1} =   N \, \cC  = K + \frac N \ell \ .
\eeq
In the last step we used \eqref{in_terms_of_integers}. Recall
that $N$ is divisible by $\ell$, so $\int _{\cC_3}  2\pi \iota_\beta \overline G_4$ is an integer.

The 3-form $ 2\pi \iota_\beta \overline G_4 $ is not exact because of the presence 
of the localized M5-brane source at $w =0$.
Indeed, we observe that $D\phi |_{w=0} = - \cC \,d\beta$.
The ``Gaussian pillbox'' that measures the charge of the M5-brane source
is defined taking $w = {\rm constant } \rightarrow 0$ and considering the directions
$\mu$, $S^2$, $D\phi$. We may regard the Gaussian pillbox as a $D\phi \propto d\beta$
fibration over $\mu$ and the $S^2$. The base of this fibration can be identified
with the 3-cycle $\cC_3$, in the limit in which the point $\mathsf Q_1$
is brought towards $\mathsf P_1$.

The non-exactness of $ 2\pi \iota_\beta \overline G_4 $
is an obstruction to the construction of $E_4$ via the ansatz \eqref{naive_ansatz}.
To proceed, we must consider a more general ansatz.

\subsubsection{Resolution of the Puzzle:   a Novel St\"uckelberg Mechanism}

The analysis of the previous subsection revealed the importance of the 
following closed but not exact 3-form on $M_6$,
\beq \label{Lambda3_def}
\Lambda_3 := -  \cC^{-1} \, 
\frac{{\rm vol}_{S^2}}{4\pi} \wedge d\alpha_{0\beta} \ , 
\eeq 
which is defined is such a way that
\beq \label{Lambda3_properties}
2\pi \iota_\beta \overline G_4 - N \, \cC \, \Lambda_3 = 0 \ , \qquad
\int_{\cC_3} \Lambda_3 = 1 \ ,
\eeq 
where $\cC_3$ is the 3-cycle in $M_6$ defined above \eqref{3cycle_check}.
We extend the ansatz for $E_4$ including not only the external $U(1)$ gauge fields $A^\chi$,
$A^\beta$, but also an external 0-form gauge field $a_0$ (a real periodic scalar, {\it i.e.}~an axion).
We think of $a_0$ as the light mode originating from fluctuations of the M-theory
3-form $C_3$ along the cohomology class defined by the closed but not exact 3-form
$\Lambda_3$. Let $f_1$ be the 1-form field strength of $a_0$. We allow for a non-trivial
Bianchi identity for $f_1$, of the form
\beq \label{f1_Bianchi}
df_1 = \sum_{I  = \chi,\beta} q_I \, F^I \ .
\eeq
The constant parameters $q_I$ will be determined momentarily.
The field strengths of the $U(1)$ gauge fields $A^\chi$, $A^\beta$ remain
standard, $F^\chi = dA^\chi$, $F^\beta = dA^\beta$.

The improved ansatz for $E_4$ reads
\beq
E_4 =  \overline G_4^{\rm g}+ \sum_{I = \chi,\beta} \frac{F^I}{2\pi} \wedge \omega_{2 I}^{\rm g}
+ \sum_{I,J= \chi,\beta}\frac{F^I}{2\pi} \wedge \frac{F^J}{2\pi} \, \sigma_{0IJ}
+ \frac{f_1}{2\pi} \wedge \Lambda_3^{\rm g} \ .
\eeq
(Since $\Lambda_3$ has no legs along $\chi$, $\beta$, the gauging prescription
`g' on $\Lambda_3$ could be dropped.) Closure of $E_4$
gives the conditions
\begin{align} \label{new_closure}
2\pi \iota_I \overline G_4 + d\omega_{2I} + q_I \, \Lambda_3 = 0 \ , \quad
2\pi \iota_{(I} \omega_{2 J)} + d\sigma_{0IJ} = 0 \ , \quad
d\Lambda_3 = 0 \ , \quad
2\pi \iota_I \Lambda_3 = 0 \ .
\end{align}
The last two conditions are satisfied, because the 3-form $\Lambda_3$ is closed
and invariant under the action of both $U(1)$ isometries,
$\pounds_I \Lambda_3 = 0$, as can be seen explicitly from its definition \eqref{Lambda3_def}.
The first condition in \eqref{new_closure} can now be solved
by setting
\beq
\omega_{2\chi} = N \, \alpha_{0\chi} \, \frac{{\rm vol}_{S^2}}{4\pi} \ , \qquad
q_\chi = 0 \ , \qquad
\omega_{2\beta} = 0 \ , \qquad
q_\beta = -N \, \cC \ ,
\eeq
as can be seen from \eqref{the_exact_one}, \eqref{Lambda3_properties}.
Notice that we can set $\omega_{2\beta}$ to zero without loss in generality:
a non-zero $\omega_{2\beta}$ could be reabsorbed by a redefinition of $\Lambda_3$
by an exact piece.
Since $\omega_{2\chi}$ has no legs along $\chi$ and/or $\beta$,
we can solve the second condition in \eqref{new_closure} simply setting $\sigma_{0IJ} = 0$.

Having identified the parameters $q_I$, the Bianchi identity \eqref{f1_Bianchi} for $f_1$ 
reads
\beq \label{final_Bianchi}
df_1  = N \, \cC \, F^\beta \ , \qquad
f_1 = da_0   - N \, \cC \, A^\beta \  .
\eeq
It demonstrates that the $U(1)$ gauge field $A^\beta$ gets massive via a St\"uckelberg
mechanism by ``eating'' the axion $a_0$. The $U(1)$ gauge group associated to $A^\beta$
is thus spontaneously broken.

There might be a non-trivial cyclic discrete subgroup that remains unbroken
after the St\"uckelberg mechanism. 
In order to determine this discrete subgroup,
it is necessary to fix the normalizations of the axion $a_0$ and the vector
$A^\beta$. The fact that the 3-form $\Lambda_3$   integrates to 1 over $\cC_3$, see \eqref{Lambda3_properties},
suggests that $a_0$ is correctly normalized (\emph{i.e.}~is a compact scalar with
period $2\pi$). Fixing the normalization of $A^\beta$ is more subtle.
It would also be interesting to identify
which operators in the dual SCFT would be charged under the
discrete subgroup left over after the St\"uckelberg mechanism.
We leave these questions for future investigation.

We conclude this section with a comparison with the
spontaneously broken $U(1)$ symmetries in M-theory
discussed in \cite{Bah:2020uev}. In that reference, the focus is on $U(1)$ $p$-form
symmetries originating from the expansion of the M-theory 3-form $C_3$ onto cohomology classes.
Some of these symmetries are spontaneously broken by topological mass terms of BF type.
A BF coupling is related to a St\"uckelberg mechanism by dualization of a $p$-form
gauge field (as reviewed for instance in \cite{Banks:2010zn}).
Thus, the main physical mechanism observed in the present setup
is the same as in \cite{Bah:2020uev}. Their 11d origin, however, is different:
while in \cite{Bah:2020uev} all BF couplings originate from the $C_3 G_4 G_4$ Chern-Simons term in the
M-theory effective action, in the solutions of this paper
the St\"uckleberg coupling originates from a non-trivial Bianchi identity,
which is required by self-consistency of $G_4$ after the Kaluza-Klein vector $A^\beta$ is 
turned on. 

%
%
%
%

\paragraph{General Formulation.}
We have uncovered an example   of the following phenomenon in M-theory reductions
to five dimensions:
\begin{quote}
A $U(1)$ gauge field associated to an Abelian isometry of the internal space $M_6$
gets massive by eating an axion originating from the expansion of the M-theory 3-form $C_3$
onto a non-trivial class in the third cohomology of $M_6$.
\end{quote}
Let us give a general formulation of the conditions for this phenomenon to happen.

Let $I$ be an index labeling the generators of the $U(1)$ factors in the isometry group of $M_6$.
Let $\xi_I$ be the Killing vector field associated to the $I$-th factor. 
We use the notation $\iota_I$, $\pounds_I$ for the interior product with the vector field $\xi_I$
and the Lie derivative along $\xi_I$, respectively.
Let $[\Lambda_{3x}]$ be a basis of the de Rham cohomology $H^3(M_6,\mathbb R)$,
$x = 1,\dots, b^3(M_6)$.

In order for the Killing vector field $\xi_I$ to be a symmetry of the full holographic
M-theory solution, we must demand $\pounds_I \overline G_4 = 0$.
As a result, the 3-form $2\pi \iota_I \overline G_4$ is necessarily closed,
as follows immediately from $d \overline G_4 = 0$ and $\pounds_I = d\iota_I + \iota_I d$.
The closed 3-form $2\pi \iota_I \overline G_4$ defines a (possibly trivial) de Rham
cohomology class,  
which can be expanded onto the basis $[\Lambda_{3x}]$ as
\beq
[2\pi \iota_I \overline G_4] + Q^x{}_I \, [\Lambda_{3x}] = 0 \ .
\eeq
The expansion coefficients $Q^x{}_I$ are  
 identified with the constants entering the Bianchi identities
for the field strengths $f_1^x$ of the axions $a_0^x$ obtained from expanding
$C_3$ onto the basis $[\Lambda_{3x}]$,
\beq \label{general_Q}
df_1^x = Q^x{}_I \, F^I  \ , \qquad
f_1^x = da_0^x + Q^x{}_I \, A^I \ .
\eeq
Here $F^I = dA^I$ is the field strength of the $U(1)$ gauge field
$A^I$ associated to the Killing vector field $\xi_I$.
Non-zero coefficients $Q^x{}_I$ indicate a non-trivial St\"uckelberg mechanism.
If the $U(1)$ gauge fields $A^I$ and the axions $a_0^x$ are correctly
normalized, the coefficients $Q^x{}_I$ are integrally quantized.
They determine to which cyclic subgroup the $U(1)$ gauge group associated to $A^I$ is
spontaneously broken by the 
St\"uckelberg mechanism.
To see this, we observe that 
the St\"uckelberg couplings encoded in \eqref{general_Q} can equivalently be
cast in the form of BF-like topological terms \cite{Maldacena:2001ss,Banks:2010zn}. This can be done
by dualizing the axions $a_0^x$ to 3-form potentials $c_{3x}$. The relevant topological
terms in the 5d supergravity effective action take the form
\beq
S_{\rm top} = \frac{1}{2\pi} \, Q^x{}_I \, \int_{\cM_5} c_{3x} \wedge F^I \ ,
\eeq
where $\cM_5$ is 5d external spacetime.
This topological action describes 5d 1-form and 3-form gauge fields with
discrete gauge group. The discrete gauge group is read off from the Smith normal form of the
matrix $Q^x{}_I$ \cite{Morrison:2020ool}.

We observed above that, in our solutions, non-exactness of
$ 2\pi \iota_\beta \overline G_4 $ is closely related to the
presence of an M5-brane source in the solution.
It is natural to ask whether smooth solutions without internal sources can be found, 
for which $ 2\pi \iota_I \overline G_4$ is non-trivial in cohomology 
for some isometry direction $I$. We aim to address this question
more systematically in the future.

As discussed in \cite{Bah:2019rgq, Hosseini:2020vgl}, 
the construction of $E_4$ from $\overline G_4$
can be phrased mathematically using the language of $G$-equivariant
cohomology (where $G$ stands for the isometry group of the internal space $M_6$).
More precisely, $\overline G_4$ is a closed invariant form on $M_6$,
and the task at hand is to construct an equivariant extension of $\overline G_4$.
Obstructions to such a construction have been discussed in the mathematical literature
\cite{WU1993381}. Our physical analysis 
detects the obstructions and offers a way to circumvent them,
by generalizing the notion of equivariant extension 
with the inclusion of the axion field. 

In our discussion so far we have implicitly modeled $p$-form
gauge fields using differential forms. While this is adequate to capture
local aspects of their dynamics (such as their curvatures), a better mathematical framework
to describe the physics of $p$-form gauge fields is differential cohomology
(reviews aimed at physicists include \cite{Bauer:2004nh, Freed:2006yc, Cordova:2019jnf}). Since we are turning on gauge fields
associated to the isometries of $M_6$, we should be actually employing
  $G$-equivariant differential cohomology \cite{equivdiff}. It would be interesting to
adopt this language to 
study the obstructions we have encountered and their resolution.

\subsection{Anomaly Inflow}

Having identified how to treat the isometry direction $\partial_\beta$,
we can complete the construction of the full form of $E_4$,
including background $\mathfrak {su}(2)$ gauge fields associated to the non-Abelian
isometry algebra of the $S^2$. This is most easily accomplished noticing that
$\overline G_4$, $\omega_{2\chi}$, and $\Lambda_3$ all have a common factor
${\rm vol}_{S^2}$. We may simply perform the replacement
\beq
\frac{ {\rm vol}_{S^2} }{4\pi} \rightarrow e_2 \ ,
\eeq
where $e_2$ is the global angular form of $SO(3)$,
which is closed, gauge-invariant, and normalized to integrate to $1$ on $S^2$.
(We refer the reader to  \eqref{e2_expr} for the explicit
expression of $e_2$.)
Notice that the non-Abelian isometry $\mathfrak{su}(2)$ cannot
participate in any non-trivial St\"uckelberg mechanism with 
the axion $a_0$.

In conclusion,
the final form of $E_4$,
including the external $U(1)$ gauge fields $A^\chi$, $A^\beta$, the axion $a_0$,
and the background $\mathfrak{su}(2)$ gauge fields,
can be written as
\begin{align}
\!\!\!\! E_4   = N \, e_2 \wedge \bigg[  d\alpha_{0\chi} \wedge \frac{(d\chi)^{\rm g}}{2\pi}
+  d\alpha_{0\beta} \wedge \frac{(d\beta)^{\rm g}}{2\pi}  \bigg]
+ N \, \alpha_{0\chi} \, e_2 \wedge \frac{F^\chi}{2\pi}
- \cC^{-1} \, \frac{f_1}{2\pi} \wedge e_2 \wedge d\alpha_{0\beta} \ . \!\!\!
\end{align}
We can verify directly the closure of $E_4$ using $de_2 = 0$,
\eqref{gauging}, and \eqref{final_Bianchi}.

It is now straightforward to compute $E_4\wedge E_4 \wedge E_4$ and fiber integrate
along $M_6$. The integral over $S^2$ is most easily performed
with the help of the Bott-Cattaneo formulae \cite{bott1999integral},
\beq
\int_{S^2} e_2 = 1 \ , \qquad
\int_{S^2} e_2^2 = 0 \ , \qquad
\int_{S^2} e_2^3 = \frac 14 \, p_1(SO(3))  \ ,
\eeq
where $p_1(SO(3))$ is the first Pontryagin class of the $SO(3)$ bundle
associated to the $S^2$ fibration over external spacetime.

We assign positive orientation to $d\beta \wedge d\chi$.
Taking into account that both these angles have periodicity $2\pi$,
we arrive at
\begin{align}
I_6^\text{SCFT, large $N$} = \frac 16 \, \int_{M_6} E_4^3 & = - \frac 18 \, N^3 \, \frac{F^\chi}{2\pi} \, p_1(SO(3)) \, \int_{B_2} d\alpha_{0\beta} \wedge d(\alpha_{0\chi}^2) \ ,
\end{align}
where $B_2$ denotes the rectangle spanned by $w$ and $\mu$.
Assigning positive orientation to $dw \wedge d\mu$, 
and recalling the definitions \eqref{alpha0_defs} of $\alpha_{0\chi}$, $\alpha_{0\beta}$,
one finds
\beq
\int_{B_2} d\alpha_{0\beta} \wedge d(\alpha_{0\chi}^2) =  \frac 43 \, \cC \, w_1^4
=  \frac{\ell \, K^2}{3 \, N \, (N + K \, \ell)}   \ ,
\eeq 
where in the second step we have used \eqref{y_range_1}, \eqref{in_terms_of_integers}.
The quantities $F^\chi$, $p_1(SO(3))$ are related to the Chern classes of the
$U(1)_r$ and $SU(2)_R$ bundles of the 4d $\cN = 2$ superconformal R-symmetry by
\beq \label{Rsymm_identifications}
\frac{F^\chi}{2\pi} = - 2 \, c_1 (U(1)_r) \ , \qquad
p_1(SO(3)) = - 4 \, c_2(SU(2)_R) \ .
\eeq
With these identifications, we get the result
\beq \label{anomaly1}
I_6^\text{SCFT, large $N$}  = - \frac{\ell \, K^2 \, N^2}{3  \, (N + K \, \ell)} \, c_1(U(1)_r) \, c_2(SU(2)_R)  \ .
\eeq
The central charges $a$, $c$ are related to the 't Hooft anomaly coefficients
for the $SU(2)_R \times U(1)_r$ R-symmetry as reviewed in appendix \ref{sec:appanomalies}. Comparing \eqref{anomaly1} with \eqref{standard_anomaly}
and using the relations \eqref{eq:change}, we verify that \eqref{anomaly1} is compatible with the holographic central charge \eqref{central_charge}.

\subsubsection{Flavor Central Charge From   Anomaly Inflow}

Expanding the M-theory 3-form $C_3$ onto the
resolution cycles of the $\mathbb R^4/\mathbb Z_\ell$ orbifold singularity at
$\mathsf P_3$, one obtains $\ell-1$ Abelian gauge fields.
The gauge group enhances to $SU(\ell)$ by virtue of
states from M2-branes wrapping the resolution cycles \cite{Gaiotto:2009gz}.
We can compute the associated flavor central charge $k_{SU(\ell)}$ 
by computing the mixed  't Hooft anomaly
between  $SU(\ell)$ and $U(1)_r$.
To this end, we can follow the methods of  \cite{Bah:2019jts}.
We turn on background gauge fields $\widehat A_\alpha$, $\alpha = 1,\dots, \ell-1$ for the Cartan
subalgebra of $SU(\ell)$. 
We include a new term in $E_4$,
\beq \label{new_term}
\Delta E_4 = \sum_{\alpha=1}^{\ell-1} \frac{\widehat F_\alpha}{2\pi} \wedge \widehat \omega_\alpha  \ .
\eeq
In the previous expression $\widehat F_\alpha = d\widehat A_\alpha$ and $\widehat \omega_\alpha$
denote the harmonic 2-forms dual to the resolution 2-cycles
of the $\mathbb R^4/\mathbb Z_\ell$ singularity (in the 4d space
spanned by $\mu$, $w$, $\phi$, $z$).
The intersection pairing of the 2-forms $\widehat \omega_\alpha$
reproduces the Cartan matrix of $\mathfrak{su}(\ell)$,
\beq
\int_{\cM_4} \widehat \omega_\alpha \wedge \widehat \omega_\beta  = - C^{\mathfrak{su}(\ell)}_{\alpha \beta} \ , \qquad \alpha, \beta =1,\dots ,\ell-1 \ .
\eeq
Intuitively speaking, we can think of $\widehat \omega_\alpha$ as being localized
at the point $\mathsf P_3$, see Figure \ref{rectangle}. The 4d space $\cM_4$ is a local
 Taub-NUT model for the resolved $\mathbb R^4/\mathbb Z_\ell$ singularity at $\mathsf P_3$. 
 
We may repeat the computation of the fiber integral of $E_4^3$ including the
new term \eqref{new_term}. We obtain an additional term in the inflow anomaly polynomial,
\beq \label{anomaly}
I_6^\text{SCFT, large $N$}  \supset   
- \frac{N \, K \, \ell}{2\, (N + K \, \ell)} \, \frac{F^\chi}{2\pi} \,   \sum_{\alpha, \beta  = 1}^{\ell-1} C_{\alpha \beta}^{ \mathfrak{su}(\ell) } \, 
\frac{\widehat F_{\alpha}}{2\pi} \wedge \frac{\widehat F_{\beta}}{2\pi} 
 \ .
\eeq
As argued above, non-perturbative M2-brane states 
enhance the $U(1)^{\ell-1}$ symmetry to $SU(\ell)$.
Correspondingly, we have\footnote{~This expression corrects a typo in equation (6.3) of \cite{Bah:2019jts}.} 
\beq
  \sum_{\alpha, \beta  = 1}^{\ell-1} C_{\alpha \beta}^{ \mathfrak{su}(\ell) } \, 
\frac{\widehat F_{\alpha}}{2\pi} \wedge \frac{\widehat F_{\beta}}{2\pi} 
\rightarrow       2   \, c_2(SU(\ell))  \ .
\eeq
Making use of \eqref{Rsymm_identifications}, we can write the new term in the inflow anomaly polynomial as
\beq  
I_6^\text{SCFT, large $N$}  \supset   
\frac{2 \, N \, K \, \ell}{N + K \, \ell} \, c_1(U(1)_r) \, c_2(SU(\ell)) 
 \ .
\eeq
Comparison with the standard presentation \eqref{standard_anomaly} of the anomaly polynomial
of a 4d $\cN = 2$ SCFT yields the flavor central charge
\begin{align} \label{holo_flavor_charge}
k_{SU(\ell)}   = \frac{2 \, N \, K \, \ell}{N + K \, \ell} \ .
\end{align}

\section{Field Theory Duals} \label{sec:fieldtheory}

\newcommand{\x}[1]{\textcolor{cyan}{[\bf #1 ]}}

We propose that the supergravity solutions presented above are dual to four-dimensional SCFTs that arise from the low-energy limit of $N$ M5-branes---whose worldvolume theory at low energies is the 6d (2,0) theory of type $A_{N-1}$---wrapped on a sphere with one irregular and one regular puncture. The 4d SCFTs of interest are of Argyres-Douglas type, meaning they are intrinsically strongly coupled and possess relevant Coulomb branch operators with fractional dimensions. In this section we review the properties of these field theories, and discuss the matching of their central charges and operators to the gravity duals. In appendix \ref{sec:AD} we further elaborate on the landscape of Argyres-Douglas SCFTs, their properties, and their geometric construction via irregular punctures.

\subsection{Properties of the $(A_{N-1}^{(N)}[k],Y_{\ell})$ Argyres-Douglas SCFTs}
\label{sec:anak}

The field theories dual to the supergravity solutions presented in section \ref{sec:spindlesolutions} are the 4d $\mathcal{N}=2$ SCFTs that are geometrically engineered by wrapping $N$ M5-branes on a sphere with one irregular puncture of type $A_{N-1}^{(N)}[k]$, and one regular puncture. The labeling of the irregular puncture  follows from the classification in \cite{Xie:2012hs,Wang:2015mra}---we refer the reader to appendix \ref{sec:AD} for a review.\footnote{~These are also denoted as Type I theories in \cite{Xie:2012hs}, and $I_{N,k}$ theories in \cite{Xie:2013jc}.} The regular puncture is labeled by a Young diagram in the shape of a rectangular box with $\ell$ columns and $N/\ell$ rows, which we denote by $Y_{\ell}$. We refer to the 4d SCFTs thus constructed as $(A_{N-1}^{(N)}[k],Y_{\ell})$. These SCFTs carry three integer labels: $N$ the number of M5-branes, $k>-N$ labeling the irregular puncture, and $\ell$ a positive integer that divides $N$ labeling the regular puncture, which contributes an $SU(\ell)$ flavor symmetry. 
The case $\ell=1$ corresponds to the ``non-puncture'', and is equivalent to having no regular puncture on the sphere. These are the $(A_{N-1},A_{k-1})$ SCFTs which can also be obtained in Type IIB string theory \cite{Cecotti:2010fi}, and throughout this section we interchangeably refer to the $\ell=1$ cases as $(A_{N-1}^{(N)}[k],Y_1)$ and  $(A_{N-1},A_{k-1})$. The case $\ell=N$ corresponds to the maximal puncture with associated $SU(N)$ flavor symmetry, known also as the $D_{p=k+N}^{b=N}( SU(N))$ theories studied in \cite{Cecotti:2012jx,Cecotti:2013lda,Giacomelli:2017ckh}.


%

\subsubsection{R-symmetry Twist}

We denote the R-symmetry of the $\CN=2$ SCFT by $SU(2)_{R} \times U(1)_{r}$, with $R=2I^3$ the Cartan generator of $SU(2)_R$ (in conventions in which $R$ has integer-valued charges), and $r$ the generator of $U(1)_r$. 
The R-symmetry that is preserved at the fixed point can be deduced from the properties of the Higgs field $\Phi$ in the Hitchin system that arises from
first compactifying the 6d $(2,0)$ theory on a circle to five dimensions, and then twisting over the sphere (see appendix \ref{sec:AD} for more details).
 The $U(1)_r$ symmetry of the 4d $\CN=2$ field theory is a combination of the $SO(2)_{\phi} \subset SO(5)$ R-symmetry that would be preserved in the absence of an irregular defect on the sphere, and a global $U(1)_{z}$ isometry of the sphere.\footnote{~Indeed, the requirement that $U(1)_{z}$ is globally defined is what leads to the restriction that the Riemann surface in the compactification must have genus zero.}  This combination can be fixed by requiring that the coefficient to the leading singularity of the Higgs field (the matrix $T_k$ in \eqref{eq:phiz}) is covariant under a $U(1)_r$ rotation. 
  For an irregular puncture of type $A_{N-1}^{(b)}[k]$, the result is to fix the $SO(2)_{\phi}$ generator to be proportional to the $U(1)_{z}$ generator, with proportionality factor $ \frac{b}{k+b}$. For the theories $(A_{N-1}^{(N)}[k], Y_{\ell})$ ({\it i.e.} for $b=N$), we thus identify the $U(1)_r$ generator as the combination
	\ba{
	r = R_\phi +  \frac{N}{k+N} R_{z} \ . \label{eq:rp}
	}

\subsubsection{Seiberg-Witten Curve and Deformations}

 The Seiberg-Witten curve of the 4d theory is identified with the spectral curve of the Hitchin system. The Seiberg-Witten curve in the conformal phase is
	\ba{
	y^2 = x^N+z^k\ ,\qquad (A_{N-1},A_{k-1})\ , \label{eq:curve1}
	}
from which it follows (by requiring that the dimension of the Seiberg-Witten differential $\lambda_{\text{SW}} = x dz$ is unity) that the scaling dimensions of $x$ and $z$ are
	\ba{
	\Delta(x) = \frac{k}{k+N}\ ,\qquad \Delta(z) = \frac{N}{k+N}\ . \label{eq:dim1}
	}
	
The possible deformations of the curve \eqref{eq:curve1} take the form $u_{ab} x^a z^b$, and are encoded in a Newton polygon. For the $(A_{N-1},A_{k-1})$  $(\ell=1)$ theories the Newton polygon consists of a single triangle in the upper right quadrant bounded by a line with (minus) slope $\rho-2=\frac{k}{N}$, where $\rho$ is the leading pole for the irregular singularity from \eqref{eq:phiz} (see \cite{Xie:2012hs}). An example Newton polygon with $k=8$ and $N=4$ is shown in Figure \ref{fig:irreg}.
An integer point on the polygon with coordinates $(a,b)$ encodes a deformation $u_{ab}$ of the curve, with dimension
	\ba{
	y^2 \supset u_{ab} x^az^b\ ,\qquad \Delta(u_{ab}) = \frac{kN-ak-bN}{k+N}\ . \label{eq:dim2}
	}
Points that lie on the lines $z^{k-1}$ and $x^{N-1}$ are excluded, since they can be removed by translation invariance of the $z$ coordinate, and by removal of the $U(N)$ trace ($\Phi$ is traceless). The remaining points fall into the following classes:

\tikzset{
    square/.style={%
        draw=none,
        circle,
        append after command={%
            \pgfextra \draw[black] (\tikzlastnode.north-|\tikzlastnode.west) rectangle 
                (\tikzlastnode.south-|\tikzlastnode.east);\endpgfextra}
    }
}

\begin{figure}
\centering
 \begin{tikzpicture}
     \begin{scope}[shift={(0,0)}]
    \draw[very thin,color=gray,step=0.5] (-0.5, -0.5) grid (2.5,4.5);
   \draw[->] (0,0) -- (0,4.5);
      \draw[->] (0,0) -- (2.5,0);
   
   \filldraw[color=blue] (0,0) circle (0.05);
   \filldraw[color=blue] (0.5,0) circle (0.05);
   \filldraw[color=blue] (1,0) circle (0.05);
   \filldraw[color=black] (1.5,0) circle (0.05);
   \filldraw[color=black] (2,0) circle (0.05);
   
      \filldraw[color=blue] (0,0.5) circle (0.05);
   \filldraw[color=blue] (0.5,0.5) circle (0.05);
   \filldraw[color=blue] (1,0.5) circle (0.05);
   \filldraw[color=black] (1.5,0.5) circle (0.05);
   
    \filldraw[color=blue] (0,1) circle (0.05);
   \filldraw[color=blue] (0.5,1) circle (0.05);
   \filldraw[color=blue] (1,1) circle (0.05);
   \filldraw[color=black] (1.5,1) circle (0.05);
   
      \filldraw[color=blue] (0,1.5) circle (0.05);
   \filldraw[color=blue] (0.5,1.5) circle (0.05);
   \filldraw[color=blue] (1,1.5) circle (0.05);
   
         \filldraw[color=blue] (0,2) circle (0.05);
   \filldraw[color=blue] (0.5,2) circle (0.05);
   \filldraw[color=blue] (1,2) circle (0.05);
   
   \filldraw[color=blue] (0,2.5) circle (0.05);
   \filldraw[color=blue] (0.5,2.5) circle (0.05);

   \filldraw[color=blue] (0,3) circle (0.05);
   \filldraw[color=blue] (0.5,3) circle (0.05);  
   
      \filldraw[color=black] (0,3.5) circle (0.05);
           \filldraw[color=black] (0,4) circle (0.05);

     \draw[densely dotted] (0,4) -- (2,0);
     \end{scope}
     \end{tikzpicture}
     \caption{The Newton polygon for the $(A_{N-1},A_{k-1})$ theory, drawn for $N=4$ and $k=2N=8$.  The black points represent excluded deformations corresponding to the $x^{N-1}$ and $z^{k-1}$ lines, as well as the bounding points $x^N$ and $z^k$.  The blue points that lie on the dotted line bounding the triangle correspond to the $N-2$ exactly marginal deformations. \label{fig:irreg}}
\end{figure}
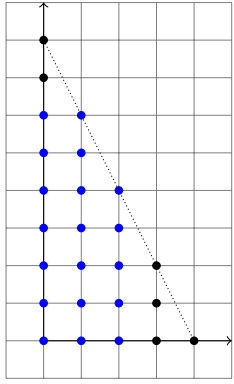

	\begin{itemize}
	
	\item Parameters $u_{ab}$ with dimensions $\Delta(u_{ab})>1$ correspond to Coulomb branch operators, which we will denote by $u_i$ with one subscript. These operators are by definition scalar primaries of the protected chiral $\CN=2$ multiplets $L\bar{B}_1[0,0]_{\frac{r}{2}}^{(0;r)}$ (in the notation of \cite{Cordova:2016rsl}) with R-charges $r(u_i) = 2 \Delta(u_i)$ and $R=0$. The total number of Coulomb branch operators is the rank of the Coulomb branch, which was computed for the $(A_{N-1},A_{k-1})$ theories from the Type IIB geometric engineering setup in \cite{Xie:2016evu} (based on proposals in \cite{Xie:2015rpa}) and then corrected in \cite{Giacomelli:2017ckh}. The result is
		\ba{
		\text{rank}(\text{CB}) = \frac{1}{2} \left( (k-1)(N-1) - \left( \text{GCD}(k,N) - 1 \right) \right)\ ,\quad \ell = 1\ .  \label{eq:rcb1}
		}
	Identifying the Coulomb branch operators from the Newton polygon is especially simple when $k=mN$ for $m$ an integer, in which case one finds a set of $\frac{1}{2}  (k-2)(N-1)$ operators $u_i$ of dimensions
		\ba{
		\Delta(u_{i}) = \frac{i}{m+1}\ ,\qquad i = m+2,\dots,l m \ ,\qquad l = 2,\dots,N\ \ ,\qquad k=mN\ . \label{eq:cbo}
		}
	
	\item For every Coulomb branch operator whose dimension satisfies $1<\Delta(u_{i})<2$, there is a corresponding coupling $\lambda_i$ with $\Delta(\lambda_i)<1$ that satisfies $\Delta(u_i)+\Delta(\lambda_i)=2$ \cite{Argyres:1995xn}.  These couplings are identified with relevant ($\CN=2$)-preserving deformations of the form $\int d^4\theta\ \lambda_i U_i$, for $U_i$ the chiral superfield whose bottom component is $u_i$.
	
	\item Deformation parameters with $\Delta(u_{ab})=1$ correspond to mass deformations, whose number is equal to the rank of the global symmetry $F$ of the SCFT. For the $(A_{N-1},A_{k-1})$ theories, this is \cite{Giacomelli:2017ckh}
		\ba{
		\text{rank}(F) = \text{GCD}(k,N)-1\ , \qquad \ell = 1\ .\label{eq:rf}
		}
		which for $k$ an integer multiple of $N$ reduces to
		$\text{rank}(F) = N-1$. 
	These deformations are paired with the moment map operators $\mu$ with dimension $\Delta(\mu) = 2$ and R-charges $(R,r)(\mu)=(2,0)$, which are the primaries of  $\CN=2$ multiplets $B_1 \bar{B}_1[0,0]_2^{(2;0)}$ containing conserved flavor currents.

	\item Exactly marginal couplings are identified with the parameters $u_{ab}$ of dimension zero. 
	The number of exactly marginal couplings is the complex dimension of the conformal manifold $\CM_{\CC}$.
	For general $k,N$, 
		\ba{
			\text{dim}_{\mathbb{C}} \CM_{\CC} = \text{GCD}(k,N)-1\ , \label{eq:conff}
		}
		as can be seen from the Newton polygon (also see \cite{Giacomelli:2020ryy}).
		
	When $k=mN$ for $m$ a positive integer, there are $N-1$ points on the bounding line of the Newton polygon in addition to the points at the tips of the triangle. $N-2$ of these points  correspond to exactly marginal deformations (since their dimension is zero), while one lies on the $x^{N-1}$ line and is thus excluded. In this case the complex dimension of the conformal manifold $\CM_{\CC}$ is then $N-2$, 
		\ba{
		\text{dim}_{\mathbb{C}} \CM_{\CC} = N-2\ ,\qquad k = mN\ . \label{eq:mmn}
		} 
	When $k=N$, the dimension is reduced further to $\text{dim}_{\mathbb{C}} \CM_{\CC} = N-3$. 

	\end{itemize}

Adding an arbitrary regular puncture with Young diagram $Y$ deforms the curve \eqref{eq:curve1} of the $(A_{N-1},A_{k-1})$ theory with terms
	\ba{
y^2\supset	\sum_{l=2}^N	\alpha_\ell(z) x^{N-l}   + \dots \ ,\qquad \alpha_l = \dots + v_1z^{-1} + \dots +v_{n_l} z^{-p_l} \ . \label{eq:rega}
	}
	Here $p_l = l - h_l$ is the {\it pole structure} of the Young diagram introduced in \cite{Gaiotto:2009we}, with $l$ (distinct from $\ell$!) labeling the boxes from $1$ to $N$ starting in the bottom left corner, and  $h_l$ the height of the $l$'th box. (Here $Y$  is arranged with column sizes decreasing from left to right and row lengths increasing from top to bottom.) 	For example, the maximal puncture $\ell=N$ consists of the diagram with $N$ boxes in a single row, so $h_l = 1$ for all $l$. The additional deformation parameters for the box puncture $Y_{\ell}$ with $\ell > 1$ are thus given by  
	\ba{
	y^2 \supset& \left( \sum_{ l=2}^{ \ell} \sum_{j=1}^{l-1}+  \sum_{x=2}^{N/\ell} \sum_{ l=(x-1)\ell + 1}^{x \ell} \sum_{j=1}^{l-x}  \right)x^{N-l} v_{l,j} z^{-j}\ ,\qquad \Delta(v_{l,j}) = \frac{k l + j N}{k+N} \ .  \label{eq:ddef}
	}
The $v$-parameters associated to the regular puncture appear as points in the lower right quadrant in the diagram of the Newton polygon, where the lowest point occurs at $(a,b)=(0,-N(1-\frac{1}{\ell}) )$.\footnote{~Note that the addition of the regular puncture also allows the deformations corresponding to the $z^{k-1}$ line in the Newton polygon to be turned on, since translational symmetry on the sphere is broken.}

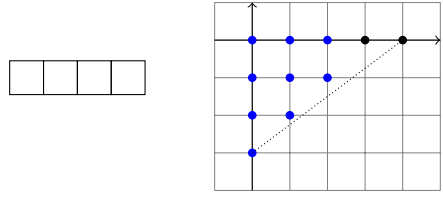
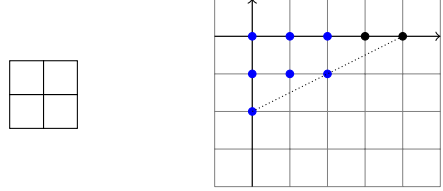
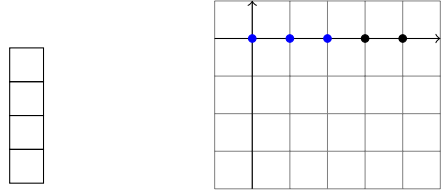
\begin{figure}
\centering
 \begin{subfigure}[b]{1\textwidth}
 \centering
 \begin{tikzpicture}
 \begin{scope}[shift={(-3,-0.5)}]
 \node [square,scale=1.2] at (0,0) (s1) { };
  \node [square,scale=1.2] at (0.45,0) (s2) { };
  \node [square,scale=1.2] at (0.9,0) (s3) { };
  \node [square,scale=1.2] at (1.35,0) (s4) { };
 \end{scope}
    \begin{scope}[shift={(0,0)}]
    \draw[very thin,color=gray,step=0.5] (-0.5,-2) grid (2.5,0.5);
   \draw[->] (0,-2) -- (0,0.5);
   \draw[->] (-0.5,0) -- (2.5,0);
   
   \filldraw[color=blue] (0,0) circle (0.05);
   \filldraw[color=blue] (0.5,0) circle (0.05);
   \filldraw[color=blue] (1,0) circle (0.05);
   \filldraw[color=black] (1.5,0) circle (0.05);
   \filldraw[color=black] (2,0) circle (0.05);
   
   \filldraw[color=blue] (0,-0.5) circle (0.05);
   \filldraw[color=blue] (0.5,-0.5) circle (0.05);
   \filldraw[color=blue] (1,-0.5) circle (0.05);
   \filldraw[color=blue] (0,-1) circle (0.05);
   \filldraw[color=blue] (0.5,-1) circle (0.05);
   \filldraw[color=blue] (0,-1.5) circle (0.05);
   
   \draw[densely dotted] (0,-1.5) -- (2,0);
   \end{scope}
%
%
%
%
%
\end{tikzpicture}
 \caption{The maximal puncture with flavor symmetry $SU(N)$. The Young tableaux consists of a single row of length $N$. $p_l = l -1$, such that $p_{l} = \{ 0, 1,2,3 \}$ for $N=4$. }
 \end{subfigure}
 %
 %
\begin{subfigure}[b]{1\textwidth}
\centering
  \begin{tikzpicture}
   \begin{scope}[shift={(-3,-1)}]
 \node [square,scale=1.2] at (0,0) (s1) { };
  \node [square,scale=1.2] at (0.45,0) (s2) { };
    \node [square,scale=1.2] at (0,0.45) (s3) { };
  \node [square,scale=1.2] at (0.45,0.45) (s4) { };
 \end{scope}
     \begin{scope}[shift={(0,0)}]
    \draw[very thin,color=gray,step=0.5] (-0.5,-2) grid (2.5,0.5);
   \draw[->] (0,-2) -- (0,0.5);
   \draw[->] (-0.5,0) -- (2.5,0);
   
   \filldraw[color=blue] (0,0) circle (0.05);
   \filldraw[color=blue] (0.5,0) circle (0.05);
   \filldraw[color=blue] (1,0) circle (0.05);
   \filldraw[color=black] (1.5,0) circle (0.05);
   \filldraw[color=black] (2,0) circle (0.05);
   
      \filldraw[color=blue] (0,-0.5) circle (0.05);
   \filldraw[color=blue] (0.5,-0.5) circle (0.05);
   \filldraw[color=blue] (1,-0.5) circle (0.05);
   \filldraw[color=blue] (0,-1) circle (0.05);
   
      \draw[densely dotted] (0,-1) -- (2,0);
     \end{scope}
%
%
%
%
%
\end{tikzpicture}
  \caption{The puncture with flavor symmetry $SU(N/2)$. The Young tableaux consists of two rows of length $N/2$. $p_{l=1,\dots,N/2} = l -1$ and $p_{l=N/2+1,\dots,N}=l-2$, such that $p_{l} = \{ 0, 1,1,2 \}$ for $N=4$. }
  \end{subfigure}
\begin{subfigure}[b]{1\textwidth}
\centering
  \begin{tikzpicture}
       \begin{scope}[shift={(-3,-1.7)}]
 \node [square,scale=1.2] at (0,0) (s1) { };
  \node [square,scale=1.2] at (0,0.45) (s2) { };
   \node [square,scale=1.2] at (0,0.9) (s3) { };
    \node [square,scale=1.2] at (0,1.35) (s4) { };
 \end{scope}
    \begin{scope}[shift={(0,0)}]
    \draw[very thin,color=gray,step=0.5] (-0.5,-2) grid (2.5,0.5);
   \draw[->] (0,-2) -- (0,0.5);
   \draw[->] (-0.5,0) -- (2.5,0);
   
   \filldraw[color=blue] (0,0) circle (0.05);
   \filldraw[color=blue] (0.5,0) circle (0.05);
   \filldraw[color=blue] (1,0) circle (0.05);
   \filldraw[color=black] (1.5,0) circle (0.05);
   \filldraw[color=black] (2,0) circle (0.05);
   \end{scope}
%
%
%
%
%
   
\end{tikzpicture}
     \caption{The non-puncture. The Young tableaux consists of $N$ rows of length 1.  $p_{l} =0$ for all $l$. }
     \end{subfigure}
     \caption{The Young tableaux and Newton polygons of some of the possible regular punctures for $N=4$. The black dots correspond to the $x^{N-1}$ line and the leading $x^N$ term, which have no corresponding deformation parameters. \label{fig:reg}}
 \end{figure}

\begin{figure}
\centering
 \begin{tikzpicture}
     \begin{scope}[shift={(0,0)}]
    \draw[very thin,color=gray,step=0.5] (-0.5, -2) grid (2.5,4.5);
   \draw[->] (0,-2) -- (0,4.5);
      \draw[->] (0,0) -- (2.5,0);
   
   \filldraw[color=blue] (0,0) circle (0.05);
   \filldraw[color=blue] (0.5,0) circle (0.05);
   \filldraw[color=blue] (1,0) circle (0.05);
   \filldraw[color=black] (1.5,0) circle (0.05);
   \filldraw[color=black] (2,0) circle (0.05);
   
      \filldraw[color=blue] (0,0.5) circle (0.05);
   \filldraw[color=blue] (0.5,0.5) circle (0.05);
   \filldraw[color=blue] (1,0.5) circle (0.05);
   \filldraw[color=black] (1.5,0.5) circle (0.05);
   
    \filldraw[color=blue] (0,1) circle (0.05);
   \filldraw[color=blue] (0.5,1) circle (0.05);
   \filldraw[color=blue] (1,1) circle (0.05);
   \filldraw[color=black] (1.5,1) circle (0.05);
   
      \filldraw[color=blue] (0,1.5) circle (0.05);
   \filldraw[color=blue] (0.5,1.5) circle (0.05);
   \filldraw[color=blue] (1,1.5) circle (0.05);
   
         \filldraw[color=blue] (0,2) circle (0.05);
   \filldraw[color=blue] (0.5,2) circle (0.05);
   \filldraw[color=blue] (1,2) circle (0.05);
   
   \filldraw[color=blue] (0,2.5) circle (0.05);
   \filldraw[color=blue] (0.5,2.5) circle (0.05);

   \filldraw[color=blue] (0,3) circle (0.05);
   \filldraw[color=blue] (0.5,3) circle (0.05);  
   
      \filldraw[color=blue] (0,3.5) circle (0.05);
           \filldraw[color=black] (0,4) circle (0.05);

     \draw[densely dotted] (0,4) -- (2,0);

   \draw[->] (-0.5,0) -- (2.5,0);
   
   \filldraw[color=blue] (0,-0.5) circle (0.05);
   \filldraw[color=blue] (0.5,-0.5) circle (0.05);
   \filldraw[color=blue] (1,-0.5) circle (0.05);
   \filldraw[color=blue] (0,-1) circle (0.05);
   \filldraw[color=blue] (0.5,-1) circle (0.05);
   \filldraw[color=blue] (0,-1.5) circle (0.05);
   
   \draw[densely dotted] (0,-1.5) -- (2,0);
     \end{scope}
     \end{tikzpicture}
     \caption{The Newton polygon for $\ell=N$, drawn for $N=4$ and $k=2N=8$.  \label{fig:irreg2}}
\end{figure}
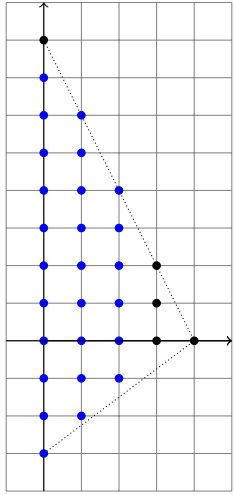

In Figure \ref{fig:reg} we give the Young tableaux and Newton polygons of the possible box-diagram regular punctures for $N=4$. 
The Newton polygons in Figure \ref{fig:reg} are reflected versions of those in Figure 7 of \cite{Gaiotto:2009we}, given in conventions such that they can be appended to the bottom of the Newton polygon of the $\ell=1$ theory (given for $N=4$, $k=8$ in Figure \ref{fig:irreg}). For example, the Newton polygon of the $\ell=N$ theory is depicted in Figure \ref{fig:irreg2} for $N=4$ and $k=8$. To summarize, the Newton polygon for general $\ell > 0$ consists of a right triangle in the first quadrant whose hypotenuse has slope $k/N$, plus a reflected right triangle below the horizontal axis whose hypotenuse has slope $(1- \frac{1}{\ell})$. 

Since all the $v_{l,j}$ in \eqref{eq:ddef} have dimension greater than unity, they  correspond to Coulomb branch operators $u_i$. The dimensions of the Coulomb branch operators then satisfy
	\ba{
	1 < \Delta(u_i) \leq N - \frac{N^2}{\ell(k+N)}\ . \label{eq:cbop}
	}
Thus, the rank of the Coulomb branch is increased from  \eqref{eq:rcb1} at $\ell=1$ to 
	\ba{
	\text{rank}(\text{CB})= \frac{1}{2} \left(  (k-1)(N-1) - (\text{GCD}(k,N) -1) + N^2 \left(1 - \frac{1}{\ell}\right)    \right) \label{eq:rcb2}
	}	
	for general $\ell$.
The rank of the flavor symmetry is now equal to \eqref{eq:rf}, plus $\ell-1$ due to the additional $SU(\ell)$ global symmetry of to the regular puncture,
	\ba{
	\text{rank}(F) =\text{GCD}(k,N)+ \ell-2 \  .
	}
The dimension of the conformal manifold is unchanged from \eqref{eq:conff} by the regular puncture.

\subsubsection{Central Charges}

The combination $2a-c$ of the $a$ and $c$ central charges can be computed using the useful formula \cite{Shapere:2008zf}\footnote{~This result uses the assumption that the Coulomb branch is freely generated.}
	\ba{
	2a-c = \frac{1}{4} \sum_i (2 \Delta(u_i)-1 )\ ,\label{eq:2ac}
	}
which relates the central charges to the dimensions of the Coulomb branch operators $u_i$. The dimensions of the $u_i$ for the $(A_{N-1}^{(N)}[k],Y_{\ell})$ theories are described around \eqref{eq:cbo} and \eqref{eq:ddef}. The quantities $a$ and $c$ can be separately extracted from another useful set of formulae derived in in  \cite{Shapere:2008zf},   
	\ba{
	a = \frac{R(A)}{4} + \frac{R(B)}{6} + \frac{5r}{24} + \frac{h}{24}\ ,\qquad c = \frac{R(B)}{3} + \frac{r}{6} + \frac{h}{12} \ , \label{eq:acr}
	}
where $R(A)$ and $R(B)$ are R-charges computed from topological field theories,  and $r,h$ are the number of free vector multiplets and hypermultiplets at a generic point on the Coulomb branch. For the $(A_{N-1}^{(N)}[k],Y_{\ell})$ theories, $r$ is equal to $\text{rank(CB)}$ given in \eqref{eq:rcb2}, and $h=0$. The quantity $R(A)$ can then be expressed
	\ba{
	R(A) = \sum_i (\Delta(u_i)-1 )\ ,\label{eq:ra}
	}
	which can be computed from the Newton polygon. The quantity
$R(B)$ was conjectured in \cite{Xie:2012hs} for the $(A_{N-1},A_{k-1})$ theories to be
	\ba{
	R(B) = \frac{ k N(N-1)  ( k - 1)}{4(N+k)}\ ,\qquad (A_{N-1},A_{k-1}) \ ,\label{eq:rb}
	}
with the conjecture confirmed and then computed for the general  $(G,G')$ Argyres-Douglas theories in  \cite{Cecotti:2013lda} (see more discussion of this generalized class in appendix \ref{sec:AD}). 

Given \eqref{eq:rb}, our knowledge of the Coulomb branch operator spectrum from the Newton polygon, and \eqref{eq:rcb2}, $a$ and $c$ can be computed for the $\ell=1$ theories as 
\ba{
\bs{
a &=  \frac{  4k^2 (N^2-1) - 5 (k+N)(N-2+\text{GCD}(k,N))  }{48 (k+N)} \\
	&\quad + \frac{N}{8(k+N)}   \sum_{j=1}^{N-1} \left\{\frac{j (k+N)}{N}  \right\} \left( 1 - \left\{\frac{j (k+N)}{N}  \right\} \right)\ ,\\
c &= \frac{ (N-1)(k-1) (kN + k + N) - (k+N) (\text{GCD}(k,N) - 1 ) }{12(k+N)} \ ,\quad \ell=1\ .} \label{eq:ccc}
}
Here $\{x\} = x - \lfloor x \rfloor$ denotes the fractional part.
Taking $k=mN$ for $m \in \mathbb{Z}_+$, the central charges \eqref{eq:ccc} reduce to
	\ba{ \bs{
	a &= \frac{ (N-1) \left( 2 N(N+1) m^2 - 5m - 5 \right)}{24(m+1)}   \ , \\
        c &= \frac{(N-1) \left( N(N+1) m^2 - 2m-2  \right)}{12 (m+1)} \ ,\qquad k=mN\ ,
	} \label{eq:ac1}}
which were first computed in this case in \cite{Xie:2012hs}.

Next we include the regular puncture associated to the diagram $Y_{\ell}$.  When $\ell=N$ and the puncture is maximal, the central charges were first computed for $k=mN$ in \cite{Xie:2013jc}, and for general $k,N$ in  \cite{Cecotti:2013lda} (also see the nice summary in \cite{Giacomelli:2020ryy}). In this case,\footnote{~In the notation of  \cite{Cecotti:2013lda}, $p=k+N$.}
\ba{\bs{
a&= \frac{1}{48}\left(10 + 4 k (N^2-1) - 5 \text{GCD}(k,N)  - 4 N + N^2 (-5 + 4 N) \right) \\
&\quad + \frac{ N}{8(k+N)}  \sum_{j=1}^{N-1} \left\{\frac{j (k+N)}{N}  \right\} \left( 1 - \left\{\frac{j (k+N)}{N}  \right\} \right)\ , \\
c &= \frac{1}{12} \left(  (k+N - 1)(N^2-1)    + 1 - \text{GCD}(k,N)   \right) \ ,\qquad \ell = N\ .
}\label{eq:aa1}}
Again, $\{..\}$ denotes the fractional part. 
The flavor central charge in this case is \cite{Xie:2016evu}
	\ba{
	k_{SU(N)} = 2 N  - \frac{2N}{k+N}\ . \label{eq:maxkf}
	}

Now consider the case of an arbitrary regular puncture labeled by $Y$ on the sphere. As described in \cite{Giacomelli:2020ryy}, a straightforward way to compute the central charges is to start with the known central charges of the $\ell=N$ theories, and then to partially close the maximal puncture by giving a nilpotent VEV to the moment map operator of the $SU(N)$ flavor symmetry. 
The central charges of the $(A_{N-1}^{(N)}[k],Y)$ theories can then be cast as those of the $(A_{N-1},A_{k-1})$ theories without the regular puncture \eqref{eq:ccc}, plus  the additional contributions \cite{Giacomelli:2020ryy}
	\ba{
	\Delta a = a_Y + \frac{6 I_{\rho_Y} - N(N^2-1)}{12} \frac{N}{k+N} \ ,\qquad \Delta c = c_Y + \frac{6 I_{\rho_Y} - N(N^2-1) }{12} \frac{N}{k+N}\ .\label{eq:genp}
	}
In \eqref{eq:genp},  $a_Y,c_Y$ denote the standard contribution of the puncture $Y$ ignoring the irregular puncture, which in this notation (following \cite{Giacomelli:2020ryy}) includes the  contribution of one puncture to the bulk 't Hooft anomalies,  $\chi\supset -1$ for $\chi$ the Euler characteristic (see appendix \ref{sec:appanomalies}). These quantities can be extracted from \eqref{eq:acbulk}, and are given below in examples.  $I_{\rho_Y}$ is the embedding index of $SU(2)$ into $SU(N)$ that labels the nilpotent VEV in the RG flow, defined in terms of the data of the Young diagram as
	\ba{
	I_{\rho_Y} = \frac{1}{6} \sum_{i=1}^{\tilde{p}} i (i^2-1) \tilde{k}_i\ ,\qquad F = S\left( \prod_{i=1}^{\tilde{p}} U(\tilde{k}_i)\right)\ . \label{eq:ir}
	}
Here $\tilde{p}$ is the number of rows in the Young tableaux, and $\tilde{k}_i = \tilde{\ell}_i - \tilde{\ell}_{i+1}$ for $\tilde{\ell}_i$ the length of the $i$'th row of the tableaux, in a convention where row lengths increase from top to bottom. Then, the Young tableaux is labeled by the partition $N = \sum_i i \tilde{k}_i$.  See appendix \ref{sec:appanomalies} for more details.

As a first example of applying \eqref{eq:genp}, consider the non-puncture $\ell=1$ given by the last diagram in Figure \ref{fig:reg}. In this case there are  $\tilde{p}=N$ rows each of length $\tilde{\ell}_{i=1,\dots,N}=1$, such that the only nonzero $\tilde{k}_i$ is $\tilde{k}_N=1$. Then, the embedding index is
	\ba{
	I_{\rho_{\text{non}}} = \frac{1}{6} N (N^2-1)\ ,	
	}
	and $\Delta a=\Delta c = 0$, as expected. 
	For the general box diagram $Y_{\ell}$ with $\ell$ columns,  $\tilde{p} = N/\ell$, $\tilde{\ell}_{i=1,\dots,N/\ell}=\ell$, and the only nonzero $\tilde{k}_i$ is $\tilde{k}_r=\ell$. The  contributions $(a_{Y_{\ell}}, c_{Y_{\ell}})$ are given by adding the contributions of \eqref{eq:boxa}, to the terms in \eqref{eq:acbulk} that are proportional to $\chi \supset -1$. For example, one computes $a_{Y_{\ell}}$ as
	\ba{
	a_{Y_{\ell}} &= \frac{1}{48} (8N^3-3N-5) + \frac{1}{24} (n_h(Y_{\ell}) + 5 n_v(Y_{\ell}) ) \\
	&= \frac{1}{48} N (-3 + 3 \ell- 8  \frac{N^2}{\ell} + 8 N^2  )\ ,\label{eq:ayy}
	}
	and similarly for $c_{Y_\ell}$. 
The embedding index of the box diagram is given by
	\ba{
	I_{\rho_{\ell}} =\frac{1}{6} N \left(\frac{N^2}{\ell^2}-1\right)\ . \label{eq:ibox}
	}
 Substituting \eqref{eq:ayy} and \eqref{eq:ibox} into \eqref{eq:genp} and adding the resulting $\Delta a$ to \eqref{eq:ccc} (and similarly for $c$), we obtain
\ba{
\bs{
a &= \frac{ 4k^2 (N^2-1)-5  (k+N) \left(  \frac{(8-3\ell) }{5}N-2+\text{GCD}(k,N)  \right)+  4N^3(1 -\frac{1}{\ell}) \left( 2 k  + N (1 - \frac{1}{\ell})\right) }{48 (k+N)}    \\
&\  + \frac{ N }{8 (k+N)}   \sum_{j=1}^{N-1} \left\{\frac{j (k+N)}{N}  \right\} \left( 1 - \left\{\frac{j (k+N)}{N}  \right\} \right)   \ ,    \\ 
c&=  \frac{k^2 (N^2-1)-(k+N) \left(N (2-\ell)-2 +  \text{GCD}(k,N)\right)   + N^3(1-\frac{1}{\ell}) (2 k  + N (1 - \frac{1}{\ell}) ) }{12 (k+N)}   \ .
 }}
 Taking $\ell=1$ reproduces \eqref{eq:ccc}, while taking $\ell=N$ reproduces \eqref{eq:aa1}.

The flavor symmetry central charge of the regular puncture is conjectured in \cite{Xie:2013jc} to be equal to two times the scaling dimension of the Coulomb branch operator of maximal dimension. Using \eqref{eq:cbop}, for the regular puncture labeled by $Y_{\ell}$ this translates into a flavor central charge $k_{SU(\ell)}$ of
	\ba{
	k_{SU(\ell)} = 2 N  - \frac{2 N^2}{\ell (k+N)}\ . \label{eq:ff}
	}
For example, $k_{SU(N)}$ reduces to \eqref{eq:maxkf} upon taking $\ell = N$.

In Table \ref{tab:tp1} we summarize the properties discussed in this subsection.

\begingroup
\renewcommand{\arraystretch}{2}
\begin{table}[h!]
\centering
\begin{tabular}{|c||c|}
\hline
 & $(A_{N-1}^{(N)}[k],Y_{\ell})$ \\ 
\hline  \hline
$a$ & $ \begin{array}{c}   \  \frac{ 4k^2 (N^2-1)-5  (k+N) \left(  \frac{(8-3\ell) }{5}N-2+\text{GCD}(k,N)  \right)+ 4N^3(1 -\frac{1}{\ell}) \left( 2 k  + N (1 - \frac{1}{\ell})\right)}{48 (k+N)}  \\
+ \frac{ N }{8 (k+N)}   \sum_{j=1}^{N-1} \left\{\frac{j (k+N)}{N}  \right\} \left( 1 - \left\{\frac{j (k+N)}{N}  \right\} \right) \end{array}$   \\ \hline
$c$ & $  \begin{array}{c} \  \frac{k^2 (N^2-1)-(k+N) \left(N (2-\ell)-2 +  \text{GCD}(k,N)\right)  +N^3(1-\frac{1}{\ell}) (2 k  + N (1 - \frac{1}{\ell}) )  }{12 (k+N)}  
  \end{array}$  \\ \hline
$a|_{N\to \infty} = c|_{N\to \infty}$ & $\frac{N^2 \left( k + N(1-\frac{1}{\ell})\right)^2}{12 (k+N)}$  \\ \hline 
$k_{SU(\ell)}$ & $ 2 N  - \frac{2 N^2}{\ell (k+N)}$ \\ \hline 
$\text{rank}(\text{CB})$ & $\frac{1}{2} \left(  (k-1)(N-1) - (\text{GCD}(k,N) -1) + N^2 (1 - \frac{1}{\ell})    \right)$ \\ \hline 
$\text{rank}(F)$ & $ \text{GCD}(k,N)+\ell-2 $ \\ \hline 
$\text{dim}(\CM_{\CC})$ & $\text{GCD}(k,N)-1$ \\ \hline 
\end{tabular}
\caption{The properties of the $(A_{N-1}[k]^{(N)},Y_{\ell})$ theories that arise from $N$ M5-branes wrapping a sphere with one irregular puncture of type $A_{N-1}^{(N)}[k]$, and one regular puncture whose Young diagram consists of $\ell$ columns and $N/\ell$ rows. Here $\{ x \} = x - \lfloor x \rfloor $ denotes the fractional part. The case $\ell=1$ yields a ``non-puncture'' on the sphere, and these  reduce to the class $(A_{N-1},A_{k-1})$. The case $\ell=N$ yields the maximal regular puncture with associated $SU(N)$ flavor symmetry. To compute the large-$N$ scaling we assume that $k$ is of order $N$.   The dimension of the conformal manifold is reduced according to \eqref{eq:mmn} for special values of $k$. \label{tab:tp1}}
\end{table}
\endgroup

\subsection{Checks of the Holographic Duality}

Let us summarize the checks of our proposed holographic duality between the field theory data of the $(A_{N-1}^{(N)}[k],Y_{\ell})$ SCFTs described in this section (and summarized in Table \ref{tab:tp1}), and the supergravity solution described in sections \ref{sec:spindlesolutions} and \ref{sec:anomaly}. 

\begin{itemize}

\item Comparing the $U(1)_r$ symmetry generator in the field theory given in \eqref{eq:rp} with \eqref{Rsym_and_flavor}, we are led to identify $\frac{N\ell}{N+ K \ell}$ with $\frac{N}{k+N}$, yielding
	\ba{
	K = k+ N\left(1 - \frac{1}{\ell}\right)\ . \label{eq:translate}
	}
	(Recall that $\ell$ divides $N$.)
	The condition $k>-N$ on the field theory side is consistent with the positivity of $K$ in the supergravity solution.
	
\item The large-$N$ limit of the $a$ and $c$ central charges is given in Table \ref{tab:tp1}, where we take $N,k\to \infty$ with $k/N$ finite.
 Using \eqref{eq:translate}, this can be rewritten
	\ba{
	a|_{N\to \infty} = c|_{N\to \infty} = \frac{\ell  N^2 K^2}{12 (  N+K \ell)}\ ,
	}
which precisely matches \eqref{central_charge}.

\item The flavor central charge $k_{SU(\ell)}$ for the $SU(\ell)$ symmetry associated to the regular puncture $Y_{\ell}$ is given in \eqref{eq:ff} in the field theory. Translating from $k$ to $K$ using \eqref{eq:translate}, we can rewrite
	\ba{
	k_{SU(\ell)} = \frac{2 N K \ell}{N + K \ell}\ . \label{eq:ks}
	}
This precisely matches the computation of $k_{SU(\ell)}$ obtained in \eqref{holo_flavor_charge}.

\item From \eqref{eq:cbop}, the largest dimension Coulomb-branch operator has scaling dimension given by $N - \frac{N^2}{\ell(k+N)}$, with R-charges satisfying $(r,R)=(2\Delta,0)$. Using \eqref{eq:translate}, we find that this precisely matches the dimension and R-charges of the operator $\CO_1$ computed in \eqref{eq:o1}, \eqref{brane_Rcharges}. We thus identify the wrapped M2-brane operator $\CO_1$ with the maximal dimension Coulomb branch operator.

\item The rank of the flavor symmetry in the field theory is $\text{GCD}(k,N) +\ell - 2$. Of the total rank, $\ell - 1$ corresponds to the $SU(\ell)$ flavor symmetry that is evident both in the field theory and gravity descriptions. The maximal remaining rank is $N-1$, which matches the maximal possible rank from the source consisting of $N$ M5-branes in the supergravity solutions (with the minus one corresponding to an overall center of mass mode). It will be interesting to further understand the dynamics of the source on the gravity side, and to explicitly see the reduction from rank $N-1$ in the maximal case, to $\text{GCD}(k,N) -1$ depending on $k$. We also expect the matching of the dimension of the conformal manifold to depend on the detailed dynamics of the source.

\end{itemize}

A dual Lagrangian description of the $(A_{N-1}^{(N)}[k], Y_1) =  (A_{N-1},A_{k-1})$ Argyres-Douglas SCFTs for $k$ an integer multiple of $N$ was obtained in \cite{Agarwal:2017roi,Benvenuti:2017bpg}.\footnote{ The cases $N=2$ with general $k$ were first obtained in \cite{Maruyoshi:2016tqk,Maruyoshi:2016aim} via RG flow from conformal SQCD.  } The RG flow of interest begins in the UV with a conformal $\CN=2$ quiver gauge theory with $N-1$ gauge nodes and non-abelian flavor symmetry group $SU(k)$, to which we couple an $\CN=1$ chiral multiplet that transforms in the adjoint representation of the $SU(k)$ flavor group. One then undergoes the nilpotent Higgsing procedure that was first considered in \cite{Heckman:2010qv,Gadde:2013fma} (see also \cite{Tachikawa:2013kta}). Upon giving a particular vev to this chiral multiplet and decoupling the massive and Nambu-Goldstone modes (utilizing \cite{Agarwal:2014rua}), one flows to the $(A_{N-1},A_{(k=mN)-1})$ theory at low energies. This RG flow is reviewed in detail in appendix \ref{sec:lagrangian}, and both the UV and IR quivers are summarized in Figure \ref{fig:quiver}. As we review in that appendix, many of the properties of the $\ell=1$ theories reviewed here are reproduced by the Lagrangian description. 

Using the Lagrangian description, we have the following additional check for $\ell=1$ and $k$ an integer multiple of $N$:

\begin{itemize}

\item One can construct $2^{N}-2$ Higgs branch operators with $R$-charges $(r,R)=(0,\Delta)$, and scaling dimensions $\Delta = k  - \frac{k}{N}$ (see \eqref{eq:hbop} in appendix \ref{sec:lagrangian}).  Using \eqref{eq:translate} with $\ell = 1$ and taking the limit  that $k,N\to \infty$ with $k/N$ finite, these become
	\ba{
	\Delta |_{N\to\infty} = K\ .
	}
This precisely reproduces the dimensions and R-charges of the operators $\CO_2^i$ computed in \eqref{dim_O2}, \eqref{brane_Rcharges}. 
Recall from the discussion around \eqref{dim_O2} that in gravity, the degeneracy of the operators $\CO_2^i$ is determined by the possible boundary conditions of the M2-brane on the M5-branes at $w=0$, leading to a degeneracy of  $2^N-1$. We thus identify all but one of the operators $\CO_2^i$ with the Higgs branch operators on the field theory side, while one mode decouples from the interacting fixed point. It would be interesting to  understand the origin of this decoupled mode, which it is natural to expect is associated to the center-of-mass mode of the stack of M5-branes. 

\end{itemize}



\section{Discussion} \label{sec:discussion}

In this work we have proposed gravity duals
for a class of 4d $\cN = 2$ SCFTs of Argyres-Douglas (AD) type, which can be engineered
by wrapping a stack of M5-branes on a sphere
with one irregular puncture and one regular puncture. 
The latter is
described by a Young diagram of   rectangular shape.\footnote{ Rectangular Young diagrams from orbifold singularities were studied in \cite{Bobev:2019ore}.}
Our solutions have been found in
7d gauged supergravity and uplifted on $S^4$.
Crucially, we find M5-brane sources in the internal
space, which model the irregular puncture.
We test the proposed holographic duality by matching
the central charges and the dimensions of suitable 
BPS operators originating from wrapped M2-brane probes.

Our results suggest several natural directions for future
investigations. It would be interesting to obtain
a more systematic understanding
of the structure of the novel
solutions to the Toda equation
of the type we have discovered. In particular,
since the solutions are axially symmetric,
one can analyze the electrostatic system obtained after
the B\"acklund transform \cite{Gaiotto:2009gz}. The goal is to identify the gravity duals of 4d $\cN = 2$ SCFTs of AD type featuring a regular puncture
with a Young diagram of arbitrary shape,
and to explore whether the field-theoretic classification
of irregular punctures can be recovered
from the gravity side.

The St\"uckelberg coupling involving the $U(1)$ gauge field
associated to the  Killing vector $\partial_\beta$
deserves further study. 
In particular, it would be interesting to
identify the leftover discrete subgroup (if any) and
the states in the dual SCFTs that are charged under it. More broadly,
one can ask whether this phenomenon   appears in other
contexts in supergravity and how it is related to the presence of internal
sources. The inclusion of external background gauge fields in the 4-form flux of M-theory can be described
using equivariant cohomology. It would be beneficial
to phrase the St\"uckelberg mechanism at hand in
this broader mathematical language.

The anomaly inflow methods of \cite{Bah:2019rgq}
can be used to extract 't Hooft anomalies beyond
the leading terms in the large-$N$ limit. 
Moreover, $\cO(N^0)$ terms could also
be accessible via a study of singleton modes in the
gravity dual. It would be interesting to 
apply these ideas to the solutions discussed
in this paper, aiming to match the known exact
't Hooft anomalies of SCFTs of AD type.

It is also natural
to study  
generalizations  of our $AdS_5$ solutions
preserving 4d $\cN = 1$ superconformal symmetry.
A class of $\cN = 1$ solutions describing M5-branes 
wrapped on a spindle have been presented in \cite{Ferrero:2021wvk}.
We would like to analyze whether $\cN = 1$ solutions
with internal M5-brane sources can be found. More generally,
a systematic analysis of $\cN = 1$ gravity solutions
can yield useful insights into the spectrum of allowed
regular\footnote{ Holographic duals of $\mathcal{N}=1$ regular punctures were studied in \cite{Bah:2013wda,Bah:2015fwa,Bah:2017wxp}. } and irregular punctures 
for 4d $\cN = 1$ SCFTs in class $\cS$ and generalizations thereof \cite{Maruyoshi:2009uk,Benini:2009mz,Bah:2011je,Bah:2011vv,Bah:2012dg,
Bah:2013qya,
Bah:2015fwa}. 

A subset of the $\cN = 2$ SCFTs discussed in this paper can be realized as low-energy fixed points
of  Lagrangian $\cN = 1$ flows \cite{Agarwal:2017roi,Benvenuti:2017bpg}. 
It would be interesting to investigate Lagrangian
realizations of the larger class of theories of AD type
studied in this work. Furthermore, our  
$AdS_5$ solutions
offer a new avenue 
 to study the holographic duals
of these  supersymmetry-enhancing 
flows.


\section*{Acknowledgments}

We are grateful to 
Nikolay Bobev, Simone Giacomelli,
 Yifan Wang
 for interesting
conversations and correspondence. 
The work of IB   is supported in part by NSF grant PHY-1820784.
FB is supported by the European Union’s Horizon 2020 Framework: ERC Consolidator Grant 682608.
FB is supported by STFC Consolidated Grant ST/T000864/1.
RM is supported in part by ERC Grant 787320-QBH Structure
and by ERC Grant 772408-Stringlandscape.
The work of EN is supported by DOE grant DE-SC0020421.


\appendix



\section{Gauged Supergravity Solutions} \label{sec:appsugra}

In this appendix we provide a derivation of the 7d gauged supergravity
solutions described in the main text in section \ref{sec_7dsols}.
The supergravity model of interest is obtained as a $U(1)^2$ truncation of the 
full 7d $\cN = 4$ $SO(5)$ gauged supergravity of  \cite{Pernici:1984xx}. We follow the notation
and conventions of \cite{Liu:1999ai}.

\subsection{Equations of Motion and BPS Equations}  \label{app_EOMs}
The bosonic equations of motion are recorded in \cite{Liu:1999ai}.
The scalar equations of motion read
\begin{align}
\nabla^2( 3 \, \lambda_1 + 2 \, \lambda_2) &=  - e^{- 4 \lambda_1} \, F^{(1)}_{\mu\nu} \, F^{(1) \mu\nu}
+ m^2 \, e^{-4 \lambda_1 - 4 \lambda_2} \, C_{\mu\nu\rho} \, C^{\mu\nu\rho}
+ \frac{m^2}{8} \, \frac{\partial \cV}{\partial \lambda_1} \ , \\
\nabla^2( 2 \, \lambda_1 + 3 \, \lambda_2) &=  - e^{- 4 \lambda_2} \, F^{(2)}_{\mu\nu} \, F^{(2) \mu\nu}
+ m^2 \, e^{-4 \lambda_1 - 4 \lambda_2} \, C_{\mu\nu\rho} \, C^{\mu\nu\rho}
+ \frac{m^2}{8} \, \frac{\partial \cV}{\partial \lambda_2} \ , \label{scalar_EOM_2}
\end{align}
where $\cV$ is the scalar potential, given as
\beq
\cV = - 8 \, e^{2 \lambda_1 + 2 \lambda_2} - 4 \, e^{- 2 \lambda_1 - 4 \lambda_2}
- 4 \, e^{- 4 \lambda_1 - 2 \lambda_2}
+ e^{- 8 \lambda_1 - 8 \lambda_2}  \ .
\eeq
The gauge field equations of motion are
\begin{align}
\nabla^\mu (  e^{- 4 \lambda_1} \, F^{(1)}_{\mu\nu}  ) &= \frac{1}{2 \, \sqrt 3} \, \epsilon_{\mu\nu}{}^{\rho_1 \dots \rho_5} \, \nabla^\mu (  F^{(2)}_{\rho_1 \rho_2} \, C_{\rho_3 \rho_4 \rho_5} ) \ , \label{EOM_vec1}  \\
\nabla^\mu (  e^{- 4 \lambda_2} \, F^{(2)}_{\mu\nu}  ) &= \frac{1}{2 \, \sqrt 3} \, \epsilon_{\mu\nu}{}^{\rho_1 \dots \rho_5} \, \nabla^\mu (  F^{(1)}_{\rho_1 \rho_2} \, C_{\rho_3 \rho_4 \rho_5} ) \ , \label{EOM_vec2}
\end{align}
where $F^{(1)}_{\mu\nu} = 2 \, \partial_{[\mu} A^{(1)}_{\nu]}$,
$F^{(2)}_{\mu\nu} = 2 \, \partial_{[\mu} A^{(2)}_{\nu]}$,
while the 
 3-form equation of motion is
\beq \label{threeform_EOM}
e^{- 4 \lambda_1 - 4 \lambda_2} \, C_{\mu_1 \mu_2 \mu_3} = \frac{1}{6\, m} \, \epsilon_{\mu_1 \mu_2 \mu_3}{}^{\nu_1 \dots \nu_4} \, \partial_{\nu_1} C_{\nu_2 \nu_3 \nu_4}
- \frac{1}{2 \, m^2 \, \sqrt 3} \, \epsilon_{\mu_1 \mu_2 \mu_3}{}^{\nu_1 \dots \nu_4} \, 
F^{(1)}_{\nu_1 \nu_2} \, F^{(2)}_{\nu_3 \nu_4}  \ .
\eeq
Finally, Einstein's equation can be written in the form
\begin{align}
R_{\mu\nu} & =  \frac{m^2}{10} \, \cV \, g_{\mu\nu}
+ 5 \, \partial_\mu (\lambda_1 + \lambda_2) \, \partial_\nu (\lambda_1 + \lambda_2)
+ \partial_\mu (\lambda_1 - \lambda_2) \, \partial_\nu (\lambda_1 - \lambda_2) \nn \\
& + 2 \, e^{- 4 \lambda_1} \, \big[  F^{(1)}_{\mu \rho} \, F^{(1)}{}_\nu {}^\rho   
- \tfrac{1}{10} \, g_{\mu\nu} \, F^{(1)}_{\rho \sigma} \, F^{(1) \rho \sigma}
 \big]
 + 2 \, e^{- 4 \lambda_2} \, \big[  F^{(2)}_{\mu \rho} \, F^{(2)}{}_\nu {}^\rho   
- \tfrac{1}{10} \, g_{\mu\nu} \, F^{(2)}_{\rho \sigma} \, F^{(2) \rho \sigma}
 \big]
 \nn \\
& - 3 \, m^2 \, e^{-4 \lambda_1 - 4 \lambda_2}  \big[  C_{\mu \rho \sigma} \, C_\nu {}^{\rho \sigma}  
- \tfrac{2}{15} \, g_{\mu\nu} \, C_{\rho_1 \rho_2 \rho_3} \, C^{\rho_1 \rho_2 \rho_3} \big] \ .
\end{align}
The constant parameter $m$ has dimensions of mass. It sets the scale of the $AdS_7$ vacuum
solution of the 7d gauged supergravity model: in these conventions, the radius of $AdS_7$ is $L_{AdS_7} = 2/m$.
The BPS equations of this supergravity model are \cite{Liu:1999ai}
\begin{align}
0 & =  
\nabla_\mu \epsilon^{\cI}   +\frac g2 \, \Big[ A^{(1)}_\mu \, (\Gamma^{12})^{\cI}{}_{\cJ}  
+ A^{(2)}_\mu \, (\Gamma^{34})^{\cI}{}_{\cJ}
\Big] \, \epsilon^{\cJ}
+ \frac m4 \, e^{-4 \lambda_1 - 4 \lambda_2} \, \gamma_\mu  \, \epsilon^{\cI}
+ \frac 12 \, \gamma_\mu \, \gamma^\nu \, \partial_\nu (\lambda_1 + \lambda_2) \, \epsilon^{\cI}
\nn \\
& \qquad +  \frac 12 \, \gamma^\nu \, e^{- 2 \lambda_1}\, F^{(1)}_{\mu\nu} \, (\Gamma^{12})^{\cI}{}_{\cJ} \,
\epsilon^{\cJ}
+  \frac 12 \, \gamma^\nu \, e^{- 2 \lambda_2}\, F^{(2)}_{\mu\nu} \, (\Gamma^{34})^{\cI}{}_{\cJ} \, \epsilon^{\cJ}
\nn \\
& \qquad - \frac{m \, \sqrt 3}{4} \, \gamma^{\nu \rho} \, e^{ - 2  \lambda_1 - 2 \lambda_2} \, C_{\mu\nu\rho} \, (\Gamma^5)^{\cI}{}_{\cJ} \, \epsilon^{\cJ}
   \ ,    \label{delta_gravitino} \\
0 & =  
\frac m4 \, (e^{2\lambda_1}  - e^{- 4\lambda_1 - 4 \lambda_2} )  \, \epsilon^{\cI}
- \frac 14 \, \gamma^\mu \, \partial_\mu (3\, \lambda_1 + 2 \, \lambda_2) \, \epsilon^{\cI}
- \frac 18 \, \gamma^{\mu\nu} \, e^{- 2 \lambda_1} \, F^{(1)}_{\mu\nu} \, (\Gamma^{12})^{\cI}{}_{\cJ} \, \epsilon^{\cJ}
\nn \\
& \qquad + \frac{m}{8 \, \sqrt 3} \, \rho^{\mu\nu\rho} \, e^{- 2 \lambda_1 - 2 \lambda_2 } \, C_{\mu\nu\rho} \, (\Gamma^5)^{\cI}{}_{\cJ} \, \epsilon^{\cJ}
 \  , \label{delta_fermion1} \\
 0 & =  
\frac m4 \, (e^{2\lambda_2}  - e^{- 4\lambda_1 - 4 \lambda_2} )  \, \epsilon^{\cI}
- \frac 14 \, \gamma^\mu \, \partial_\mu (2\, \lambda_1 + 3 \, \lambda_2) \, \epsilon^{\cI}
- \frac 18 \, \gamma^{\mu\nu} \, e^{- 2 \lambda_2} \, F^{(2)}_{\mu\nu} \, (\Gamma^{34})^{\cI}{}_{\cJ} \,
\epsilon^{\cJ}
\nn \\
& \qquad + \frac{m}{8 \, \sqrt 3} \, \rho^{\mu\nu\rho} \, e^{- 2 \lambda_1 - 2 \lambda_2 } \, C_{\mu\nu\rho} \, (\Gamma^5)^{\cI}{}_{\cJ} \, \epsilon^{\cJ}
 \  .  \label{delta_fermion2}
\end{align}
The constant $g$ is the gauge coupling of the 7d gauged supergravity model, related to $m$ as
$g = 2\, m$.
The supersymmetry parameter $\epsilon$ is a 7d Dirac spinor, but we do not indicate explicitly its spacetime spinor index. It also carries a  index $\cI = 1, \dots, 4$ associated to the 
 $\mathbf 4$ representation of $SO(5)_c \cong USp(4)_c$, which is the composite
$SO(5)_c$ symmetry of the scalar coset of the full 7d $\cN = 4$ $SO(5)$ gauged supergravity \cite{Pernici:1984xx}. The index $\cI$ on $\epsilon^{\cI}$ is acted upon
by $SO(5)_c$ gamma matrices $\Gamma^1$, \dots, $\Gamma^5$; we have introduced
$\Gamma^{12} = \Gamma^1 \, \Gamma^2$ and $\Gamma^{34} = \Gamma^3 \, \Gamma^4$.
The 7d spacetime gamma matrices $\gamma_\mu$ commute with the $SO(5)_c$ gamma matrices
$\Gamma^1$, \dots, $\Gamma^5$.

\subsection{Ansatz}

The ansatz for the 7d line element reads
\beq \label{metric_ansatz}
ds^2_7 = f(w) \, ds^2(AdS_5) + g_1(w) \, dw^2  + g_2(w) \, dz^2 \ ,
\eeq
where $ds^2(AdS_5)$ denotes the unit-radius metric on $AdS_5$, $w$ parametrizes
an interval, and $z$ is an angular coordinate, whose periodicity will be fixed later.
The gauge field $A^{(1)}$ takes the form
\beq
A^{(1)}  = A_z (w) \, dz \ ,
\eeq
while $A^{(2)}$ and the 3-form $C$ are set to zero.
The scalar fields $\lambda_1$, $\lambda_2$ are given in terms of a single function of $w$,
\beq
\lambda_1 = \lambda(w) \ , \qquad
\lambda_2 =  - \frac 23 \, \lambda(w)  \ .
\eeq
These choices guarantee that the equations of motion for the 3-form, one of the scalars, and one
of the vectors are 
 automatically satisfied, see \eqref{threeform_EOM}, \eqref{scalar_EOM_2}, \eqref{EOM_vec2}.

The metric \eqref{metric_ansatz} suggests an obvious vielbein,
with the flat directions $0, \dots,4$ associated to $AdS_5$,
and the flat directions $5$, $6$ associated to $w$, $z$, respectively. 
The 7d gamma matrices are decomposed according to
\beq
\gamma^{\alpha} = \rho^\alpha \otimes \sigma^3 \ , \qquad \alpha = 0,1,2,3,4 \ , \qquad
\gamma^5 =   \mathbb I_4 \otimes \sigma^1 \ , \qquad
\gamma^6 =   \mathbb I_4 \otimes \sigma^2  \ ,
\eeq
where $\alpha$ are (flat) 5d spacetime indices,
$\rho^\alpha$ are 5d $4 \times 4$ gamma matrices satisfying $\{ \rho^\alpha , \rho^\beta \} = 2 \, \eta^{\alpha \beta}$ with signature $(-,+,+,+,+)$, and $\sigma^{1,2,3}$ are the standard Pauli matrices.
The 7d supersymmetry parameter $\epsilon$ is written in the form
\beq
\epsilon^\cI = n^\cI \,  \vartheta \otimes \eta \ .
\eeq
The quantity $\vartheta$ is a 4-component Killing spinor on $AdS_5$, satisfying
\beq
\nabla^{ AdS_5}_\alpha\,  \vartheta = \frac 12 \, s \, \rho_\alpha \, \vartheta \ ,
\eeq
where $s\in \{ \pm 1\}$ is an arbitrary sign. The quantity $\eta$ is a 2-component spinor
depending on the coordinates $w$ and $z$. 
The quantities $n^\cI$
are the components of a constant object in the $\mathbf 4$ representation of $SO(5)_c$. 
They are subject to the  projection condition
\beq
(\Gamma^{12})^{\cI}{}_{\cJ} \, n^\cJ = i \, n^\cI \ .
\eeq
(Flipping the sign of the gauge field $A^{(1)}$ we could have equivalently
written $-i$ on the RHS.)
Notice that we do not impose any additional projection condition on $n^\cI$ with $\Gamma^{34}$.
In other words, two out of the four  components $n^\cI$ are retained by the projection.
This ensures that, if a 2-component spinor $\eta$ can be found that satisfies
the BPS conditions spelled out below, the system automatically preserves 4d
$\cN = 2$ superconformal symmetry.

The $AdS_5$, $w$, and $z$ components of the BPS equation \eqref{delta_gravitino}  give respectively
\begin{align}
0 & =  \frac 12 \, s \, \eta
+ \frac 12 \, g_1^{-1/2} \, f^{1/2} \,  \bigg[\frac 12 \, \frac{ f'}{f}  
+ \frac 13 \,  \lambda'  
  \bigg] \, (i \, \sigma^2 \, \eta)
  + \frac m4 \, e^{- \frac 43 \lambda} \, f^{1/2} \, (\sigma^3 \, \eta) \ , 
  \label{gravitino_alpha} \\
0 & = 
\partial_w \eta
+ \frac 16 \, \lambda' \, \eta
+ \frac 12 \, g_2^{-1/2} \, 
 e^{- 2 \lambda} \,   A_{z} '
\, (i \, \sigma^2 \, \eta)   +  \frac m4 \, e^{- \frac 43 \lambda } \, g_1^{1/2} \, (\sigma^1 \, \eta )\ , 
  \label{gravitino_y} \\
  0 & = \bigg[ 
 - i \, \partial_z   + \frac 12 \, g \,    A_z  
   \bigg] \, \eta
 - \frac 12 \, g_1^{-1/2} \,  
 e^{- 2 \lambda} \,  A_{z} ' 
  \, (\sigma^1 \, \eta) \nn \\
  &  
 - \frac m4 \, e^{-\frac 43 \lambda }  \, g_2^{1/2}
\, (i \, \sigma^2 \, \eta)
 - \frac 12 \, g_1^{-1/2} \, g_2^{1/2}   \,  \bigg[
\frac 12 \, \frac{g_2'}{g_2}
+ \frac 13 \,  \lambda'  
  \bigg] \, (\sigma^3 \, \eta) \ .
    \label{gravitino_z}   
\end{align}
The BPS condition \eqref{delta_fermion1} yields
\begin{align}
0 & = m \, (e^{2\lambda} - e^{-\frac 43 \lambda}  ) \, \eta
-  \frac 53 \,   \lambda' \, g_1^{-1/2} \, (\sigma^1 \, \eta)
+   \,   e^{- 2 \lambda} \,  A_z '  \, g_1^{-1/2}\, g_2^{-1/2} \, (\sigma^3 \, \eta) 
    \label{fermion_1}     \ ,
\end{align}
while the BPS equation \eqref{delta_fermion2} is trivially satisfied.
Here and in the following a prime denotes differentiation
with respect to $w$.

We assume that the   spinor $\eta$ has a definite charge under
the $U(1)_z$ isometry, \emph{i.e.}
\beq
\eta(w,z) = e^{i n z} \, \widehat \eta(w) \ ,
\eeq
where $n$ is a constant. We notice that 
$\partial_z \eta$ enters the BPS equations only in the combination
$(- i \,\partial_z + \tfrac 12 \, g \, A_z)\eta = (n+ \tfrac 12 \, g \, A_z)\eta$.
The quantity $(- i \,\partial_z + \tfrac 12 \, g \, A_z)\eta$
is invariant under a combined transformation of $\eta$, $A^{(1)}$ of the form
\beq \label{spinor_and_gauge}
A^{(1)} \mapsto A^{(1)}  - \frac{2 \, \alpha_0}{g} \, dz \ , \qquad \eta \mapsto e^{i \alpha_0 z} \, \eta \ ,
\eeq
where $\alpha_0$ is an arbitrary constant.
It is thus convenient to define $\widehat A_z$ via
\beq \label{hatAz_def}
\tfrac 12 \, g \, \widehat A_z = n+ \tfrac 12 \, g \, A_z \ .
\eeq
Notice that $\widehat A_z' =A_z'$. 


\subsection{Analysis}

We can solve the equation of motion \eqref{EOM_vec1} for the gauge field $A^{(1)}$ by writing
\beq \label{the_extra_eq}
A_z '= b \, e^{4\lambda} \, g_1^{1/2} \, g_2^{1/2} \, f^{-5/2} \ ,
\eeq
where $b$ is an unspecified integration constant. We plug this expression for 
$ A_z'$ in the BPS equations \eqref{gravitino_alpha}, 
\eqref{gravitino_z}, \eqref{fermion_1}, and, after some rearrangements (including multiplying
from the left by suitable Pauli matrices), we arrive at
the following algebraic BPS conditions,
\begin{align}
0 & =   s \, f^{-1/2} \, \eta
+  g_1^{-1/2} \,    \bigg[\frac 12 \, \frac{ f'}{f}  
+ \frac 13 \,   \lambda'
  \bigg] \, (i \, \sigma^2 \, \eta)
  + \frac m2 \, e^{- \frac 43 \lambda } \,  (\sigma^3 \, \eta) 
  \ , \label{algBPS1} \\
  0 & =   
 g \, g_2^{-1/2} \,   \widehat A_z  
  \, (\sigma^1 \,  \eta)
 -  f^{-5/2} \, 
 e^{ 2 \lambda} \, b  
\, \eta   +   g_1^{-1/2}    \,  \bigg[
\frac 12 \, \frac{g_2'}{g_2}
+ \frac 13 \,  \lambda'
  \bigg] \, (i \, \sigma^2 \, \eta) 
   + \frac m2 \, e^{- \frac 43 \lambda}  
\, (  \sigma^3 \, \eta) \ , 
\label{algBPS2} \\
0 & =  m \, (e^{2\lambda} - e^{-\frac 43 \lambda}  ) \,(\sigma^3 \, \eta)
- \frac 53 \,    \lambda' \, g_1^{-1/2} \, (i \, \sigma^2 \, \eta)
+   f^{-5/2} \, e^{  2 \lambda} \, b  \,     \eta  \ .
\label{algBPS3} 
\end{align}

The three algebraic equations above are of the form $M^{(i)} \, \eta= 0$,
where $M^{(i)}$, $i=1,2,3$, are three $2 \times 2$ matrices,
which can be parametrized as
\beq
M^{(i)} = X_0^{(i)} \, \mathbb I_2 + X_1^{(i)}\, \sigma^1 
+ X_2 ^{(i)}\, (i \, \sigma^2) + X_3^{(i)} \, \sigma^3 \ .
\eeq
For each matrix $M^{(i)}$, let us define the column 2-vectors
\beq
v^{(i)} = \begin{pmatrix}
X_1^{(i)} + X_2^{(i)} \\
- X_0^{(i)} - X_3^{(i)}
\end{pmatrix} \ , \qquad 
w^{(i)} = \begin{pmatrix}
X_0^{(i)} - X_3^{(i)} \\
- X_1^{(i)} + X_2^{(i)}
\end{pmatrix} \ ,
\eeq
and the quantities
\beq
\mathcal A^{ij} = \det (v^{(i)} | w^{(j)}) \ , \qquad
\mathcal B^{ij} = \det (v^{(i)} | v^{(j)})  \ , \qquad
\mathcal C ^{ij} = \det (w^{(i)} | w^{(j)}) \ ,
\eeq
where $(a|b)$ denotes the $2\times 2$ matrix obtained
by juxtaposition of the column 2-vectors $a$ and $b$.
We seek a non-trivial solution 
to the algebraic BPS equations in which $\eta$ is not identically zero.
Let us therefore pick a fixed but generic value of $y$ for which 
$\eta \neq \binom 0 0$. 
At such a value of $y$, the quantities
$\cA^{ij}$, $\cB^{ij}$, $\cC^{ij}$ are zero.
This can be seen for instance as follows
(we keep the dependence on the chosen point $y$ implicit throughout the argument).
Since $\eta \neq \binom 00$, we can choose a basis of 
$\mathbb C^2$
consisting of $\eta$ and some other linearly independent 2-component spinor $\xi$.
Let $G$ be the $GL(2,\mathbb C)$ matrix that implements the change of basis
from the standard basis in $\mathbb C^2$ to the new basis $\{ \eta , \xi\}$.
Since $\eta$ is annihilated by $M^{(i)}$,
the first column of the matrix  $M^{(i)}$ in the new basis in zero, which means that
we can write
\beq
M^{(i)}= G^{-1} \, \begin{pmatrix}
0 & u^{(i)} \\
0 & v^{(i)}
\end{pmatrix} \, G \ ,
\eeq
where $u^{(i)}$, $v^{(i)}$ are unspecified. 
After parametrizing $G =\left( \begin{smallmatrix}
a & b \\ c & d
\end{smallmatrix} \right)$,
we can extract the quantities $X_{1,2,3,4}^{(i)}$ 
in terms of $u^{(i)}$, $v^{(i)}$, $a$, $b$, $c$, $d$,
and verify explicitly that
$\cA^{ij}$, $\cB^{ij}$, and $\cC^{ij}$ vanish.

The vanishing of $\cA^{ij}$, $\cB^{ij}$, and $\cC^{ij}$
at a generic point in $y$ where $\eta \neq \binom 00$ gives us 
a   number of necessary conditions for the existence of a non-trivial
solution, which facilitate the analysis. 
The vanishing of the  
diagonal components $\cA^{ii}$ gives the conditions
\begin{align}
0 & = \frac 1f + \frac{1}{g_1} \, \bigg(   \frac 12 \, \frac{f'}{f} + \frac 13 \, \lambda' \bigg)^2
- \frac 14 \, m^2 \, e^{- \frac 83 \lambda} \ ,  \label{first_eq} \\
0 &  = \frac{b^2 \, e^{ 4\lambda}}{f^5}
+  \frac{1}{g_1} \, \bigg(   \frac 12 \, \frac{g_2'}{g_2} + \frac 13 \, \lambda' \bigg)^2
- \frac 14 \, m^2 \, e^{- \frac 83 \lambda}
- \frac{g^2 \, \widehat A_z^2}{g_2} \ , \\
0 & = \frac{25 \, (\lambda')^2}{9 \, g_1}  + \frac{b^2 \, e^{4\lambda}}{f^5} 
- m^2 \, \Big( e^{2\lambda} - e^{ - \frac 43 \lambda}  \Big)^2 \ .
\label{third_det}
\end{align}
The vanishing of the  off-diagonal symmetrized components $\cA^{ij} + \cA^{ji}$  ($i\neq j$) yields
\begin{align}
0 & = 
\frac{2}{g_1} \,  \bigg(   \frac 12 \, \frac{f'}{f} + \frac 13 \, \lambda' \bigg) \, 
\bigg(   \frac 12 \, \frac{g_2'}{g_2} + \frac 13 \, \lambda' \bigg)
- \frac{2 \, s \, b \, e^{2\lambda}}{f^3}
- \frac 12 \, m^2 \, e^{- \frac 83 \lambda} \ ,  \label{first_symm} \\
0 & = \frac{2 \, s \, b \, e^{2\lambda}}{f^3}
- \frac{10}{3 \, g_1} \, \lambda' \,  \bigg(   \frac 12 \, \frac{f'}{f} + \frac 13 \, \lambda' \bigg)
- m^2 \, e^{- \frac43 \lambda} \, \Big( e^{2\lambda} - e^{- \frac 43 \lambda}  \Big)
\ ,   \label{second_symm}   \\
0 & = \frac{2 \, b^2 \, e^{4\lambda}}{f^5}
+ \frac{10}{3 \, g_1} \, \lambda' \, 
\bigg(   \frac 12 \, \frac{g_2'}{g_2} + \frac 13 \, \lambda' \bigg)
+ m^2 \, e^{- \frac43 \lambda} \, \Big( e^{2\lambda} - e^{- \frac 43 \lambda}  \Big) \ ,
\end{align}
while setting to zero the off-diagonal antisymmetrized components $\cA^{ij} - \cA^{ji}$ gives  
\begin{align}
0 & = \frac{m \, b \, e^{\frac 23 \lambda}}{f^{5/2}}
+ \frac{m \, s \, e^{- \frac 43 \lambda}}{\sqrt f}
+ \frac{2 \, g}{\sqrt{g_1} \, \sqrt{g_2}} \, \widehat A_z \,  \bigg(   \frac 12 \, \frac{f'}{f} + \frac 13 \, \lambda' \bigg) 
\label{nice_no1}
\ , \\
0 & = \frac{m \, b \, e^{\frac 23 \lambda}}{ f^{5/2}} - \frac{2 \, m \, s}{\sqrt f} \, 
 \Big( e^{2\lambda} - e^{- \frac 43 \lambda}  \Big)
 \label{f_vs_lambda}
 \ , \\
0 & =  \frac{m \, b \, e^{\frac 23 \lambda}}{f^{5/2}}
+ \frac{2 \, m \, b \, e^{2\lambda}}{f^{5/2}} \, 
 \Big( e^{2\lambda} - e^{- \frac 43 \lambda}  \Big)
 - \frac{10 \, g  \, \widehat A_z \, \lambda'  }{ 3 \, \sqrt{g_1} \, \sqrt{g_2} } \ .
 \label{nice_no2}
\end{align}
The vanishing of all components of $\cB^{ij} + \cC^{ij}$ yields
\begin{align}
0 & = \frac{2 \, b \, e^{2\lambda}}{ f^{5/2} \, \sqrt{ g_1}} \, 
 \bigg(   \frac 12 \, \frac{f'}{f} + \frac 13 \, \lambda' \bigg)
 + \frac{2 \, s}{\sqrt f \, \sqrt{g_1}} \, 
  \bigg(   \frac 12 \, \frac{g_2'}{g_2} + \frac 13 \, \lambda' \bigg)
  +\frac{m\, g \, e^{- \frac 43 \lambda} \, \widehat A_z }{\sqrt{g_2}} \ , \\
  0 & = \frac{2 \, b \, e^{2\lambda} \, }{f^{5/2} } \, 
  \bigg(   \frac 12 \, \frac{f'}{f} + \frac 13 \, \lambda' \bigg)
 + \frac{10 \, s \, \lambda' }{3 \, \sqrt f }   \ , \\
 0 & = \frac{10 \, b \, e^{2\lambda} \, \lambda'  }{3 \, f^{5/2} \, \sqrt{g_1}}  
 - \frac{2 \, b \, e^{2\lambda}}{f^{5/2} \, \sqrt{g_1}} \, 
   \bigg(   \frac 12 \, \frac{g_2'}{g_2} + \frac 13 \, \lambda' \bigg)
   - \frac{2 \, m \, g \, \widehat A_z}{\sqrt{g_2}} \, 
    \Big( e^{2\lambda} - e^{- \frac 43 \lambda}  \Big) \ .
\end{align}
Finally, the vanishing of all  components of $\cB^{ij} - \cC^{ij}$ gives
\begin{align}
0 & = \frac{m \, e^{- \frac 43 \lambda}}{\sqrt{g_1}} \,     
 \bigg(   \frac 12 \, \frac{g_2'}{g_2} + \frac 13 \, \lambda' \bigg)
 -  \frac{m \, e^{- \frac 43 \lambda}}{\sqrt{g_1}} \,     
 \bigg(   \frac 12 \, \frac{f'}{f} + \frac 13 \, \lambda' \bigg)
 + \frac{2 \, s \, g}{\sqrt f \, \sqrt{g_2}} \, \widehat  A_z \ , \\
 0 & = \frac{5 \, m \, e^{- \frac 43 \lambda} \, \lambda' }{3 \, \sqrt{g_1}}
 + \frac{2 \, m}{\sqrt{g_1} } \, 
   \bigg(   \frac 12 \, \frac{f'}{f} + \frac 13 \, \lambda' \bigg) \,
       \Big( e^{2\lambda} - e^{- \frac 43 \lambda}  \Big) \ , \\
0 & = \frac{5 \, m \, e^{- \frac 43 \lambda} \, \lambda' }{3 \, \sqrt{g_1}}
        + \frac{2 \, m}{\sqrt{g_1} } \, 
   \bigg(   \frac 12 \, \frac{g_2'}{g_2} + \frac 13 \, \lambda' \bigg) \,
       \Big( e^{2\lambda} - e^{- \frac 43 \lambda}  \Big)
       + \frac{2 \, b \, g \, e^{2\lambda} \, \widehat A_z}{ f^{5/2} \, \sqrt{g_2}} \ . \label{the_last_eq}
\end{align}

The relation \eqref{f_vs_lambda} is an algebraic relation for $f$ in terms of
$\lambda$. In particular, it implies that $\lambda$ is a constant if and only if $f$ is a constant.
The case of interest for this paper is $\lambda$ non-constant; the case of constant $\lambda$
does not yield new solutions. Let us therefore assume that $\lambda$ is not a constant.
We can   solve  \eqref{f_vs_lambda}  for $f$,
\beq
f = \frac{2 \, B \, m^{-2} \, e^\lambda} { \sqrt{  \kappa \, (1 - e^{\frac{10}{3} \lambda}) }} \ ,
\eeq
where $\kappa$ is a sign and $B>0$ is a constant such that
\beq
8 \, s \, B^2 \, m^{-4} + \kappa \, b = 0 \ .
\eeq
In what follows, we express $b$ in terms of $B$ using the above relation.
Next, we solve \eqref{third_det} for $g_1$,
\beq
g_1 = \frac{25 \,   B \, m^{-2}   \, e^{\frac 83 \lambda}}{9 \, (1 - e^{\frac{10}{3} \lambda  }  )^2  \,
\Big[   B  - 2 \, e^{\frac 53 \lambda} \, \sqrt{ \kappa \, (1 - e^{\frac{10}{3} \lambda  }  )    }  \Big] }  \, (\lambda')^2 \ .
\eeq
We proceed by considering \eqref{nice_no1}, or equivalently
\eqref{nice_no2}. These equations give an expression for $\sqrt{g_1}\sqrt{g_2}$
in terms of $\widehat A_z$,
\beq \label{the_first_way}
\sqrt{g_1} \, \sqrt{g_2} =  
\frac{  5 \, \kappa \,s  \, g \, \sqrt{2 \, B \, m^{-2}} \, e^{\frac{11}{6} \lambda}  }{
3\, m \, (1 - 2 \, e^{\frac{10}{3} \lambda  })  \, \Big[ \kappa \, (1 - e^{\frac{10}{3} \lambda  }  )  \Big]^{\frac 54}
}  \,\widehat  A_z \, \lambda'   \ .
\eeq
On the other hand, \eqref{the_extra_eq} gives an expression for $\sqrt{g_1}\sqrt{g_2}$
in terms of $A'_z =  \widehat A'_z$,
\beq \label{the_second_way}
\sqrt{g_1} \, \sqrt{g_2} =  
- \frac{\kappa \, s \, \sqrt{2 \, B \, m^{-2}} \, e^{- \frac 32 \lambda}  }{  
 \Big[ \kappa \, (1 - e^{\frac{10}{3} \lambda  }  )  \Big]^{\frac 54}
   }  \, A_z' \ .
\eeq
Comparing \eqref{the_first_way} and \eqref{the_second_way} we get a  simple ODE for $ \widehat A_z$,
which is solved by
\beq \label{hatAz_expr}
\widehat A_z =- m^{-1} \,  \cC \, \left (  e^{\frac{10}{3} \lambda  } - \frac 1 2 \right)   \ ,
\eeq
where $\cC$ is a real constant.
From the definition \eqref{hatAz_def} of $\widehat A_z$ and the expression \eqref{hatAz_expr}
we conclude that
\beq \label{Az_expr}
A_z = - m^{-1} \, \left[  \cC\, \left (  e^{\frac{10}{3} \lambda  } - \tfrac 1 2 \right) + n   \right]  \ .
\eeq
Having determined $\widehat A_z$, we also have an expression for $g_2$,
\beq
g_2 = \frac{2  \, \cC^2 \, m^{-2} \, e^\lambda \, \Big[  B - 2 \, e^{\frac 53 \lambda} \, \sqrt{ \kappa \, (1 - e^{\frac{10}{3} \lambda  }  )    }  \Big] }{    \sqrt{ \kappa \, (1 - e^{\frac{10}{3} \lambda  }  )    }  }  \ .
\eeq
We can now verify that all equations 
\eqref{first_eq} to \eqref{the_last_eq} together with \eqref{the_extra_eq} are   satisfied,
provided that the signs of $\cC$ and $\lambda'$ are related as 
\beq \label{sign_constraint}
\sign (\cC) =   \kappa \, s \, \sign(\lambda') \ .
\eeq

We have exploited the necessary conditions originating
from the vanishing of the quantities $\cA^{ij}$, $\cB^{ij}$, $\cC^{ij}$.
We can now verify directly that a 2-component spinor $\eta$ can be found,
which satisfies the original algebraic BPS equations 
 \eqref{algBPS1}-\eqref{algBPS3}.
This spinor is
\beq
\eta = e^{inz} \, Q(y) \, \begin{pmatrix}
\phantom{- \kappa \, \sign (\lambda') \,}
\sqrt{    \sqrt{ B}  - \sqrt 2 \, s \, e^{\frac 56  \lambda}   \, \Big[ \kappa \, (1 - e^{ \frac{10}{3} \lambda })\Big]^{\frac 14}    } \\
- \kappa \, \sign (\lambda') \, \sqrt{     \sqrt{ B}  + \sqrt 2 \, s \, e^{\frac 56  \lambda}   \, \Big[ \kappa \, (1 - e^{ \frac{10}{3} \lambda })\Big]^{\frac 14}    }
\end{pmatrix}
\eeq
To write $\eta$ we have exploited the factorization
\beq
B - 2 \, e^{\frac 53 \lambda} \, \sqrt{  \kappa \, (1 - e^{\frac{10}{3} \lambda })  }
= \Big(  \sqrt B + \sqrt 2 \, s \, e^{\frac 56 \lambda} \, [\kappa \, (1 - e^{\frac{10}{3} \lambda })]^{\frac 14} \Big) \,
 \Big(  \sqrt B - \sqrt 2 \, s \, e^{\frac 56 \lambda} \, [\kappa \, (1 - e^{\frac{10}{3} \lambda })]^{\frac 14} \Big)  \ .
\eeq
The LHS must be positive to ensure positivity of $g_1$.
On the RHS, one of the two factors is a sum of positive quantities,
hence is automatically positive, and therefore the other factor must be positive, too.
Finally, we consider the BPS equation \eqref{gravitino_y} to determine $Q$ as a function of $w$.
We get a simple ODE, which is solved by
\beq
Q = Q_0 \, \frac{e^{\frac 14 \lambda}}{ \Big[  \kappa \, (1 - e^{ \frac{10}{3} \lambda }) \Big]^{\frac 18} }
 \ .
\eeq
To finish, we verify that all bosonic equations of motion
are satisfied.

Notice that we have not determined $\lambda$ as a function of $w$.
This is in accordance with the general covariance of the BPS equations and equations of motion.
We find it convenient to fix the ambiguity in reparametrizations of $w$ by choosing
\beq
\lambda(w) = \frac 35 \, \log w \ .
\eeq
This choice requires $w>0$.
The line element takes the form
\beq
m ^2 \, ds^7_2 = \frac{2\,B \, w^{3/5}}{ \sqrt{\kappa \, (1-w^2)}} \, ds^2(AdS_5)
+  \frac{B \, w^{-2/5}}{h(w) \, (1-w^2)^2} \, dw^2
+ \frac{2 \, \cC^2 \, w^{3/5} \, h(w)}{ \sqrt{\kappa \, (1-w^2)} } \, dz^2 \ ,
\eeq
where we have defined
\beq
h(w) = B - 2 \, w \, \sqrt{\kappa \, (1-w^2)} \ .
\eeq
The gauge fields and scalars are given as
\beq
\lambda_1 = \frac 35 \, \log w \ , \qquad
\lambda_2 = - \frac 25 \, \log w \ , \qquad
A^{(1)} = - m^{-1} \, \left[ \cC \left (  w^2 - \tfrac 1 2 \right) + n   \right] \, dz \ .
\eeq
The spinor $\eta$ takes the more explicit form
\beq
\eta = Q_0 \, e^{inz} \, \frac{w^{3/20}}{ \Big[  \kappa \, (1 -  w^2) \Big]^{\frac 18} } \, \begin{pmatrix}
\phantom{- \kappa  \,}
\sqrt{    \sqrt{ B}  - \sqrt 2 \, s \, w^{1/2}   \, \Big[ \kappa \, (1 - w^2)\Big]^{\frac 14}    } \\
- \kappa   \, \sqrt{     \sqrt{ B}  + \sqrt 2 \, s \, w^{1/2}   \, \Big[ \kappa \, (1 -  w^2)\Big]^{\frac 14}    }
\end{pmatrix} \ .
\eeq

\subsection{Regularity of the Gauge Field and Killing Spinor}  \label{app_regularity}

Let us study the regularity of $A^{(1)}$ and $\eta$ for the choice of
parameters \eqref{y_range_1}, repeated here for convenience,
\beq  
\kappa = 1 \ , \qquad 0 < B < 1 \ , \qquad 0 < w < w_1 :=  \sqrt{ \tfrac 12 \, \Big(1 - \sqrt{1-B^2} \Big) }  \ .
\eeq
From \eqref{seven_metric} we see that   line element near $w = w_1$ can be written as
\begin{align} \label{nearw1}
r^2 := w_1-w \ , \quad m^2 \, ds^2_7 \stackrel{r \rightarrow 0}{=} \frac{2 \, B \, w_1^{3/5}}{\sqrt{1-w_1^2}} \, \bigg[
ds^2(AdS_5)
+ \frac{2\, \Big[ 
dr^2 + \cC^2 \, (1-B^2) \, r^2 \, dz^2
\Big]}{- h'(w_1) \, w_1 \, (1-w_1^2)^{3/2}} 
\bigg] \ ,
\end{align}
which confirms that the $w$, $z$ directions near 
the point $w = w_1$
are locally an $\mathbb R^2/\mathbb Z_\ell$ orbifold
if we impose \eqref{orbifold_condition}.
Since the $z$ circle shrinks at $w = w_1$, the quantity $A_z$
must vanish at $w = w_1$. This requirement fixes the constant $n$,
\beq
n = 
\frac 12 \, \cC \, \sqrt{1-B^2} =
\frac{1}{2\,\ell} \ , \qquad A^{(1)} = - m^{-1} \, \cC \, (w^2 - w_1^2) \, dz \ .
\eeq
We have made use of \eqref{orbifold_condition}
and we have fixed 
${\rm sign}(\cC) =+1$,
which is the choice made in the main text when discussing
the uplift of the 7d solution. From \eqref{sign_constraint} we see that we must select $s = 1$.
The Killing spinor $\eta$ is therefore given by
\beq \label{final_spinor}
\eta = Q_0 \, e^{\frac{iz}{2\ell}} \, \frac{w^{3/20}}{   (1 -  w^2) ^{\frac 18} } \, \begin{pmatrix}
\phantom{-    \,}
\sqrt{    \sqrt{ B}  - \sqrt 2   \, w^{1/2}   \,   (1 - w^2)^{\frac 14}    } \\
-     \, \sqrt{     \sqrt{ B}  + \sqrt 2   \, w^{1/2}   \, (1 -  w^2)^{\frac 14}    }
\end{pmatrix} \ .
\eeq
From this expression we see that
\beq \label{spinor_near_w1}
\lim_{w \rightarrow w_1} \eta = - Q_0 \, \frac{w_1^{3/20}}{   (1 -  w_1^2) ^{\frac 18} } \, \sqrt{     \sqrt{ B}  + \sqrt 2   \, w_1^{1/2}   \, (1 -  w_1^2)^{\frac 14}    } \, \begin{pmatrix}
0 \\ 1
\end{pmatrix} \, e^{\frac{iz}{2\ell}} \ .
\eeq

Let us now argue that $\eta$ is well-defined near $w = w_1$,
using arguments similar to those presented in \cite{Ferrero:2020twa}.
We see from \eqref{nearw1} that a good set of polar coordinates 
near the point $w = w_1$ is furnished by the angle $z$ and $r =\sqrt{w_1-w}$.
The flat metric on the orbifold $\mathbb R^2/\mathbb Z_\ell$
can   be written in various equivalent forms,
\beq
ds^2 = dr^2 + \frac{1}{\ell^2} \, r^2 \, dz^2 = dx^2 + dy^2 = (e^1)^2 + (e^1)^2
= (e'^1)^2 + (e'^1)^2 \ ,
\eeq
where the Cartesian coordinates $x$, $y$ and the 1-forms $e^{1,2}$, $e'^{1,2}$ are defined as
\beq
x  = r \, \cos\frac z \ell \ , \quad y = r \, \sin \frac z \ell \, \quad e^1= dx \ , \quad e^2 = dy \ , \quad e'^1 = dr \ , \quad e'^2 = \frac 1 \ell \, r \, dz \ .  
\eeq
The vielbeins $e^{a=1,2}$ and $e'^{a=1,2}$ are related by a local rotation,
\beq
e'^a = \Lambda^a{}_b \, e^b \ , \qquad \Lambda^a{}_b =( \exp \lambda)^a{}_b \ , \qquad
\lambda^a{}_b= \begin{pmatrix}
0 & - z/\ell \\
z/\ell & 0
\end{pmatrix} \ .
\eeq
Let $\psi$ be a 2-component Dirac spinor in the frame $e^a$,
and let $\psi'$ denote its components in the frame $e'^a$.
The transformation relating $\psi'$ to $\psi$ is
\beq \label{spinor_rotation}
\psi' = S \, \psi \ , \qquad S = \exp \Big (\tfrac 14 \,\gamma_{ab} \, \lambda^{ab}  \Big) = \begin{pmatrix}
e^{-\frac{iz}{2\ell}} & 0 \\
0 & e^{ \frac{iz}{2\ell}}
\end{pmatrix} \ ,
\eeq
where we have chosen the 2d Euclidean gamma matrices to be $\gamma^1 =\sigma^1$, $\gamma^2 = \sigma^2$, with $\sigma^{1,2}$ standard Pauli matrices.
The transformation \eqref{spinor_rotation} shows that, if the spinor $\psi$ in the Cartesian frame
is constant, its components $\psi'$ in the polar frame acquire a $z$-dependence
through the phase factors
$e^{\mp \frac{iz}{2\ell}}$.
In particular, if $\psi$ has negative chirality
(its only non-zero component is the lower one),
it acquires a phase factor $e^{\frac{iz}{2\ell}}$.
This is exactly the dependence found in \eqref{spinor_near_w1},
demonstrating that our Killing spinor is well-defined near $w = w_1$,
because it corresponds to a constant spinor in the Cartesian frame.

\subsection{Alternative Possibilities   for the Range of $w$} \label{app_others}

The allowed possibilities for the range of $w$ depend on the values
of the constant parameters $\kappa$ and $B$. We find six possibilities:
\begin{itemize}
\item Case I: $\kappa = 1$, $0< B < 1$, $0 < w < w_1$ where
\beq
w_1 = \sqrt{ \tfrac 12 \, \Big(1 - \sqrt{1-B^2} \Big) } = \tfrac 12 \, \Big( \sqrt{1+B} -  \sqrt{1-B}\Big) \ .
\eeq
\item Case II: $\kappa = 1$, $0< B < 1$, $w_2 < w < 1$ where
\beq
w_2 = \sqrt{ \tfrac 12 \, \Big(  1 + \sqrt{1-B^2}  \Big)  } = \tfrac 12 \, \Big( \sqrt{1+B} +  \sqrt{1-B}\Big) \ .
\eeq
\item Case III: $\kappa = 1$, $B=1$, $0<w<1/\sqrt 2$.
\item Case IV: $\kappa = 1$, $B=1$, $1/\sqrt 2<w<1$.
\item Case V: $\kappa = 1$, $B>1$, $0<w<1$.
\item Case VI: $\kappa = -1$, $B>0$, $1<w<w_3$ where
\beq
w_3 = \sqrt{ \tfrac 12 \, \Big(      1 + \sqrt{1+B^2}  \Big)  }   \ .
\eeq
\end{itemize}
Case I has been   discussed in the main text and in the previous
subsection. Let us describe here the salient features of the line element in the other cases.

\begin{figure}
\centering
\includegraphics[width = 7 cm]{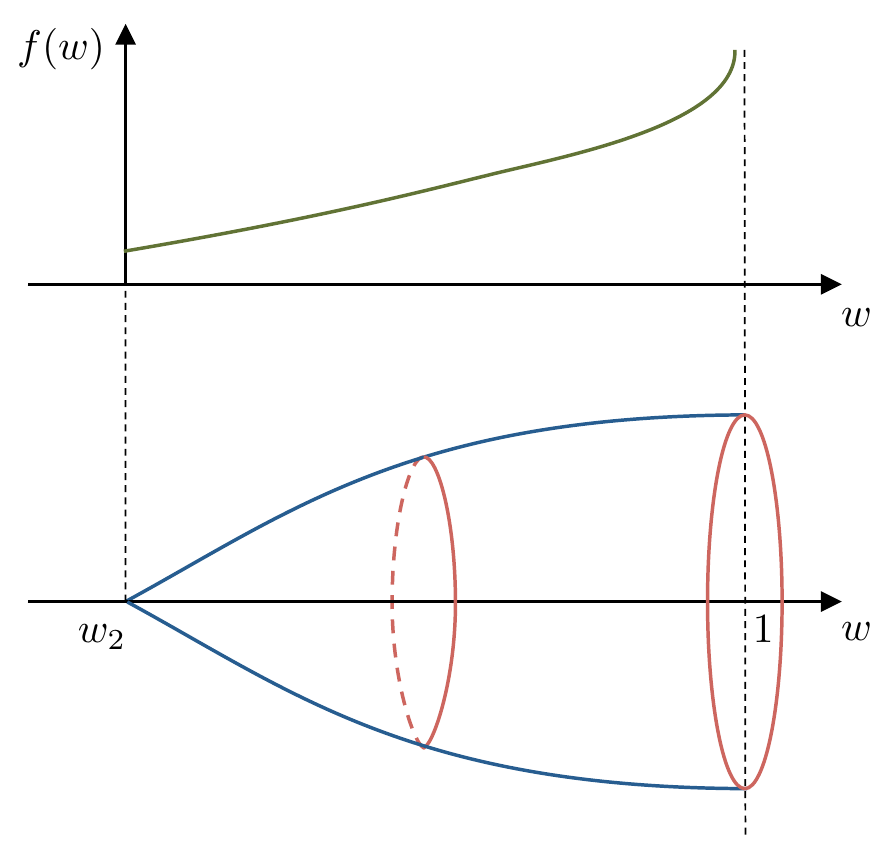}
\caption{
A schematic depiction of the internal geometry in Case II. 
The  $z$ circle is fibered  over the $w$ interval to yield $\Sigma$. In the metric
$ds^2(\Sigma)$ in 
\eqref{seven_metric}, $\Sigma$ has the topology of a disk with a $\mathbb Z_\ell$ orbifold
singularity at the center.
We also depict the qualitative behavior of the $AdS_5$ warp function $f(w) = 2\,B \, w^{3/5}/\sqrt{1-w^2}$.
}
\label{Case_II_fig}
\end{figure}

\paragraph{Case II.} The $z$ circle shrinks near the endpoint $w = w_2$. 
If we impose the same condition as \eqref{orbifold_condition} above, we get an orbifold point
$\mathbb R^2/\mathbb Z_\ell$. As we approach $w = 1$, the $AdS_5$ warp factor goes to infinity,
see   Figure \ref{Case_II_fig}.
If we set $w= 1- B^2 \, \cC^2 \, r^4/8$ and consider the limit $r \rightarrow 0^+$, the line element takes the form
\beq
m^2 \, ds^2_7 \approx \frac{4}{r^2} \, \bigg[ \frac{1}{\cC^2 } \, ds^2(AdS_5) +    dr^2 +  dz^2 \bigg] \ ,
\qquad r \rightarrow 0^+ \ ,
\eeq
from which we see that the 7d metric is given approximately by a conformal rescaling of the
direct product of $AdS_5$ and a cylinder.
The space $\Sigma$ in the metric $ds^2(\Sigma)$ in \eqref{seven_metric}
has the topology of a disk, as in Case I. The regularity of $A^{(1)}$ and of the Killing spinor
can be analyzed in a similar way as done above in Case I.

\paragraph{Case III.} This case is similar to Case I, except that the function $h(w)$
has a double zero at $w = 1/\sqrt 2$. As a result the $z$ circle does not shrink smoothly for
any value of the constant $\cC$. Instead, if we set $w = \tfrac{1}{\sqrt 2} - \frac 1R$,
as we consider $R \rightarrow +\infty$ the 7d line element is approximately given by
\beq
m^2 \, ds^2_7 \approx 2^{6/5} \, ds^2(AdS_5) + \frac{1}{2^{4/5}} \, \frac{dR^2  + 32 \, \cC^2 \, dz^2}{R^2} \ ,
\qquad R \rightarrow + \infty \ .
\eeq
The behavior near $w= 0$ is similar to Case I.

\paragraph{Case IV.} The behavior as we approach $w= 1/\sqrt 2$ from the right is similar to the behavior
of Case III approaching $w= 1/\sqrt 2$ from the left. The behavior in the limit $w \rightarrow 1^-$ is
as in Case II, with a decompactification of the $z$ circle.

\paragraph{Case V.} This case combines the features of Case I near $w = 0$ and the features of Case II near $w = 1$.

\paragraph{Case VI.} As $w \rightarrow 1^+$, the $z$ circle decompactifies and the $AdS_5$ warp factor diverges. If we set $w = 1 + B^2 \, \cC^4 \, r^4/8$ and consider $r \rightarrow 0^+$, the line element
takes the form
\beq
m^2 \, ds^2_7 \approx \frac{4}{r^2} \, \Big[ \frac{1}{\cC^2} \,  ds^2(AdS_5) 
+  dr^2
+ dz^2
\Big] \ , \qquad r \rightarrow 0^+ \ .
\eeq
As we approach $w = w_3$, the $z$ circle shrinks. If we impose
\beq
|\cC| = \frac{1}{ \ell \, \sqrt{1+B^2}}   \ , \qquad \ell = 1,2,3,\dots  \ ,
\eeq
we get an $\mathbb R^2/\mathbb Z_\ell$ orbifold point.



\section{Solutions in Canonical $\cN =2$ Form} \label{sec:appLLM}

In this appendix we describe in greater detail the change of variables
that brings the uplifted solution \eqref{11d_metric}, \eqref{G4flux} into canonical
$\cN = 2$
LLM form \eqref{LLM}. We also review some general facts about Killing spinors and
spinor bilinears for LLM setups and their connections with the most general
supersymmetric $AdS_5$ solution of 11d supergravity discussed in \cite{Gauntlett:2004zh}.

\subsection{Change of Variables to LLM Form} \label{app_LLM_change}

All solutions discussed in this paper fall into a subclass
of the canonical LLM form \eqref{LLM}
with an enhanced $U(1)$ isometry.
In terms of the polar coordinates $(r,\beta)$
in the $(x_1,x_2)$ plane introduced in \eqref{LLMpolar},
the function $D$ depends on $y$ and $r$ only,
and therefore the same holds true for the warp factor $\widetilde \lambda$,
determined by $D$ via \eqref{lambda_and_D}. It follows that $\partial_\beta$ is a Killing vector.
When the function $D$ is independent of $\beta$,
the Toda equation \eqref{Toda_equation} and the expression \eqref{v_def} for the 1-form $v$
take a simpler form,
\beq \label{axialToda}
\bigg( \partial_r^2 + \frac 1r \, \partial_r \bigg) D + \partial_y^2 e^D  = 0 \  , \qquad 
v= - \frac 12 \, r\, \partial_r D \,  d\beta \ .
\eeq
The relation between the LLM angular variables $\chi$, $\beta$
and the angular variables $\phi$, $z$ in \eqref{11d_metric}
was given in \eqref{angular_vars}, repeated here for convenience as
\beq    \label{angles_again}
\begin{pmatrix}
d\chi \\
d\beta
\end{pmatrix}
=  \begin{pmatrix}
1 + \cC^{-1} & -1  \\
-\cC^{-1} & 1
\end{pmatrix}\, 
\begin{pmatrix}
d\phi \\
dz
\end{pmatrix} \ , \qquad
\begin{pmatrix}
\partial_\chi \\
\partial_\beta
\end{pmatrix}
= \begin{pmatrix}
1 &  \cC^{-1} \\
 1 & 1 + \cC^{-1}
\end{pmatrix} \,
\begin{pmatrix}
\partial_\phi \\
\partial_z
\end{pmatrix} \ .
\eeq
The matrix that implements this linear change of coordinates
has determinant 1, consistent with the fact that all these four angular variables 
have period $2\pi$.
As anticipated in the main text, 
the LLM coordinates $y$, $r$ are related to the coordinates $\mu$, $w$ 
in \eqref{11d_metric} via
\begin{align}  \label{y_and_r_bis}
y = \frac{4\, B \, w\, \mu}{  \sqrt{\kappa\, (1-w^2) }  } \ , \qquad
r = (1-\mu^2)^{-\frac{1}{2\cC} } \, \cG(w) \ ,
\end{align}
where the function $\cG(w)$ 
is determined up to an overall constant normalization
and is a solution to the ODE
\beq \label{cG_ODE}
\frac{\cG'(w)}{\cG(w)} = \frac{-B \, w }{\cC \, 
\left(1-w^2 \right)   \left[ B-2 \,  w \,  \sqrt{\kappa \,(1 -   w^2)}\right] } \ .
\eeq
For the choice of parameters and range of $w$ specified
in \eqref{y_range_1},  $\cG(w)$ is given explicitly as
\begin{align} \label{explicit_G}
\cG(w) & = \cG_0 \, \exp \bigg\{
- \frac{1}{2 \, \cC} \, \log (1-w^2)
+ \frac{1 - \cB}{2 \,  \cC \, \cB} \, \log(1 - \cB - 2 \, w^2)
  \\
& - \frac{1 + \cB}{4 \,  \cC \, \cB} \, \log \Big[
(\sqrt 2  - \sqrt{1-w^2} \, \sqrt{1- \cB} - w \, \sqrt{1+ \cB }   ) \, (
\sqrt 2  + \sqrt{1-w^2} \, \sqrt{1- \cB} + w \, \sqrt{1+ \cB }  
)
 \Big]
 \nn \\
& -  \frac{1 - \cB}{ 4 \, \cC \, \cB} \, \log \Big[
(\sqrt 2  + \sqrt{1-w^2} \, \sqrt{1+ \cB} - w \, \sqrt{1- \cB }   ) \, (
\sqrt 2  - \sqrt{1-w^2} \, \sqrt{1+ \cB} + w \, \sqrt{1- \cB }  
)
 \Big] \bigg\} \ , \nn
\end{align}
where $\cG_0$ is an  integration constant and we have introduced 
 the shorthand notation
\beq
\cB = \sqrt{1-B^2} \ .
\eeq
The quantity $D$, expressed in terms of $w$ and $\mu$, is given as
\beq \label{expDbis}
e^D =
\frac{16\, B  \,  \cC^2 \,  \left(1-\mu ^2\right)^{1 + 1 / \cC} \,  \left [ B - 2  \,  w \,  \sqrt{\kappa \,(1  - 
   w^2) }\right] }{     \kappa \,\left(1 - w^2\right)   \, \cG(w)^2 }  \ .
\eeq
Using the chain rule, \eqref{expDbis}, \eqref{y_and_r_bis}, \eqref{cG_ODE}
one computes the derivatives of $D$ with respect to $y$ and $r$.
The expressions
for $\partial_y D$, $\partial_r D$,
and the warp factor, computed from \eqref{lambda_and_D}, are
\begin{align}
- \partial_yD &= \frac{\kappa \, (1-w^2) \, \mu}{2\, B \, \Big[ \mu^2 \, h + B \, w^2 \, (1-\mu^2)  \Big]} \ ,
\qquad e^{- 6 \widetilde \lambda} = \frac{ [\kappa \, (1-w^2) ]^{3/2}}{8 \, B^3 \, w\, \cH} \  , 
 \nn \\
r\, \partial_r D & = \frac{
B \, (1-\mu^2) \, \Big[ \cC - 2w^2 ( \cC+1)   \Big] - h \, ( \cC + 2\mu^2 + \cC   \mu^2)
}{ \mu^2 \, h + B \, w^2 \, (1-\mu^2)  }
\end{align}
The second derivatives of $D$ are computed in a similar way and can be used to verify
that $D$ satisfies the Toda equation in the form \eqref{axialToda}.

In order to verify that the $G_4$ flux \eqref{G4flux} matches
with the LLM expression, it is convenient to observe that
the LLM $G_4$ flux is $G_4 =1/(4m^2) \, {\rm vol}_{S^2} \wedge \Omega_2 $,
where
\begin{align}
\Omega_2 & :=   D\chi \wedge d(y^3 \, e^{- 6 \widetilde \lambda})
+ y \, (1 - y^2  \, e^{-6\widetilde \lambda}) \, dv
- \frac 12 \, \partial_y e^D \, dx_1 \wedge dx_2 \nn \\
& = d \Big[  - y^3 \, e^{- 6 \widetilde \lambda} \, D\chi - y \, v \Big] + \frac 12 \, (r \, \partial_r D \, dy 
- \partial_y e^D \, r \, dr) \wedge d\beta \ .
\end{align}
The 1-form $(r \, \partial_r D \, dy 
- \partial_y e^D \, r \, dr)$ is closed by virtue of the Toda equation,
hence locally exact. For the solutions we are discussing, one verifies indeed that
\beq
r \, \partial_r D \, dy 
- \partial_y e^D \, r \, dr = d \cF \ , \qquad \cF := \frac{ - 8 \, B \, w \, \mu \, (\cC+1)}{\sqrt{\kappa \, (1-w^2)}}
+ 8 \, \cC \, \mu \ .
\eeq
The above relations can be used to check that
\beq \label{final_Omega2}
\Omega_2 = d \Big[  - y^3 \, e^{- 6 \widetilde \lambda} \, D\chi - y \, v
 + \tfrac 12 \, \cF \, d\beta \Big] = -4 \, d \bigg[
\frac{ \mu^3}{\mu^2 + w^2 \, (1-\mu^2)} \, D\phi
\bigg] \ .
\eeq
In the second step we have used 
the definition \eqref{v_def} of $D\chi$,
the definition \eqref{Dphi_def} of $D\phi$,
and the change of variables \eqref{angles_again}.
The expression \eqref{final_Omega2} for $\Omega_2$
shows that the $G_4$ flux in \eqref{LLM} matches
exactly with \eqref{G4flux}.

\subsection{Killing Spinors and Calibration} \label{app_spinors_and_calibration}

In this section we review some facts about Killing spinors
for the general LLM solution \eqref{LLM}. In particular,
we are interested in establishing a precise map with the Killing spinors
and bilinears of the most general supersymmetric $AdS_5$ solution
of 11d supergravity analyzed in Gauntlett-Martelli-Sparks-Waldram (GMSW) \cite{Gauntlett:2004zh}.
Our aim is to study calibration conditions for wrapped M2-branes
in the   solutions described in this work.

\subsubsection{Killing Spinors and Spinor Bilinears in LLM}

Killing spinors and their bilinears for LLM solutions are described in detail
in the appendices of the original paper \cite{Lin:2004nb}.
The spinor bilinear analysis of \cite{Lin:2004nb} is performed
for stationary 11d solutions containing an $S^5$ and an $S^2$ factor.
They are related to the $AdS_5$ solutions \eqref{LLM}
by a double analytic continuation of $S^5$ to $AdS_5$ and time to the angle $\chi$.
Instead of following the analytic continuation, we find it convenient to 
  repeat the steps of \cite{Lin:2004nb}
directly for the $AdS_5$ case. 
Our main objective is to set a consistent set of conventions and notation,
so we will be brief and refer to the original paper for further 
explanations of some aspects of the Killing spinor analysis.

\paragraph{Split of $M_6$ Into $S^2$ and $M_4$.}
The parametrization of the 11d line element and $G_4$ flux is
\beq \label{11d_split}
ds^2_{11} = m^{-2} \, e^{2 \widetilde \lambda }  \, \Big[ ds^2(AdS_5) + ds^2(M_6)\Big] \ ,
\qquad
G_4 = m^{-3} \, \cG_4 \ ,
\eeq
where $ds^2(AdS_5)$ has unit radius and $\widetilde \lambda$
is the warp factor. This is the same parametrization as in 
\cite{Gauntlett:2004zh}, except that we have factored out the overall scale $m^{-2}$.
The quantity $\cG_4$ is a closed form on $M_6$.
As in \cite{Gauntlett:2004zh}
the Killing spinor of 11d supergravity is decomposed as
$\psi_{\rm AdS} \otimes e^{\widetilde \lambda/2} \,\xi$,
where $\psi_{\rm AdS}$ is a Killing spinor on $AdS_5$
and $\xi$ is a Dirac (non-chiral) spinor on $M_6$.
The BPS equations for $\xi$ are derived in \cite{Gauntlett:2004zh}.
We write them with $m=1$ because we have factored out the overall scale.
They read
\begin{align} \label{GMSW_eqs}
\Big[  \nabla_m + \tfrac i2  \, \gamma_m \, \gamma_7 - \tfrac{1}{24} \, e^{-3 \widetilde \lambda} \, \cG_{mn_1 n_2 n_3} \, \gamma^{n_1 n_2 n_3} \Big] \, \xi & = 0 \ , \nn \\
\Big[  \gamma^m \, \nabla_m \widetilde \lambda  + \tfrac{1}{144} \, e^{-3 \widetilde \lambda }   \, \cG_{n_1 n_2 n_3 n_4} \, \gamma^{n_1 n_2 n_3 n_4} - i  \,\gamma_7 \Big] \, \xi & = 0 \ . 
\end{align}
The indices $m, n = 1, \dots, 6$ are curved indices on $M_6$, which are 
raised/lowered with the metric $ds^2(M_6)$ defined by \eqref{11d_split}. 
Throughout this appendix, flat indices are underlined to distinguish them 
from curved indices. Thus   the 6d gamma matrices with flat indices
are denoted $\gamma^{\underline m}$. They are Hermitian and obey the Clifford
algebra $\{ \gamma^{\underline m} , \gamma^{\underline n}  \} = 2 \, \delta^{\underline m \underline n}$.
The chirality matrix $\gamma_7$ is defined as $\gamma_7 = \gamma^{\underline 1} \dots \gamma^{\underline 6}$,  is anti-Hermitian, and satisfies $\gamma_7^2 = -1$.
As usual, a gamma matrix with several indices denotes the product of 
gamma matrices totally antisymmetrized with weight 1.

To specialize the general BPS equations \eqref{GMSW_eqs} to LLM setups,
we write the internal metric $ds^2(M_6)$ in the form
\beq \label{S2_apart}
ds^2(M_6) = e^{2A} \, ds^2(S^2) + ds^2(M_4)  \ ,
\eeq
where $ds^2(S^2)$ is the metric on a unit-radius round sphere,
$ds^2(M_4)$ is a Riemannian metric on a 4d space $M_4$, and 
$A$ is a function on $M_4$.
The $G_4$ flux is parametrized as
\beq \label{cG4_factorized}
\cG_4 =  F \wedge {\rm vol}_{S^2} \ ,
\eeq
where ${\rm vol}_{S^2}$ is the volume form of the metric $ds^2(S^2)$
and $F$ is a closed 2-form on $M_4$.
The 6d index $m$ is split as $m = (x, \alpha)$,
with $x = 1,2$ for the $S^2$ directions and $\alpha =1,2,3,4$ for the $M_4$
directions. Following \cite{Lin:2004nb}, we split the 6d gamma matrices $\gamma^m$ as
\beq \label{decompose_gammas}
\gamma^{\underline x} = \sigma^{\underline x} \otimes \Gamma_5 \ , \qquad
\gamma^{\underline \alpha} = 1 \otimes \Gamma^{\underline \alpha} \ ,
\eeq
where $\sigma^{\underline x = 1,2}$ are the standard Pauli matrices
and the 4d gamma matrices $\Gamma^{\underline \alpha}$ are Hermitian
and satisfy $\{ \Gamma^{\underline \alpha} , \Gamma^{\underline \beta} \} = 2 \, \delta^{\underline \alpha \underline \beta}$. The 4d chirality matrix $\Gamma_5$ is defined
as $\Gamma_5 = \Gamma^{\underline 1} \dots \Gamma^{\underline 4}$, is Hermitian,
and satisfies $\Gamma_5^2 = 1$.

We decompose the 6d Dirac spinor $\xi$  
that enters the BPS conditions \eqref{GMSW_eqs} into Killing spinors on $S^2$
and Dirac spinors on $M_4$. More precisely, we write 
\beq \label{two_xis}
\xi^{\cI} = \vartheta^{\cI} \otimes \epsilon_+ + (i\, \sigma^{3} \, \vartheta^{\cI}) \otimes \epsilon_-   \ .
\eeq
In the above relation, $\epsilon_\pm$ are Dirac spinors on $M_4$ while $\vartheta^\cI$ is a basis of independent Killing 2-component spinors
on $S^2$, satisfying 
\beq \label{S2_Killing}
\nabla_x^{S^2} \vartheta^\cI = + \frac i2 \, \sigma_x \, \vartheta^\cI  \ .
\eeq
The index $\cI =1,2$ is \emph{not} a spinor index, but rather labels the two linearly independent
solutions to the above equation. We can regard $\cI$ as a fundamental index
of the isometry algebra $\mathfrak{su}(2) \cong \mathfrak{so}(3)$ of the round $S^2$.
The fact that the 6d spinor $\xi$ carries a label $\cI = 1,2$ is simply
the statement that we are seeking solutions preserving 4d $\cN = 2$ superconformal
symmetry.
Notice that the 
 spinor $(i\, \sigma^{3} \, \vartheta^\cI)$ satisfies the Killing equation \eqref{S2_Killing}
with $+i$ on the RHS replaced with $-i$. 

The spinors $\epsilon_\pm$ are not independent, but rather related as (see \cite{Lin:2004nb}
and also \cite{OColgain:2010ev})
\beq
\epsilon_- = - a \, \Gamma_5 \, \epsilon_+ \ , \qquad a =\pm 1 \ .
\eeq
As a result, we can also write the expression \eqref{two_xis} for $\xi$ as
\beq \label{xi_spinors}
\xi ^\cI = (1 - a \, \gamma_7) \, (\vartheta^\cI \otimes \epsilon) \ , \qquad \epsilon \equiv \epsilon_+ \ .
\eeq
The BPS equations \eqref{GMSW_eqs} imply the following conditions on $\epsilon$,
\begin{align} \label{BPS_for_epsilon}
 \nabla_{  \alpha} \widetilde \lambda \,  \Gamma^{  \alpha}\,  \epsilon
 +   \tfrac{a}{ 12 } \,e^{-3 \widetilde \lambda -2A}\, F_{  \alpha   \beta}  \,     \Gamma^{  \alpha   \beta}  \,\Gamma_5 \, \epsilon 
 - i \, a \,     \epsilon  & = 0   \ , \nn \\
 \nabla_{  \alpha} \epsilon   
+ \tfrac{i \, a}{2} \,      \Gamma_{  \alpha} \,  \epsilon 
   - \tfrac{a}{4} \,   e^{- 3 \widetilde \lambda -2A}     \, F_{  \alpha   \beta}  \, 
    \Gamma^{  \beta} \, \Gamma_5\,  \epsilon & = 0  \ ,  \nn \\
 \nabla_{  \alpha} A \, \Gamma^{  \alpha} \, \epsilon 
 + i \, e^{-A} \, \Gamma_5\,  \epsilon 
+  i \, a \,  \epsilon 
- \tfrac{a}{4}\, e^{-3 \widetilde  \lambda - 2A} \, F_{  \alpha   \beta}  \,     \Gamma^{  \alpha   \beta}  \,\Gamma_5 \, \epsilon  &  = 0   \ .
\end{align}

\paragraph{Dirac Bilinears.}
We are now in a position to retrace the steps of \cite{Lin:2004nb}
to study bilinears constructed with $\epsilon$
and projection conditions on $\epsilon$.
Dirac bilinears are constructed with $\bar \epsilon \equiv \epsilon^\dagger$.
The scalar bilinear is a constant: we choose the normalization
\beq
\bar \epsilon \, \epsilon = 1 \ .
\eeq
The pseudo-scalar bilinear defines a non-trivial 0-form on $M_4$.
The Fierz rearrangements
\beq \label{myFierz}
(\bar \epsilon \, \Gamma^\alpha \, \epsilon)^2 = - (\bar \epsilon \, \Gamma^\alpha \, \Gamma_5 \epsilon)^2
= (\bar \epsilon \, \epsilon)^2 - (\bar \epsilon \, \Gamma_5 \, \epsilon)^2  \ , 
\eeq
together with the fact that $\bar \epsilon \, \Gamma^\alpha \, \epsilon$ is real
and $\bar \epsilon \, \Gamma^\alpha \, \Gamma_5 \epsilon$ purely imaginary,
 show that $|\bar \epsilon \, \Gamma_5 \, \epsilon| \le |\bar \epsilon \, \epsilon|=1$.
Therefore, we can   
 parametrize  $\bar \epsilon \, \Gamma_5 \, \epsilon$ as
\beq \label{sign_in_pseudoscalar}
\bar \epsilon \, \Gamma_5 \, \epsilon = - \sin \zeta  \ ,
\eeq
with $\zeta \in [-\pi/2, \pi/2]$.
The quantity $\zeta$ is related to the canonical coordinate $y$ in LLM form 
\eqref{LLM} via
\beq \label{y_from_zeta}
y = e^{3 \widetilde \lambda} \, \sin \zeta  \ .
\eeq
The vector bilinear defines a Killing vector, which is identified
with the canonical angular direction $\chi$ in the standard LLM form
\eqref{LLM}. More precisely, if $y^\alpha$ are local coordinates on $M_4$,
we can write
\beq \label{vector_bilinear}
\bar \epsilon \, \Gamma^\alpha \, \epsilon \, \frac{\partial}{\partial y^\alpha} = \frac{\partial}{\partial \chi}  \ .
\eeq
The pseudo-vector bilinear
$\bar \epsilon \, \Gamma_\alpha \, \Gamma_5 \, \epsilon$
turns out to be given in terms of the warp factor
and the derivative of the pseudo-scalar bilinear
$\bar \epsilon \, \Gamma_5 \, \epsilon$.
More precisely, we can write
\beq
\bar \epsilon \, \Gamma_\alpha \, \Gamma_5 \, \epsilon \, dy^\alpha = - \frac{i \, a}{2} \, e^{- 3 \widetilde  \lambda} \, dy \ .
\eeq
The Fierz rearrangements \eqref{myFierz}, together with the additional Fierz identity
\beq \label{orthogonality}
(\bar \epsilon \Gamma^\alpha  \epsilon ) (\bar \epsilon \Gamma_\alpha \Gamma_5 \epsilon) = 0 \ ,
\eeq
can also be used to constrain the form of the metric on $M_4$.
Together with the coordinates $\chi$, $y$ defined in \eqref{vector_bilinear}, \eqref{y_from_zeta},
we have two coordinates $x^i$, $i =1,2$.
The line element reads
\beq \label{M4_metric}
ds^2(M_4) = \cos ^2 \zeta \, (d\chi + v_i \, dx^i)^2 + \frac{1}{4\,e^{6\widetilde \lambda}\, \cos^2 \zeta} \,
(dy^2 + \gamma_{ij} \, dx^i \, dx^j) \ .
\eeq
Notice how \eqref{orthogonality} is exploited by choosing the coordinate $y$
not to have any mixed metric component with any other coordinate.
The quantities $\zeta$, $v_i$, $\gamma_{ij}$ and the warp factor
depend on $y$, $x^i$, but not on $\chi$.
A natural vielbein for $ds^2(M_4)$ is
\beq
e^{\underline \chi} = \cos \zeta \, D\chi \equiv \cos \zeta \, (d\chi + v_i \, dx^i) \ , \qquad
e^{\underline y} =  \frac{1}{2\,e^{3\widetilde \lambda}\, \cos \zeta} \ , \qquad
e^{\underline i} =  \frac{1}{2\,e^{3\widetilde \lambda}\, \cos \zeta} \, \widehat e^{\underline i} \ ,
\eeq
where  $ \widehat e^{\underline 1}$, $ \widehat e^{\underline 2}$ is a vielbein
for   $\gamma_{ij}$.
We choose the orientation $i=1$, $i=2$, $\chi$, $y$,
so that
$\epsilon_{\underline 1 \underline 2 \underline \chi \underline y} = 1$
and $\Gamma_5 = \Gamma_{\underline 1} \, \Gamma_{\underline 2} \, \Gamma_{\underline \chi} \, \Gamma_{\underline y}$.


Following the steps detailed in appendix F.2 of \cite{Lin:2004nb},
one can manipulate the BPS conditions \eqref{BPS_for_epsilon} on $\epsilon$
to determine the function $A$ and the 2-form $F$.
The function $A$ is  
\beq \label{A_expression}
e^A = - \frac a2 \, \sin \zeta \ .
\eeq
The coordinate $y$ in LLM form is non-negative.
Comparing \eqref{y_from_zeta} and \eqref{A_expression} we see that we have to 
choose the sign $a$ to be
\beq
a = -1 \ .
\eeq
The sign of  $a$ is correlated to our choice
\eqref{sign_in_pseudoscalar}. We could perform the redefinition $\zeta \rightarrow - \zeta$
and change the sign of $a$.
The  2-form $F$ is given by
\begin{align} \label{F_expression}
F & =   \frac{3 }{4  }  \,  \frac{y^2 \, e^{-6 \widetilde \lambda}  }{ 1 -  y^2 \, e^{-6 \widetilde \lambda}  } \, 
*_3 d\widetilde \lambda
+ \frac{1  }{4}  \, D\chi \wedge d( y^{3} \, e^{-6\widetilde \lambda} )   \ ,
\end{align}
where $*_3$ denotes the Hodge star
operation associated to the line element
$dy^2 + \gamma_{ij} \, dx^i \, dx^j$,
with ordering   $i=1$, $i=2$, $y$.

\paragraph{Projection Conditions.}
We may now plug the values of $A$ and $F$
in the first and third BPS equation in \eqref{BPS_for_epsilon}.
The resulting algebraic conditions on the spinor $\epsilon$
can be manipulated to take the form of projection conditions.
More precisely, one finds
\beq \label{projections}
\Big[ 1 - i \, \Gamma^{\underline 1 \underline 2}  \Big] \epsilon = 0 \ , \qquad
\Big[ \cos \zeta + i \, \sin \zeta \, \Gamma^{\underline y}
+ i \,  \Gamma^{\underline y} \, \Gamma_5 \Big] \epsilon = 0 \ .
\eeq
These relations take a simpler form in terms of the rescaled spinor
\beq  \label{new_epsilon}
\widetilde \epsilon = e^{\frac i2 \, \zeta \, \Gamma^{\underline y}} \, \epsilon  \ ,
\eeq
since they can be written as
\beq
\Big[ 1 - i \, \Gamma^{\underline 1 \underline 2}  \Big] \widetilde \epsilon = 0 \ , \qquad 
\Big[ 1 + i \, \Gamma^{\underline y  } \, \Gamma_5 \Big] \widetilde \epsilon = 0 \ .
\eeq
These relations demonstrate that $\widetilde \epsilon$ has only one independent component.
Moreover we can always use a local Lorentz rotation to
make sure that $\widetilde \epsilon$ is independent of the coordinates
$y$, $x^1$, $x^2$. (Its $\chi$ dependence will be fixed momentarily.)

\paragraph{Majorana Bilinears.}
To proceed we follow \cite{Lin:2004nb}
and construct a Majorana 1-form bilinear.
The 4d charge conjugation matrix $C$ satisfies
\beq
(\Gamma^\alpha)^T = - C \, \Gamma^\alpha \, C^{-1} \ , \qquad C^T = -C \ .
\eeq 
The bilinear of interest is
\beq
\omega_1 = \epsilon^T \, C \, \Gamma_\alpha \, \epsilon \, dy^\alpha   \ .
\eeq
Since the matrices $C$, $C \, \Gamma_5$, and $C\, \Gamma^\alpha \, \Gamma_5$
are antisymmetric, and the spinor $\epsilon$ is Grassmann even,
the bilinears $\epsilon^T \, C \, \epsilon$, $\epsilon^T \, C \, \Gamma_5 \, \epsilon$,
and $\epsilon^T \, C\, \Gamma^\alpha \, \Gamma_5 \, \epsilon$ are all identically zero.
This information, combined with the projection conditions 
\eqref{projections}, allow one to conclude that the 1-form $\omega_1$
has only one independent component,
because
\beq
 \epsilon^T \, C \, \Gamma^{\underline y} \,   \epsilon = 0 \ , \qquad
  \epsilon^T \, C \, \Gamma^{\underline \chi} \,   \epsilon = 0 \ , \qquad
   \epsilon^T \, C \, \Gamma^{\underline 2} \,   \epsilon =   -i \,  \epsilon^T \, C \, \Gamma^{\underline 1} \,   \epsilon \ .
\eeq
The definition \eqref{new_epsilon}  implies
\beq
 \epsilon^T \, C \, \Gamma^{\underline 1} \,   \epsilon = 
\cos \zeta \, \widetilde \epsilon^T \, C \, \Gamma^{\underline 1} \, \widetilde \epsilon \ .
\eeq
Using this relation, the 1-form $\omega_1$ can
 be cast in the form
\beq \label{explicit_omega1}
\omega_1 = \frac 12 \, e^{- 3\widetilde \lambda} \, (\widetilde \epsilon^T \, C \, \Gamma^{\underline 1} \, \widetilde \epsilon) \, \big(  \widehat e^{\underline 1} - i \, \widehat e^{\underline 2} \big) \ ,
\eeq
where  $ \widehat e^{\underline 1}$, $ \widehat e^{\underline 2}$ is a vielbein
for the $\gamma_{ij}$ metric in \eqref{M4_metric}.
The 2-form $d\omega_1$ can be computed making use of the BPS
equations \eqref{BPS_for_epsilon}, with the result
\beq
d\omega_1 = \frac i4 \, (\epsilon^T \, C \, \Gamma_{\alpha \beta} \, \epsilon) \, dy^\alpha \wedge dy^\beta
- 3 \, d \widetilde \lambda \wedge \omega_1  \ .
\eeq
The components of the bilinear 
$\epsilon^T \, C \, \Gamma_{\alpha \beta} \, \epsilon$ are greatly constrained
by the projection conditions \eqref{projections}. 
Combining them with
the fact that $\epsilon^T \, C\, \Gamma^\alpha \, \Gamma_5 \, \epsilon$
is identically zero, one verifies 
\begin{align}
& i \, \epsilon^T \, C \, \Gamma_{\underline y \underline 2} \, \epsilon =  \epsilon^T \, C \, \Gamma_{\underline y \underline 1} \, \epsilon
=  i \, \tan \zeta \, \epsilon^T \, C \, \Gamma_{  \underline 1} \, \epsilon  \  , \qquad
i \, \epsilon^T \, C \, \Gamma_{\underline \chi \underline 2} \, \epsilon = \epsilon^T \, C \, \Gamma_{\underline y \underline 1} \, \epsilon
= - \frac{1}{\cos \zeta} \, \epsilon^T \, C \, \Gamma_{ \underline 1} \, \epsilon \ , \nn \\
& \epsilon^T \, C \, \Gamma_{\underline 1 \underline 2} \, \epsilon =  0 
=    \epsilon^T \, C \, \Gamma_{\underline \chi \underline y} \, \epsilon \ .
\end{align}
These relations allow one to recast the equation for $d\omega_1$ in the form
\beq \label{omega_derivative}
d\omega_1 = \bigg[
- \frac{y \, e^{-6\widetilde \lambda}}{2 \, (1 - y^2 \, e^{-6\widetilde \lambda })} \, dy
-i \, D\chi
- 3 \, d\widetilde \lambda
 \bigg] \wedge \omega_1 \ .
\eeq
As explained in \cite{Lin:2004nb}, using
\eqref{explicit_omega1} and 
 the fact that
the rescaled spinor $\widetilde \epsilon$ is independent of $y$, $x^1$, $x^2$,
one can use \eqref{omega_derivative} to argue that the vielbein for the metric 
$\gamma_{ij}$ can be chosen to be
\beq
\widehat e^1 = e^{D/2} \, dx^1 \ , \qquad
\widehat e^2 = e^{D/2} \, dx^2 \ ,
\eeq
for some function $D$ of $y$, $x^1$, $x^2$
(which is ultimately identified with the function $D$ in the canonical LLM form).
Once this choice for $\widehat e^1$, $\widehat e^2$ is made,
the various components of the equation \eqref{omega_derivative}
can be studied separately and yield the following results.
Firstly, the warp factor is determined by the function $D$ 
by the expression \eqref{lambda_and_D}.
Secondly, the $\chi$ dependence of the rescaled spinor $\widetilde \epsilon$
is determined to be
\beq
\partial_\chi \, \widetilde \epsilon =- \frac i2 \,  \widetilde \epsilon  \ .
\eeq
Finally, the quantities $v_i$ in the metric \eqref{M4_metric}
are fixed in terms of the function $D$ as
\beq \label{v_equations}
v_1 = + \frac 12 \, \partial_2 D \ , \qquad v_2 = - \frac 12 \, \partial_1 D \ .
\eeq

\paragraph{Toda Equation and Final Form of the 2-form $F$.}
The relation \eqref{vector_bilinear}
and the form of the metric \eqref{M4_metric} imply that  
\beq \label{alpha_expr}
\alpha_1 := (\bar \epsilon \, \Gamma_\alpha\, \epsilon) \, dy^\alpha = \cos^2 \zeta \, D\chi
= (1 - y^2 \, e^{- 6 \widetilde \lambda}) \, D\chi \ .
\eeq
On the one hand, the 2-form $d\alpha_1$ is readily computed from this expression for $\alpha_1$.
On the other hand, the BPS equations \eqref{BPS_for_epsilon}
imply the following equation for $d\alpha_1$,
\beq \label{alpha_der}
d \alpha_1 = -2 \, i  \cdot    \tfrac 12 \, (\bar \epsilon \, \Gamma_{\alpha \beta} \, \epsilon)\,
dy^\alpha \wedge dy^\beta  
-   e^{-3 \widetilde \lambda - 2A} \, (\bar \epsilon \, \Gamma_5 \, \epsilon) \, F  \ .
\eeq
The components of $\bar \epsilon \, \Gamma_{\alpha \beta} \, \epsilon$ are constrained
by the projection conditions \eqref{projections}.
For example,
we have $\Gamma^{\underline 2} \, \epsilon =-  i \, \Gamma^{\underline 1} \, \epsilon$
and $\bar \epsilon \, \Gamma^{\underline 1} = -  i \, \bar \epsilon\, \Gamma^{\underline 2}$,
and therefore
\beq
\bar \epsilon \, \Gamma^{\underline y} \, \Gamma^{\underline 1} \, \epsilon
= - (\bar \epsilon \, \Gamma^{\underline 1}) \, \Gamma^{\underline y} \, \epsilon
= i \, \bar \epsilon \, \Gamma^{\underline 2} \, \Gamma^{\underline y} \, \epsilon
= - i \,  \bar \epsilon \, \Gamma^{\underline y} \, (\Gamma^{\underline 2} \, \epsilon)
- \bar \epsilon \, \Gamma^{\underline y} \, \Gamma^{\underline 1} \, \epsilon \ ,
\eeq
which shows that $\bar \epsilon \, \Gamma^{\underline y} \, \Gamma^{\underline 1} \, \epsilon=0$.
By similar arguments, one verifies that the only independent non-zero components
of $\bar \epsilon \, \Gamma_{\alpha \beta} \, \epsilon$ are
\beq
\bar \epsilon \, \Gamma_{\underline 1 \underline 2} \, \epsilon = -i \ , \qquad
\bar \epsilon \, \Gamma_{\underline \chi \underline y} \, \epsilon  = - i \, \sin \zeta \ .
\eeq
This information, combined with the expressions \eqref{A_expression}, \eqref{F_expression}
for $A$ and $F$, allows us to make a direct comparison
between the relation \eqref{alpha_der} for $d\alpha_1$
and the actual value of $d\alpha_1$ as computed from \eqref{alpha_expr}.
The $dx^1 \wedge dx^2$ piece of the resulting 2-form equation,
using \eqref{v_equations}, 
implies the Toda equation for $D$ \eqref{Toda_equation}.

Having established that $D$ satisfies the Toda equation,
we can revisit the expression \eqref{F_expression} for $F$ and make use of the identity 
\begin{align}
3\, \frac{y^2 \, e^{-6\lambda}  }{ 1 -  y^2 \, e^{-6\lambda}  } \, *_3 d \widetilde \lambda  
& = y \, (1 - y^2 \, e^{-6\lambda} ) \, dv - \frac 12 \, \partial_y e^D \, dx^1 \wedge dx^2 \nn \\
& + \frac{y}{2 \, (1- y \, \partial_y D)} \, (\partial_1^2 D + \partial_2^2 D + \partial_y^2 e^D) \ .
\end{align}
Since the term on the second line is zero, we can rewrite $F$ as
\beq
F  =  \frac 14 \, \bigg[ 
y \, (1 - y^2 \, e^{-6\lambda} ) \, dv - \frac 12 \, \partial_y e^D \, dx^1 \wedge dx^2
+  D\chi \wedge d( y^{3} \, e^{-6\widetilde \lambda} ) 
\bigg]  \ ,
\eeq
Using \eqref{11d_split} and \eqref{cG4_factorized}, this expression for $F$
implies that $G_4$ is given by the expression quoted in \eqref{LLM}.

\subsubsection{The Calibration 2-form $Y'$} \label{app_calibration_2form}

The calibration 2-form for supersymmetric M2-branes 
was identified in  \cite{Gauntlett:2006ai} for 
the most general $AdS_5$ solution preserving 4d $\cN = 1$ superconformal
symmetry, as classified by Gauntlett, Martelli, Sparks, and Waldram (GMSW) \cite{Gauntlett:2004zh}.
In order to apply the results of \cite{Gauntlett:2006ai} to a solution preserving 4d $\cN = 2$ superconformal
symmetry,
we select a linear combination $\xi$ of the two LLM spinors $\xi^\cI$  in \eqref{xi_spinors},
and we identify $\xi$ with the Killing spinor of the general GMSW solution,
\beq \label{our_combo}
\xi = c_\cI \, \xi^\cI = (1 + \gamma_7) \, (c_\cI \, \vartheta^\cI \otimes \epsilon) \ ,
\eeq
where we have used $a = -1$ and $c_\cI$ are complex constants.
Before proceeding, it is useful to collect some identities
about bilinears of the $S^2$ Killing spinors $\vartheta^\cI$.
We set $\bar \vartheta_\cI = (\vartheta^\cI)^\dagger$.
A basis $\vartheta^\cI$ of solutions to \eqref{S2_Killing} can always be chosen,
in such a way that\footnote{In checking these relations, we have
adopted the explicit expressions for Killing spinors on spheres
of \cite{Lu:1998nu}.}
\begin{align} \label{S2_bilinears}
\bar\vartheta_\cI \, \vartheta^\cJ &= \delta^\cJ_\cI \  , &
\bar\vartheta_\cI \, \sigma^3\,  \vartheta^\cJ &= \hat y^A \, (\sigma_A)^\cJ{}_\cI \ , \nn \\
\bar \vartheta_\cI \, \sigma_x \, \vartheta^\cJ &= - \cK_x{}^A \, (\sigma_A)^\cJ{}_\cI \ ,  &
\bar \vartheta_\cI \, \sigma_x \, \sigma^3 \, \vartheta^\cJ &=i \, \partial_x \hat y^A\, (\sigma_A)^\cJ{}_\cI \ .
\end{align}
In the previous relations $x=1,2$ is a curved index on $S^2$.
The Pauli matrices on the LHSs play the role of gamma matrices on $S^2$.
The Pauli matrices on the RHSs are invariant tensors of the $\mathfrak{su}(2) \cong \mathfrak{so}(3)$ isometry algebra of $S^2$, connecting the indices $\cI, \cJ = 1,2$ in the fundamental
representation of $\mathfrak{su}(2)$ to the indices $A = 1,2,3$ in the vector
representation of $\mathfrak{so}(3)$.
The three quantities $\hat y^A$ are real scalars in $S^2$, identified
with the Cartesian coordinates of $\mathbb R^3$ in the standard embedding $S^2 \subset \mathbb R^3$.
The metric on $S^2$ is given in terms of $\hat y^A$ by
\beq
g_{xy} = \partial_x \hat y^A \, \partial_y \hat y_A  \ .
\eeq
(The $A$  indices in   are raised/lowered with $\delta$.)
The 1-forms $\cK_x{}^A$ are defined as
\beq \label{Killing_vecs}
\cK_x{}^A = \epsilon^{ABC} \, \hat y_B \, \partial_x \hat y_C \ ,
\eeq
and yield the standard Killing vectors on $S^2$ after raising their
curved index $x$ with the $S^2$ metric.
The 1-forms $\cK_x{}^A$ can also be written as Hodge duals of the gradients
of $\hat y^A$, because $\epsilon_{xy} \, \cK^{xA} = \partial_x \hat y^A$,
where $\epsilon_{12} = \sqrt{\det g_{xy}}$.

In order to preserve the normalization condition $\overline \xi\, \xi = 2$ of \cite{Gauntlett:2004zh}, 
the constants $c_\cI$ and their complex conjugates $\bar c^{\cI}$ should satisfy
$c_\cI \, \bar c^{\cI} = 1$. A choice of $c_\cI$ determines a vector $n^A$ 
in $\mathbb R^3$ via the formula
\beq
n^A = c_\cJ \, (\sigma^A)^\cJ{}_{\cI} \, \bar c^{\cI} \ . 
\eeq
Without loss of generality, we can select $c_\cI = (1,0)$ in such a way that
the 3-vector $n^A$ points in the direction $A = 3$.
(Any other choice is related by the action of the isometry group of $S^2$.)
With this choice   we have
\beq
 \hat y^A \, n_A = \hat y^{A = 3} \equiv \tau \ .
\eeq
The quantity $\tau$ lies in $[-1,1]$.
The other two real   scalars $\hat y^{A=1}$, $\hat y^{A=2}$
are parametrized as
\beq
\hat y^{A=1} =\sqrt{1-\tau^2} \,  \cos \varphi   \ , \qquad
\hat y^{A=2} =\sqrt{1-\tau^2} \,  \sin \varphi   \ ,
\eeq
where $\varphi$ is an angle of periodicity $2\pi$.
The metric on $S^2$ is written in terms of $\tau$ and $\varphi$
in~\eqref{break_up_S2}.

We are now in a position to discuss the calibration 2-form.
It is given as
\beq
Y' = \frac 12 \, Y'_{mn} \, dy^m \wedge dy^n \ , \qquad Y' _{mn}= \frac 12 \, \bar \xi \, \gamma_{mn} \, \gamma_7 \, \xi   \ .
\eeq
Making use of \eqref{our_combo}  this 2-form can be written
in LLM setups as
\begin{align}
Y' _{mn}& = (\bar c^\cJ \, \bar \vartheta_{\cJ} \otimes \bar \epsilon) \, \gamma_{mn} \, \gamma_7 \, (c_\cI \, \vartheta^\cI \otimes \bar \epsilon) \ .
\end{align}
Using \eqref{S2_apart}, \eqref{decompose_gammas}, \eqref{S2_bilinears}, we can write the 2-form $Y'$ as
\begin{align}
Y' & =  
- e^{2A} \, (\bar c^\cI \, c_\cI) \, (\bar \epsilon \, \Gamma_5 \, \epsilon) \,  {\rm vol}_{S^2}
 + e^A \,  d(\hat y^A \, n_A) \wedge (\bar \epsilon \, \Gamma_\alpha \, \epsilon) \, dy^\alpha
\nn \\
& + \frac i2 \, (\hat y^A \, n_A)  \, (\bar \epsilon \, \Gamma_{\alpha \beta} \, \Gamma_5 \, \epsilon) \, dy^\alpha \wedge dy^\beta
 \ .
\end{align}
To treat the last term, it is convenient to use the identity
\beq
\Gamma_{\alpha \beta} \, \Gamma_5 = - \frac 12 \, \epsilon_{\alpha \beta \gamma \delta} \, \Gamma^{\gamma \delta} \ .
\eeq
We have already established the non-zero independent components
of the bilinear $\bar \epsilon \, \Gamma_{\alpha\beta} \, \epsilon$.
Combining our previous results, and specializing to our choice
of $c_\cI$ such that $\hat y^A \, n_A = \tau$, we obtain
\begin{align}
Y ' & = \frac{1}{4 }  \, y^3 \, e^{-9\lambda} \, {\rm vol}_{S^2}
 + \frac{1    }{2 } \, y \, e^{-3\lambda} \, (1 - y^2 \, e^{-6\lambda}) \, d\tau \wedge D\chi
\nn \\
&  - \frac{  1 }{2  } \, \tau \, e^{-3\lambda }\,  D\chi \wedge dy
 - \frac{ 1  }{4} \,  
\frac{  y \, e^{-9\lambda} \, \tau \, e^D }{1 - y^2 \, e^{-6\lambda}}
  \, dx^1 \wedge dx^2  \ .
\end{align}
We can also write $Y'$ in terms of $d\phi$ and the  1-form $Dz$ defined in \eqref{puncture_metric},
\begin{align} \label{ugly_Y}
Y' & = \frac{\mu^3  \, w^{3/2} \, [\kappa(1-w^2)]^{3/4}}{\sqrt 2\, B^{3/2}  \, \cH^{3/2}} \,  {\rm vol}_{S^2}
+ \frac{ \cC \, \kappa \,   \mu  \, [\kappa \, (1-w^2)]^{-3/4}}{   \sqrt 2 \, \sqrt B \, \sqrt w \, \sqrt \cH } \, \tau\, dw \wedge Dz 
\nn \\
& - \frac{\kappa \, \mu \, (1-\mu^2) \, (2w^2-1)   \, [\kappa(1-w^2)]^{-1/4}  }{\sqrt 2 \, \sqrt B \, \sqrt w\,  \sqrt \cH \, \Big[ 2 \, B \, w\, \cH - (\mu^2 -1 + 4w^2) \, \sqrt{\kappa (1-w^2)}      \Big]} \, \tau \, dw \wedge d\phi
\nn \\
& + \frac{\cC \, [\kappa(1-w^2)]^{-1/4}  \, \Big[ B \, w - \sqrt{\kappa (1-w^2)}  \Big]}{   \sqrt 2 \, B^{3/2} \, \sqrt w \, \sqrt \cH } \, \tau \, d\mu \wedge Dz  
\nn \\
& + \frac{\cC \, \mu \, \sqrt w  \, [\kappa (1-w^2)]^{1/4} \, \Big[  B \, \cH - w \, (1+\mu^2) \,\sqrt{\kappa(1-w^2)} \Big]   }{   \sqrt 2 \, B^{3/2} \, \cH^{3/2} } \, d\tau \wedge Dz
\nn \\
& + \frac{ \sqrt w \, (1+\mu^2) \, h \, [\kappa(1-w^2)]^{3/4} }{   \sqrt 2 \, B^{3/2}  \, \sqrt \cH \, 
\Big[ 2 \, B \, w\, \cH - (\mu^2 -1 + 4w^2) \, \sqrt{\kappa (1-w^2)}      \Big]  } \, \tau \, d\mu \wedge d\phi
\nn \\
& + \frac{ \sqrt w \, \mu \, (1-\mu^2) \, h \, [\kappa(1-w^2)]^{3/4} }{
 \sqrt 2 \, B^{3/2}   \, \sqrt \cH \, 
\Big[ 2 \, B \, w\, \cH - (\mu^2 -1 + 4w^2) \, \sqrt{\kappa (1-w^2)}      \Big] 
} \, d\tau \wedge d\phi \ .
\end{align}

\subsubsection{R-symmetry Charges of Wrapped M2-branes} \label{app_brane_charges}

The R-symmetry charges of wrapped M2-brane operators can be extracted using the Wess-Zumino coupling
of M2-branes to $C_3$, as in \cite{Gauntlett:2006ai}. To this end, we need to introduce external
background gauge fields in $G_4$. This is accomplished in the construction
of $E_4$ in section \ref{sdsd}. The final result is repeated here for convenience,
\begin{align}
\!\!\!\! E_4   = N \, e_2 \wedge \bigg[  d\alpha_{0\chi} \wedge \frac{(d\chi)^{\rm g}}{2\pi}
+  d\alpha_{0\beta} \wedge \frac{(d\beta)^{\rm g}}{2\pi}  \bigg]
+ N \, \alpha_{0\chi} \, e_2 \wedge \frac{F^\chi}{2\pi}
- \cC^{-1} \, \frac{f_1}{2\pi} \wedge e_2 \wedge d\alpha_{0\beta} \ . \!\!\!
\end{align}
The 0-forms $\alpha_{0\chi}$, $\alpha_{0\beta}$ are defined in \eqref{alpha0_defs}.
If all external gauge fields are turned off, $E_4$ reduces to $\overline G_4$,
which is related to the background $G_4$ by the relation \eqref{G4_rescaling}.
As a result, we can write
\beq
G_4^{\rm tot} = - (2\pi \ell_p)^3 \, E_4 \ ,
\eeq
where $G_4^{\rm tot}$ denotes the background $G_4$-flux dressed with
external background gauge fields, written in the same normalization
used in the main text in giving the solution, cfr.~\eqref{G4flux}.
To proceed, let us write
\beq
G_4^{\rm tot} = G_4 + d \delta C_3 \ ,
\eeq
where $\delta C_3$ collects all terms that contain external gauge fields.
For the computation at hand, we only need to retain terms
linear in the external gauge fields inside $\delta C_3$.
In the conventions we are adopting, the coupling of an M2-brane to $C_3$ fluctuations
is given by
\beq \label{C3_coupling}
S_{\rm M2} \supset T_{\rm M2} \, \int_{\cW_3} \delta C_3 \ , \qquad 
T_{\rm M2} = \frac{2\pi}{(2\pi \ell_p)^3} \ ,
\eeq
where $\cW_3$ denotes the worldvolume of the M2-brane.

The expression of $d\delta C_3$ is extracted from $E_4 -\overline G_4$.
A useful identity for the $SO(3)$ global angular form $e_2$ is 
\beq \label{e2_expr}
e_2 = \frac{\rm {vol}_{S^2}}{4\pi} - \frac{d(\hat y^A \, A_A)}{4\pi} \ , \qquad  A = 1,2,3 \ , 
\eeq
in conventions in which $D\hat y^A = d\hat y^A - A^{AB} \, \hat y_B = 
 d\hat y^A -  \epsilon^{ABC} \, A_C \,   \hat y_B
$.
Making use of the expression \eqref{final_Bianchi} for $f_1$,
we find the following result for $d\delta C_3$,
\begin{align}
 - \frac{d\delta C_3}{(2\pi\ell_p)^3} & =  - \frac{d(\hat y^a \, A_a)}{4\pi} \,  
N \, \bigg[  d\alpha_{0\chi} \, \frac{ d\chi }{2\pi}
+  d\alpha_{0\beta} \, \frac{ d\beta }{2\pi}  \bigg]
  + N \, \frac{ {\rm vol}_{S^2} }{ 4\pi } \,  \bigg[  d\alpha_{0\chi} \,  \frac{A^\chi}{2\pi}
 +  d\alpha_{0\beta} \,  \frac{A^\beta}{2\pi}  \bigg]
 \nn \\
& + N \, \frac{ {\rm vol}_{S^2} }{ 4\pi }  \, \alpha_{0\chi} \,   \frac{dA^\chi}{2\pi}
- \cC^{-1} \, \frac{da_0 - N \, \cC \, A^\beta}{2\pi} \,  \frac{ {\rm vol}_{S^2} }{ 4\pi } \, d\alpha_{0\beta} \ .  
\end{align}
An antiderivative of the above quantity is readily extracted,
\begin{align}
 - \frac{\delta C_3}{(2\pi\ell_p)^3} & =  - \frac{ \hat y^a \, A_a }{4\pi} \,  
N \, \bigg[  d\alpha_{0\chi} \, \frac{ d\chi }{2\pi}
+  d\alpha_{0\beta} \, \frac{ d\beta }{2\pi}  \bigg]
  + N \, \frac{ {\rm vol}_{S^2} }{ 4\pi }  \, \alpha_{0\chi} \,   \frac{A^\chi}{2\pi} 
- \cC^{-1} \, \frac{a_0  }{2\pi} \,  \frac{ {\rm vol}_{S^2} }{ 4\pi } \, d\alpha_{0\beta}
\ .  
\end{align}
Plugging this expression in \eqref{C3_coupling} we can write the relevant couplings
for the two supersymmetric M2-brane probes discussed in the main text.
For the probe associated to $\cO_2^i$, it is convenient to write $\delta C_3$
in terms of $d\phi$, $Dz$. The results are
\begin{align}
\cO_1 \; : \;\; S_{\rm M2}   \supset   
  - \frac{ N \, K \, \ell}{N + K \, \ell} \,   \int_{\cW_1}   A^\chi \ , \qquad  \qquad 
  \cO_2^i \; : \;\; S_{\rm M2}   \supset       
\frac 12 \, K  \,   \int_{\cW_1}   \hat y_* ^a \, A_a \ .
\end{align}
We have used $\cW_1$ to denote the worldline of the M2-brane
in external spacetime. In the computation for $\cO_2^i$, 
we have assigned positive orientation to $dw \wedge Dz$.
The quantities $\hat y_*^a$ are the $\mathbb R^3$ embedding coordinates
of the point on $S^2$ where the M2-brane probe sits.
As verified in section \ref{sec:wrappedM2s}, the brane sits at the north pole of $S^2$, hence
\beq
\hat y_*^A = (0,0,1) \ , \qquad \hat y_*^a \, A_A = A^{A=3} \ .
\eeq
In what follows, we turn off the gauge fields $A^{A=1,2}$, leaving only a non-zero
$A^{a=3}$.
Using $\hat y^1 = \sqrt{1-\tau^2}\, \cos \varphi $, $\hat y^2 = \sqrt{1-\tau^2} \, \sin \varphi$,
$\hat y^3 = \tau$, we verify that the gauging prescription
$ d\hat y^A \rightarrow D\hat y^A  = 
 d\hat y^A -  \epsilon^{ABC} \, A_C \,   \hat y_B$ is equivalent
to
\beq
d\varphi \rightarrow d\varphi +A^\varphi \ , \qquad A^\varphi \equiv A^{A=3} \ .
\eeq
This relation confirms the identification of the Killing vector $\partial_\varphi$
with the Cartan generator of $\mathfrak{so}(3)_R \cong \mathfrak{su}(2)_R$,
and states our normalization for the associated $U(1)$ gauge field $A^\varphi$.

To match with the normalization conventions
of \eqref{brane_Rcharges}, we define appropriately rescaled versions
of $A^\chi$, $A^\varphi$,  denoted $A^r$, $A^R$,
\beq
A^r = - \frac 12 \, A^\chi \ , \qquad
A^R = \frac 12 \, A^\varphi  \ , \qquad
r = - 2 \, \partial_\chi \ , \qquad
R = 2 \, \partial_\varphi \ .
\eeq 
We have also given the identification between the generators $r$, $R$
and the Killing vectors $\partial_\chi$, $\partial_\varphi$
(by slight abuse of notation, we use the symbols $r$, $R$ both for the
abstract generators and for the charges of a given operator).
Notice that the relation between $A^r$ and $A^\chi$
is compatible with  \eqref{Rsymm_identifications} since $c_1(U(1)_r) = dA^r/(2\pi)$.
The charges $r$, $R$ of the operators $\cO_1$, $\cO_2^i$
are now read off from the M2-brane action,
$S_{\rm M2} \supset \int_{\cW_1} (r \, A^r + R \, A^R)$.
We reproduce the charges given in \eqref{brane_Rcharges} in the main text.

\section{Anomaly Polynomial in Class $\CS$} \label{sec:appanomalies}

In this section we review the anomaly polynomial for four-dimensional $\mathcal{N}=2$ SCFTs that belong to class $\mathcal{S}$. 

Consider an $\CN=2$ SCFT with $U(1)_{r}\times SU(2)_R$ R-symmetry, and flavor symmetry $F$. Denote the R-symmetry generators by $r$ and $I^a$, with $I^3$ the generator of the Cartan subgroup of  $SU(2)_R$. 
The anomaly polynomial for the theory takes the form 
	\ba{ \label{standard_anomaly}
	I_6 = (n_v-n_h) \left( \frac{(c_1^r)^3}{3} - \frac{c_1^r p_1(T^4)}{12}\right) - n_v c_1^r c_2^R - k_F c_1^r \text{ch}_2(F)\ .
	} 
This form follows from the $\mathcal{N}=2$ superconformal algebra \cite{Kuzenko:1999pi}. 
 In this expression, $c_1^r$ is the first Chern class for the $U(1)_r$ bundle,  $c_2^R$ is the second Chern class for the $SU(2)_R$ bundle, and $\text{ch}_2(F)$ is the two-form part of the Chern character for the flavor symmetry bundle, given for an $SU(m)$ flavor group as $\text{ch}_2(SU(m))) = - c_2(SU(m))$. The parameters $n_v$ and $n_h$ represent an effective number of vector and hypermultiplets respectively, coinciding with the actual number when the theory is weakly coupled. These are given in terms of the anomaly coefficients and $a,c$ central charges as
	\ba{
	\tr r^3 = \tr r &= 2(n_v-n_h) = 48 (a-c)\ ,\\
	\tr r I^aI^b &= \delta^{ab}\frac{n_v}{2} =\delta^{ab}  2 (2a-c)\ .
	}
	One can equivalently express the central charges in terms of $n_v,n_h$ as
	\ba{
	a = \frac{1}{24} (n_h+5n_v)\ ,\qquad c = \frac{1}{12} (n_h + 2 n_v)\ .\label{eq:change}
	}
The flavor central charge $k_F$ is defined in terms of the flavor symmetry group generators $T^a$ as
	\ba{
	k_{F}\delta^{ab} = - 2 \tr r T^a T^b\ .
	}

Let us now review the contributions to the anomaly polynomial for the theories of class $\CS$ of $A_{N-1}$ type that is obtained by wrapping $N$ M5-branes on a Riemann surface $\Sigma_{g,n}$ with genus-$g$ and $n$ regular punctures. The contributions can be split into a ``bulk'' term which we denote with a $\Sigma_{g,n}$ argument, and a ``local'' term associated to each regular puncture on the Riemann surface labeled by the Young diagram $Y_{\alpha}$, as \cite{Bah:2018gwc}
	\ba{
	I_6= I_6(\Sigma_{g,n}) + \sum_{\alpha=1}^n I_6(Y_{\alpha})\ .
	}
The bulk contribution is proportional to the Euler characteristic $\chi(\Sigma_{g,n})$ of the Riemann surface, and is given (using the parameterization \eqref{standard_anomaly}) by  \cite{Benini:2009mz,Alday:2009qq}
\ba{
n_v(\Sigma_{g,n}) =- \frac{1}{6} \chi(\Sigma_{g,n}) (4 N^3 - N - 3  ) \ ,\qquad n_h(\Sigma_{g,n}) = -\frac{2}{3} \chi(\Sigma_{g,n}) (N^3-N)\ .
}
These can be equivalently written in terms of the $a$ and $c$ central charges using \eqref{eq:change}, as
\ba{
a(\Sigma_{g,n}) = - \frac{1}{48} \chi(\Sigma_{g,n}) (8N^3 - 3N - 5)\ ,\qquad c(\Sigma_{g,n}) = -\frac{1}{12} \chi(\Sigma_{g,n}) (2N^3 - N - 1)\ . \label{eq:acbulk}
}

The local contributions due to regular punctures on the Riemann surface can be given 
 in terms of the data of the Young tableaux that labels the puncture. Our notation follows \cite{Bah:2019jts}. Let $\tilde{\ell}_i$ denote the lengths of the $i=1,\dots,\tilde{p}$ rows, with $\tilde{p}$ the total number of rows. Denote $\tilde{k}_i=\tilde{\ell}_i-\tilde{\ell}_{i+1}$ the differences between adjacent rows, where here we are using a convention in which the lengths of rows increases from top to bottom of the diagram, so $\tilde{\ell}_1\geq \tilde{\ell}_2 \geq \dots$. We also define $\tilde{N}_i=\sum_{j=1}^i \tilde{\ell}_j$, such that $\tilde{N}_{\tilde{p}}=N$. The associated flavor symmetry is $S( \prod_i U(\tilde{k}_i))$. 
 In terms of this data, the contribution to the central charges of the regular puncture labeled by $Y$ is given as \cite{Gaiotto:2009gz,Chacaltana:2010ks}
	\ba{\bs{
	n_v(Y) &= - \sum_{i=1}^{\tilde{p}} (N^2 - \tilde{N}_i^2 ) - \frac{1}{2} N^2 + \frac{1}{2}\ ,\qquad n_h(Y) = n_v(Y) + \frac{1}{2} \sum_{i=1}^{\tilde{p}} \tilde{N}_i \tilde{k}_i - \frac{1}{2}\ .
	}
	}
For example, the maximal puncture has $\tilde{p}=1$, $\tilde{\ell}_1=\tilde{k}_1=\tilde{N}_1=N$, contributing an $SU(N)$ flavor symmetry and yielding
	\ba{\bs{
n_v(Y_{\text{max}}) &= - \frac{1}{2}(N^2-1)\ ,\quad \ \  \   n_h(Y_{\text{max}}) = 0 \\ 
a(Y_{\text{max}}) &= - \frac{5}{48} (N^2-1) \ ,\qquad c(Y_{\text{max}}) = - \frac{1}{12} (N^2-1) \ , }\label{eq:acpmax}
}
The minimal puncture has $\tilde{p}=N-1$, $\tilde{\ell}_1=2$, $\tilde{\ell}_{i=2,\dots,N-1}=1$, $\tilde{k}_1=1$, $\tilde{k}_{2,\dots,N-2}=0$, $\tilde{k}_{N-1}=1$, contributing a $U(1)$ flavor symmetry and contributing
\ba{\bs{
n_v(Y_{\text{min}}) &= - \frac{1}{6}(4N^3-6N^2-N+3)\ ,\quad\ \ \   n_h(Y_{\text{min}}) = -\frac{1}{3} (  2N^3 - 3 N^2 - 2 N )  \\
a(Y_{\text{min}})& = -\frac{1}{48} ( 8N^3 - 12N^2 - 3N + 5 ) \ ,\quad c(Y_{\text{min}}) =- \frac{1}{12} ( 2N^3 - 3N^2 - N +1 )\ .} \label{eq:acpmin}
}
Another example that we need in the main text is the rectangular box diagram with $N/\ell$ rows and $\ell$ columns, which we denote throughout by $Y_{\ell}$. (The maximal puncture corresponds to $\ell=N$.) In this case, $\tilde{p}=N/\ell$, $\tilde{\ell}_{1,\dots,N/\ell} = \ell$, $\tilde{k}_{1,\dots,N/\ell-1}=0$, $\tilde{k}_{N/\ell}=\ell$, and $\tilde{N}_i=i\ell$. Then, we have that
	\ba{
	\bs{
	n_v(Y_{\ell}) &= \frac{1}{6} \left( 3 + \ell N - 4 \frac{N^3}{\ell} \right)\ , \qquad n_h(Y_{\ell}) = \frac{2}{3}N  \left(\ell  - \frac{N^2}{\ell}\right)\ .
	} \label{eq:boxa}
	}



\section{Landscape of Argyres-Douglas Theories} \label{sec:AD}

In this appendix we review the landscape of four-dimensional Argyres-Douglas theories, and their construction via geometric engineering. 

\label{sec:ovirr}

\subsection{Construction in Class $\CS$} 

A large class of four-dimensional quantum field theories known as class $\mathcal{S}$ are obtained by compactifying the 6d $\CN=(2,0)$ SCFTs on a genus-$g$ Riemann surface with punctures, while implementing a partial topological twist in order to preserve some supersymmetry in four dimensions. 4d $\CN=2$ SCFTs engineered in this way were first studied and classified in \cite{Gaiotto:2009we,Gaiotto:2009hg}, building on \cite{Witten:1997sc}. The parent (2,0) theories are labeled by an algebra $\mathfrak{g}$ which follows an ADE classification, and which upon circle compactification reduces to the gauge algebra of the low-energy 5d supersymmetric Yang-Mills theory.\footnote{~Below we use $\mathfrak{g}$ and $G$ interchangeably to refer to the QFT labeled by $\mathfrak{g}$.} The case relevant to this work is $\mathfrak{g}=\mathfrak{su}(N)$ -- namely, the theories of type $G=A_{N-1}$ which arise on the worldvolume of $N$ M5-branes \cite{Witten:1995zh,Strominger:1995ac}.  
 
The choice of punctures on the Riemann surface leads to a great variety of possible 4d SCFTs that can be geometrically engineered in this manner. From the perspective of the 6d (2,0) theory, punctures are 1/2-BPS codimension-two defects that extend over 4d spacetime. They correspond to singular boundary conditions of the Hitchin equations on the Riemann surface -- which arise as the BPS equations of the (2,0) theory compactified on a circle and then twisted over the Riemann surface -- and are classified by consistent boundary conditions of the Higgs field in the Hitchin system. 
There is a distinction between regular punctures (or {\it tame} defects) for which the pole of the Higgs field is simple, and irregular punctures (or {\it wild} defects) with higher order poles, to be reviewed in more detail below. The classification of regular punctures was studied for $\mathfrak{g}=\mathfrak{su}(N)$ in \cite{Gaiotto:2009we,Gaiotto:2009gz}, with  possibilities labeled by partitions of $N$, or equivalently Young Tableaux with $N$ boxes. More generally, regular punctures are labeled by an embedding of $\mathfrak{su}(2)$ into $\mathfrak{g}$, corresponding to nilpotent orbits in $\mathfrak{g}$ (e.g. see \cite{Chacaltana:2010ks}).  The contributions to the $a$ and $c$ central charges of these regular punctures was obtained in \cite{Gaiotto:2009gz,Chacaltana:2010ks} for the $A_{N-1}$ case, and in \cite{Chacaltana:2012zy} for the general ADE case.  The holographic duals of the 4d $\CN=2$ SCFTs of type $A_{N-1}$ in class $\CS$ involving only regular punctures are known, and are given by the 11d supergravity solutions of Gaiotto and Maldacena \cite{Gaiotto:2009gz}.

The classification of consistent irregular-type singularities of the Higgs field in the Hitchin system was undertaken in \cite{Xie:2012hs} (for $A_{N-1}$) and \cite{Wang:2015mra}  (for types $D$ and $E$), building on \cite{Gaiotto:2009hg}.  These are labeled by the group $G$ with algebra $\mathfrak{g}$, as well as by two integers $b >0$ and $k>-b$, and so following \cite{Xie:2016evu} we denote the irregular puncture with these labels by $G^{(b)}[k]$. In order to engineer a superconformal field theory by wrapping the (2,0) theory on a Riemann surface with irregular punctures, there are precisely two possibilities \cite{Xie:2012hs,Wang:2015mra}: the Riemann surface is a sphere with a single irregular puncture, or the Riemann surface is a sphere with one irregular puncture and one regular puncture (see also \cite{Gaiotto:2009hg,Cecotti:2011rv,Bonelli:2011aa} for earlier constructions involving $A_1$). This is in contrast to geometric engineering with only regular punctures, for which an almost unlimited number and variety of regular punctures can decorate a Riemann surface of any genus (with caveats at low genus $g=0,1$) and flow to an SCFT at low energies.

\subsection{Survey of Generalized Argyres-Douglas SCFTs} 

The Argyres-Douglas SCFTs are intrinsically strongly-coupled 4d $\CN=2$ SCFTs with Coulomb branch operators of fractional scaling dimension, that are also notable for possessing relevant deformations. The original theory of this type was discovered by Argyres and Douglas in \cite{Argyres:1995jj}, where it was obtained as a special point on the moduli space of $\CN=2$ pure $SU(3)$ gauge theory where mutually non-local BPS states simultaneously become massless.\footnote{~See also \cite{Argyres:1995xn} for a construction involving $SU(2)$ $N_f=1$ SQCD, and generalizations for $N_f>1$.}  This original Argyres-Douglas SCFT has rank-one, with a single Coulomb branch operator of  dimension $\Delta=\frac{6}{5}$ and no flavor symmetry. 

A larger class of SCFTs of this type were obtained in \cite{Eguchi:1996vu} at the maximal conformal point on the moduli space of $\CN=2$ pure $SU(k)$ super Yang-Mills with $k\geq 3$. These generalized Argyres-Douglas theories are denoted $(A_1,A_{k-1})$, with the case $k=3$ corresponding to the original Argyres-Douglas theory (also sometimes denoted by $H_0$ in the literature). The case $k=4$  -- also called $H_1$ in the literature -- is also notable for being rank-one, with a single Coulomb branch operator of dimension $\frac{4}{3}$. The theories in this series with odd $k$ have no flavor symmetry, while those with even $k$ possess a $U(1)$ global symmetry which is enhanced to $SU(2)$ for $k=4$ \cite{Argyres:2012fu}.\footnote{~This even versus odd difference is also apparent from \eqref{eq:rf}.}

The $(A_1,A_{k-1})$ theories belong to a more general set of 4d $\CN=2$ SCFTs which can be obtained via Type IIB string theory on a class of isolated hypersurface singularities labeled by $(G,G')$  (with the theories of type $(G,G')$ are equivalent so those of type $(G',G)$) \cite{Cecotti:2010fi}. The IIB background takes the form of an arbitrary closed four-manifold times a non-compact Calabi-Yau threefold with an isolated singularity given by the sum of two singularities $P_{\mathfrak{g}}(x,y) + P_{\mathfrak{g}'}(w,z)$=0, where each $(\mathfrak{g},\mathfrak{g}')$ is of ADE type. (Note that the cases $G=A_1$ with $G'=DE$ were first studied in \cite{Eguchi:1996vu,Eguchi:1996ds}.) Another special case is the class $(A_1,D_k)$ that arises from the maximal conformal point on the moduli space of $\CN=2$ $SO(2k)$ gauge theory \cite{Eguchi:1996vu}, which can also be obtained via a relevant deformation to the maximal superconformal point of the $SU(k-1)$ theory with two fundamental hypermultiplets \cite{Bonelli:2011aa}. The $(A_1,D_{k})$ theories with $k$-odd have an $SU(2)$ flavor symmetry, while those with $k$-even have an $SU(2)\times U(1)$ flavor symmetry that is enhanced to $SU(3)$  for $k=4$ \cite{Argyres:2012fu}.

A subset of the $(G,G')$ theories can be engineered in class $\CS$ using irregular punctures.  One set with this property is the class $(G,A_{k-1})$, which can be engineered from the  6d (2,0) theory of type $\mathfrak{g}$ wrapped on a Riemann surface with an irregular puncture of type $G^{(b=h_G)}[k]$, where $h_G$ is the dual Coxeter number of $G$. For example, taking $G=A_{N-1}$ with $h_{G}=N$ yields the class $(A_{N-1},A_{k-1})$, which can be engineered from $N$ M5-branes wrapping a sphere with one irregular puncture of type $A_{N-1}^{(N)}[k]$ \cite{Xie:2012hs,Wang:2015mra}, which is the case of interest in the present work. 
	
%
%

\subsection{Classification of Irregular Singularities}

Let us now review pertinent aspects of the classification of irregular singularities given in \cite{Xie:2012hs,Wang:2015mra}---also see \cite{Wang:2018gvb} for a nice review of these properties. 

We consider a 4d $\CN=2$ theory engineered by twisting the 6d (2,0) theory of type $\mathfrak{g}$ over a Riemann surface $\mathcal{C}$ with punctures.  Let $z$ denote a local holomorphic coordinate on $\CC$. By further compactifying on a circle, at low energies one obtains a 3d theory with $\CN=4$ supersymmetry. One can instead reverse the order of the compactification, first reducing the 6d theory on a circle to obtain 5d $\CN=2$ super Yang-Mills, and then twisting the 5d theory over $\CC$. The BPS equations of this configuration are the Hitchin equations for the holomorphic $(1,0)$-form Higgs field $\Phi = \Phi_z dz$ that comprises two of the adjoint scalars of 5d maximally supersymmetric Yang-Mills, and the gauge field $A = A_zd{z} + A_{\bar{z}} d\bar{z}$, both of which are valued in the Lie algebra $\mathfrak{g}$ \cite{Hitchin:1986vp}.  The space of solutions to these equations modulo gauge transformations is the Hitchin moduli space, which is identified by mirror symmetry with both the Higgs branch of the 5d theory, and the Coulomb branch of the 3d theory from the reverse-order compactification. The interplay of the 4d $\CN=2$ theory and the Hitchin system was studied in detail in \cite{Gaiotto:2009hg}. 

At punctures the fields $\Phi$ and $A$ are singular, and one must specify their boundary conditions. $\Phi=\Phi_z dz$ can be put into semisimple form by a gauge transformation, 
\begin{align}
\Phi_z(z) = \sum_{m=0}^{k+b} \frac{T_{m-b}}{z^{1 + \frac{m}{b}}} \ +\  \text{(non-divergent)}\ , \label{eq:phiz}
\end{align}
where in \eqref{eq:phiz} we have placed the defect at $z=0$ on $\CC$. The defining data of the defect is the set $(T,k,b)$, where the $\{T_{-b},\dots,T_{k}\}$ are semisimple elements of the Lie algebra $\mathfrak{g}$, $b$ is a positive integer, and $k$ is an integer satisfying $k>-b$. As indicated in Section \ref{sec:ovirr}, the defect with these labels is denoted $G^{(b)}[k]$. Evidently the order $\rho$ of the leading pole in \eqref{eq:phiz} is $\rho=2 + \frac{k}{b}$. A regular puncture has a simple pole $\rho=1$ with $k+b=0$, in which case the puncture is characterized the choice of $T$, which is a nilpotent element of the Lie algebra $\mathfrak{g}$. By contrast, the case of a higher order pole $k+b>0$ corresponds to an irregular puncture. In a nice class of solutions, the $T_{\ell}$ are {\it regular} semisimple elements of the Lie algebra, with restricted values of $b$ that are one-to-one with the three-fold isolated quasi-homogeneous singularities of compound du Val type (see Table 1 of \cite{Wang:2015mra}). For the $G=A_{N-1}$ case, this translates into a choice of $b=N$ or $N-1$, denoted respectively as Type I and Type II in the notation of \cite{Xie:2012hs,Wang:2015mra}.\footnote{~There is also a Type III singularity, which is a special case of Type I characterized by a nested Young Tableaux structure in $T$ that we will not discuss  here---also see \cite{Witten:2007td}.} 

The resulting 4d SCFTs preserve $\CN=2$ supersymmetry when the Riemann surface is a sphere \cite{Xie:2012hs}. One can also add a single regular puncture to the sphere while preserving supersymmetry, resulting in the theories of Type IV in the notation of \cite{Xie:2012hs}, which are also denoted by their two punctures as $(G^{(b)}[k], Y)$ with $Y$ the Young diagram labeling the regular puncture. The examples pertinent to this note arise from $N$ M5-branes wrapping a sphere with an irregular puncture of type $A_{N-1}^{(b=N)}[k]$ -- which alone correspond to the  $(A_{N-1},A_{k-1})$ SCFTs discussed above -- with an additional regular puncture whose Young diagram consists of a box with $\ell$ columns and $N/\ell$ rows, contributing an $SU(\ell)$ flavor symmetry. We denote the resulting field theories by $(A_{N-1}^{(N)}[k],Y_{\ell})$.  One example of this class is to take the regular puncture to be maximal, with Young diagram consisting of a single row of length $N$ ($\ell=N$) contributing an $SU(N)$ flavor symmetry. These are also known as the $D_{p=k+N}^{b=N} (SU(N))$ theories, and were studied in  \cite{Cecotti:2012jx,Cecotti:2013lda,Giacomelli:2017ckh}. 
The case $\ell =1$ is the ``non-puncture'', reducing to the $(A_{N-1},A_{k-1})$ class with no regular puncture on the sphere. 



\section{Lagrangian Description of the $(A_{N-1}^{(N)}[k],Y_{1})$ SCFTs} \label{sec:lagrangian}

In this appendix, we review the RG flow described in \cite{Agarwal:2017roi,Benvenuti:2017bpg}  between quiver Lagrangians that ends at the  $(A_{N-1},A_{(k=mN)-1})$ $(\ell=1)$ theories at low energies.

One begins with an $\CN=2$ conformal quiver gauge theory with gauge group $\prod_{\ell=1}^{N-1} SU(l m)$. The quiver is depicted at the top of Figure \ref{fig:quiver}.\footnote{~The dictionary with \cite{Agarwal:2017roi} is: $m_{\text{them}} = N$, and $N_{\text{them}} = m$. The dictionary with \cite{Benvenuti:2017bpg} is: $k_{\text{them}} = N-1$, $N_{\text{them}}=m$, $\alpha_r = M_{j=mN-1-r}$, and they refer to the $(Q,\tilde{Q})$ bifundamentals as $(b,\tilde{b})$.}
The quiver has $N-1$ nodes corresponding to the $l = 1,\dots,N-1$ gauge groups, and one final node associated to an $SU(mN=k)$ flavor symmetry. Each gauge node has an $\CN=2$ vector multiplet with associated $\CN=1$ chiral multiplet $\phi_l$ that transforms in the adjoint representation of the $SU(l m)$ gauge group. Bifundamental hypermultiplets $H_{l} = ( Q_l,\tilde{Q}_l)$ connect the nodes, with $Q_l$ transforming in the $(\Box, \overline{\Box})$ and $\tilde{Q}_l$ in the $(\overline{\Box},\Box)$ of the adjacent $SU(l m)\times SU( (l +1)m)$ gauge groups. At the final gauge node, $mN$ hypermultiplets $H_{N-1}$ transform in the fundamental representation of the $SU( m(N-1))$ gauge group, and resulting in an $SU(mN)$ flavor symmetry. The quiver is thus balanced, since the number of colors $n_l = l m$ satisfies $2 n_l - n_{l-1} - n_{l + 1} = 0$ at each node except the last, where $2 ( m (N-1) )- m(N-2) = mN$ the number of fundamental hypermultiplets at that node. Indeed this is by construction, since the quiver is built by successively adding $(l+1) m$ hypermultiplets to the $l$'th gauge node, and gauging the resulting $SU((l+1) m)$ flavor symmetry. 

	\usetikzlibrary{shapes.geometric}
	\usetikzlibrary{arrows, decorations.markings}
		
		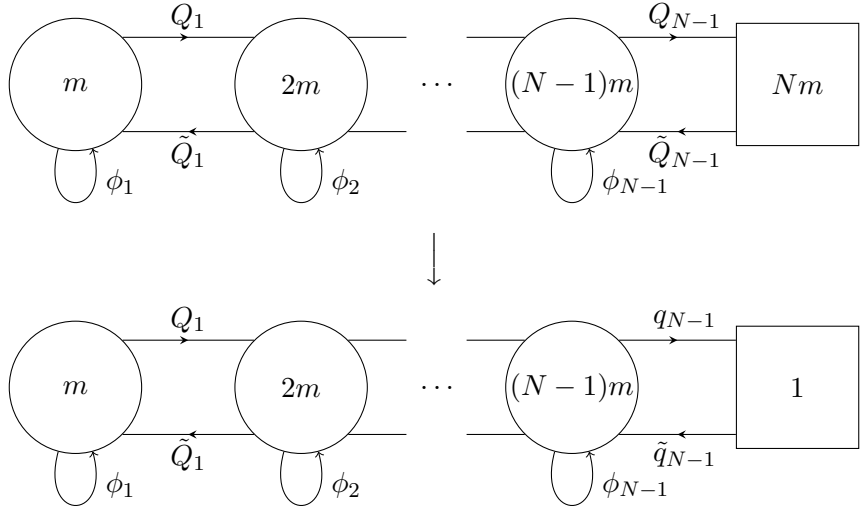
\begin{figure}
	\centering
	
 		\begin{tikzpicture}[square/.style={regular polygon,regular polygon sides=4}]

\tikzset{
  ->-/.style={decoration={markings, mark=at position 0.5 with {\arrow{stealth}}},
              postaction={decorate}},
}
		
	\node[circle,draw=black,minimum size=1.76cm,fill=none]  (1) at (0,0) { };
	\node at (0,0) {$m$};
        \draw[->] (1) to [out=-105,in=-75,looseness=5.5] (1);
        \node at (0.6,-1.3) {$\phi_1$};
	
	\node[circle,draw=black,minimum size=1.76cm,fill=none]  (2) at (3,0) { };
	\node at (3,0) {$2m$};
	 \draw[->] (2) to [out=-105,in=-75,looseness=5.5] (2);
	\node at (3.6,-1.3) {$\phi_2$};
	
	\node[circle,draw=black,minimum size=1.76cm,fill=none]  (3) at (6.6,0) { };
	\node at (6.6,0) {$(N-1)m$};
	 \draw[->] (3) to [out=-105,in=-75,looseness=5.5] (3);
	 \node at (7.45,-1.3) {$\phi_{N-1}$};
	
	\node[square,draw=black,minimum size=2.3cm,fill=none]  (4) at (9.6,0) { };
	\node at (9.6,0) {$Nm$};
	
	\draw[->-] (1.north east) -- (2.north west);
	\draw[->-] (2.south west) -- (1.south east);
	\node at (1.5,0.9) {${Q}_1$};
	\node at (1.5,-0.9) {$\tilde{Q}_1$};
	
	\draw (2.north east) -- (4.4,0.627);
	\draw (2.south east) -- (4.4,-0.627);
	
	\node at (4.84,0) {$\dots$};
	
	\draw (5.2,0.627) -- (3.north west);
	\draw (5.2,-0.627) -- (3.south west);
	
	\draw[->-] (3.north east) -- (8.78,0.627);
	\draw[->-] (8.78,-0.627) -- (3.south east);
	\node at (8.1,0.9) {${Q}_{N-1}$};
	\node at (8.1,-0.9) {$\tilde{Q}_{N-1}$};
  
		\end{tikzpicture}
		
		\begin{tikzpicture}
		
		\node at (4.8,0) {$\Big\downarrow$};
		\end{tikzpicture}
		
	 	\begin{tikzpicture}[square/.style={regular polygon,regular polygon sides=4}]

\tikzset{
  ->-/.style={decoration={markings, mark=at position 0.5 with {\arrow{stealth}}},
              postaction={decorate}},
}
		
	\node[circle,draw=black,minimum size=1.76cm,fill=none]  (1) at (0,0) { };
	\node at (0,0) {$m$};
        \draw[->] (1) to [out=-105,in=-75,looseness=5.5] (1);
        \node at (0.6,-1.3) {$\phi_1$};
	
	\node[circle,draw=black,minimum size=1.76cm,fill=none]  (2) at (3,0) { };
	\node at (3,0) {$2m$};
	 \draw[->] (2) to [out=-105,in=-75,looseness=5.5] (2);
	   \node at (3.6,-1.3) {$\phi_2$};
	
	\node[circle,draw=black,minimum size=1.76cm,fill=none]  (3) at (6.6,0) { };
	\node at (6.6,0) {$(N-1)m$};
	 \draw[->] (3) to [out=-105,in=-75,looseness=5.5] (3);
	  \node at (7.45,-1.3) {$\phi_{N-1}$};
	
	\node[square,draw=black,minimum size=2.3cm,fill=none]  (4) at (9.6,0) { };
	\node at (9.6,0) {$1$};
	
	\draw[->-] (1.north east) -- (2.north west);
	\draw[->-] (2.south west) -- (1.south east);
	\node at (1.5,0.9) {${Q}_1$};
	\node at (1.5,-0.9) {$\tilde{Q}_1$};
	
	\draw (2.north east) -- (4.4,0.627);
	\draw (2.south east) -- (4.4,-0.627);
	
	\node at (4.84,0) {$\dots$};
	
	\draw (5.2,0.627) -- (3.north west);
	\draw (5.2,-0.627) -- (3.south west);
	
	\draw[->-] (3.north east) -- (8.78,0.627);
	\draw[->-] (8.78,-0.627) -- (3.south east);
	\node at (8.1,0.9) {${q}_{N-1}$};
	\node at (8.1,-0.9) {$\tilde{q}_{N-1}$};
	
		\end{tikzpicture}		
	\caption{The upper figure is the UV quiver with matter content summarized in Table \ref{tab:atab2x}, and the lower figure is the IR quiver with matter content summarized in Table \ref{tab:atab22}. The circles are special unitary gauge groups and the squares are flavor symmetry groups, where the upper square denotes $SU(Nm)$ and the lower square is meant to denote a global $U(1)$ symmetry. \label{fig:quiver}}
	\end{figure}

Denote the scalar chiral primary operators at the bottom of the $SU(l m)$ would-be flavor current multiplets  -- i.e., the $SU(l m)$ moment map operators --  by $(\mu_{l}, \tilde{\mu}_{l-1})$, with the $\mu_{l}$ formed from the $H_l$ hypermultiplets and the $\tilde{\mu}_{l-1}$ formed from the $H_{l-1}$ hypermultiplets at that node. For example, we are treating $Q_l$ as an $l m\times (l + 1)m$ matrix and $\tilde{Q}_l$ as an $ (l + 1)m\times l m$ matrix, with $\mu_{l} =  Q_{l} \tilde{Q}_{l} - \frac{1}{l m} \tr Q_l \tilde{Q}_l$. 
There is an $\CN=2$ preserving superpotential that couples the vector multiplets and moment map operators as
	\ba{
	W_{\CN=2}  = \sqrt{2} \sum_{l =1}^{N-1} \tr \phi_l (\mu_l - \tilde{\mu}_{l -1} )\ ,\label{eq:wn2}
	}
	where in writing \eqref{eq:wn2} we have defined $\tilde{\mu}_0=0$.

The charges of the fields are listed in Table \ref{tab:atab2x}. The UV $\CN=2$ SCFT has an R-symmetry $SU(2)_{R_{\text{UV}}}\times U(1)_{r_{\text{UV}}}$, whose Cartan generators $(I^3_{\text{UV}},r_{\text{UV}})$ we denote 
	\ba{
	J_+ = 2 I^3_{\text{UV}}\ , \qquad J_-=r_{\text{UV}}\ . 
	}
(We reserve the labels $(I^3,r)$ without subscripts for the R-symmetry of the Argyres-Douglas SCFTs at the end of the flow.)
Note that each hypermultiplet $H_l$ comes with a baryonic $U(1)_{l}$ global symmetry under which the $(Q_l,\tilde{Q}_l)$ have charges $\pm 1$, since only the $SU(ml)$ part of the $U(m l)$ global symmetry acting on the hypermultiplets has been gauged in the construction of the quiver. 

\def\arraystretch{1.2}
\begin{table}
\centering
\begin{tabular}{|c | c | c   | c | c  |}
\hline
                      &     $SU(l m)$    &     $SU((l+1)m)$       &       $U(1)_l$           &       $(J_+,J_-)$  \\ \hline   \hline              
 $\phi_{l}$   &           adj            &            ${\bf 1} $                     &               0              &   $(0,2)$ \\ 
 $Q_l$        &        $\Box$         &      $\overline{\Box}$     &        1                           &  $(1,0)$ \\
  $\tilde{Q}_l$  &   $\overline{\Box}$   &     ${\Box}$            &        $-1$                      &  $(1,0)$ \\
  \hline
\end{tabular} 
\caption{ $l = 1,\dots,N-1$. The first two columns denote the gauge group factors, except that the last $l = N-1$ entry $SU(m N)$ is a flavor symmetry group. The $U(1)_l$ are baryonic flavor symmetries acting on the hypermultiplets.   \label{tab:atab2x} }
\end{table}

Now introduce an $\CN=1$ chiral multiplet $M$ that transforms in the adjoint representation of the $SU(mN)$ flavor symmetry group, and couple it to the moment map operator $\tilde{\mu}_{N-1}$ of $SU(mN)$ via the superpotential
	\ba{
	\delta W = \tr \tilde{\mu}_{N-1} M\ .
	}
	This superpotential breaks the $\CN=2$ supersymmetry to $\CN=1$, with the $\CN=1$ R-symmetry corresponding to the subalgebra\footnote{~One can in general fix an $\CN=1$ subalgebra in the $\CN=2$ algebra, with $\mathcal{N}=1$ R-symmetry generated by
	\ba{
	R_{\CN=1} = \frac{1}{3} \left( r + 2 R\right)\ . \label{eq:n1sub}
	}
The dimensions of chiral primary operators satisfy $\Delta = \frac{3}{2} R_{\CN=1}$. }
	\ba{
	R_{\CN=1} = \frac{1}{3} (2 J_+ + J_-)\ . \label{eq:embed}
	}
	The moment map operator $\tilde{\mu}_{N-1}\sim Q_{N-1}\tilde{Q}_{N-1}$ has charges $(J_+,J_-)=(2,0)$, and $M$ has charges $(0,2)$, such that the superpotential $W$ has charge $R_{\CN=1}(W)=2$. 
	
	Next give $M$ a nilpotent VEV $\langle M \rangle$ which corresponds to the principal embedding of $\mathfrak{su}(2)$ into the flavor symmetry algebra $\mathfrak{su}(mN)$, completely breaking the $SU(mN)$ global symmetry. Explicitly, $\langle M \rangle$ is given by the 
 $mN\times mN$ matrix with 1's along the entire upper diagonal. Using results from  \cite{Agarwal:2014rua} (based on the methods of \cite{Gadde:2013fma}), one can show that many of the modes decouple in the IR, including Nambu-Goldstone modes corresponding to broken flavor symmetry generators, and chiral multiplets that become massive due to the VEV, resulting in a ``fan'' superpotential. Decomposing the adjoint indices of the operators $M$ and $\tilde{\mu}_{N-1}$ in terms $(j,m)$ indices of the principal embedding of  $\mathfrak{su}(2)$, and denoting by $\tilde{M}_{j,m}$ the fluctuations about the vev $\langle M \rangle$ in this basis, the result is that the only modes $\tilde{M}_{j,m}$ that remain coupled at low energies are those with lowest weight $m=-j$ and $j=1,\dots, mN-1$. The remaining superpotential takes the form
\ba{
\delta W = (\tilde{\mu}_{N-1})_{1,-1} + \sum_{j=1}^{mN-1} \tilde{M}_{j} \hat{\mu}_j\ ,\qquad \tilde{M}_j \equiv \tilde{M}_{j,m=-j}\ ,\qquad \hat{\mu}_j \equiv (\tilde{\mu}_{N-1})_{j,m=j}\ .
}
 Due to the first term, the $J_-$ charge shifts to $J_-' = J_- - 2 \rho(\sigma_3)$ while the $J_+$ charge remains unshifted, $J_+'=J_+$, such that the superpotential has $(J_+',J_-')=(2,2)$. Then, $(J_+', J_-' )(\tilde{M}_j)= (0,2+2j)$, and $(J_+', J_-')( \hat{\mu}_j)=  (2,-2j)$.
 
 Of the $mN$ pairs of fundamental quarks (antiquarks) in the fundamental (anti-fundamental) representation of the $SU(m(N-1))$ gauge symmetry, all but one pair which we denote $(q,\tilde{q})$ receive a mass due to the VEV $\langle M \rangle$. The charges of $(q,\tilde{q})$ are $(J_+',J_-')(q,\tilde{q}) = (1,1-mN)$. The remaining $mN-1$ components $\hat{\mu}_j$ of the $SU(mN)$ moment map correspond to traces of products of $q\tilde{q}$ with powers of the vector multiplet $\phi_{N-1}$,
 	\ba{
	\hat{\mu}_j = \tr q \phi^{mN-1-j}_{N-1} \tilde{q}\ ,\qquad j=1,\dots,mN-1\ .
	}
 The field content and charges after removing all of the decoupled modes and massive fields is summarized in Table \ref{tab:atab22}, and the IR quiver is depicted at the bottom of Figure \ref{fig:quiver}.
 
 After Higgsing, the $\CN=1$ theory flows to a fixed point whose superconformal R-symmetry is given by a linear combination of $J_+'$ and $J_-'$ that is determined by $a$-maximization \cite{Intriligator:2003jj},
 	\ba{
	R_{\CN=1}(\epsilon) = \frac{1}{2} \left( (1 + \epsilon) J_+' + (1-\epsilon) J_-'\right)\ .
	}
In the UV before nilpotent Higgsing, $\epsilon_{\text{UV}} = \frac{1}{3}.$ Performing $a$-maximization, the authors of \cite{Agarwal:2017roi,Benvenuti:2017bpg} find that various gauge-invariant operators seemingly violate the unitarity bound, and thus decouple as free fields acted on by an accidental $U(1)$ global symmetry \cite{Kutasov:2003iy}. The operators that decouple are $\tr \phi_l^i$ with $i=2,\dots, m+1$ and $l = 2,\dots,N-1$, and $\tr \phi_{l=1}^i$ with $i=2,\dots,m$, along with the gauge singlets $\tilde{M}_{j}$ with $j=1,\dots,m$. After decoupling all the necessary fields and repeating the $a$-maximization procedure, $\epsilon_{\text{IR}}$ is determined as
	\ba{
	\epsilon_{\text{IR}} = \frac{3m+1}{3(m+1)}\ ,\qquad \frac{2}{3}  \leq \epsilon_{\text{IR}} < 1\ . \label{eq:eir}
	}
 The dimensions of chiral operators at the IR fixed point are thus given by $\Delta(\epsilon_{\text{IR}}) = \frac{3}{2} R_{\CN=1}(\epsilon_{\text{IR}})$. Computing the  central charges at the fixed point, the authors of \cite{Agarwal:2017roi,Benvenuti:2017bpg} find agreement with \eqref{eq:ac1}.

 \def\arraystretch{0.7}
The remaining fields are coupled together in a superpotential
\ba{\bs{
W =& \sum_{l = 1}^{N-1} \tr  \phi_l (Q_{l}\tilde{Q}_{l} -\tilde{Q}_{l-1} Q_{l - 1})  + \sum_{j=m+1}^{m N  - 1} \tilde{M}_j \tr \left( {q} \phi^{mN-1-j}_{N-1} \tilde{q} \right) \\
&+  \sum_{i=2}^m \beta_{1,i} \tr \phi_1^i + \sum_{l = 2}^{N-1}\sum_{ i = 2}^{m+1 } \beta_{l,i} \tr \phi_l^i\ .
}}
One can verify that every term has $(J_+',J_-')=(2,2)$ and thus $R_{\CN=1}= 2$. We have included the flipping fields $\beta_{l,i}$ that enforce decoupling of the  operators $\tr \phi_\ell^i$ that become free \cite{Benvenuti:2017lle}, whose charges are listed in Table \ref{tab:atab22}.

The IR fixed point is identified with the $(A_{N-1},A_{(k=mN)-1})$ theories whose properties are reviewed in Section \ref{sec:anak} (also see Table \ref{tab:tp1} for a summary). In particular, the R-symmetry at the fixed point is expected to enhance to $SU(2)_R\times U(1)_r$, with Cartan generators $(I^3=R/2, r)$ identified as
 	\ba{
	r = \frac{1}{m+1} \left( m J_+' + J_-' \right)\ ,\qquad   R = 2 I^3 ={J_+'} \ .
	}
	One can verify that these charges satisfy the analogue of \eqref{eq:embed}, 
	\ba{
	R_{\CN=1}(\epsilon_{\text{IR}}) = \frac{1}{3} \left( 2 R + r \right) \ ,
	}
	where $\epsilon_{\text{IR}}$ is given in \eqref{eq:eir}.
Then, the global symmetry of the IR SCFT is
	\ba{
	\prod_{l=1}^{N-1} U(1)_l \times SU(2)_R\times U(1)_{r}\ ,
	}
where the $U(1)_{l}$ are baryonic.
 
\def\arraystretch{1.2}
\begin{table}
\centering
\begin{tabular}{|c | c | c |  c   || c | c | c  |}
\hline
                      &     $SU(l m)$    &     $SU((l+1)m)$    &     $SU( (N-1) m)$     &       $U(1)_l$        &      $U(1)_{N-1}$    &       $(J_+',J_-')$  \\ \hline   \hline              
 $\phi_{l}$   &           adj            &            ${\bf 1} $        &           ${\bf 1}  $         &        0                      &               0              &   $(0,2)$ \\ 
 $\phi_{N-1}$   &    ${\bf 1} $         &            ${\bf 1} $        &          adj                    &        0                      &               0              &   $(0,2)$ \\ 
 $Q_l$        &        $\Box$         &      $\overline{\Box}$  &             ${\bf 1}  $       &        1                      &               0             &  $(1,0)$ \\
  $\tilde{Q}_l$  &   $\overline{\Box}$   &     ${\Box}$      &             ${\bf 1}  $       &        $-1$                 &               0             &  $(1,0)$ \\
  $q$             &        ${\bf 1} $         &            ${\bf 1} $        &          $\Box$            &       0                       &                1             &  $(1,1-mN )$  \\
 $\tilde{q}$    &        ${\bf 1} $         &            ${\bf 1} $        &  $\overline{\Box}$    &       0                       &             $ -1$          &  $(1,1-mN )$  \\
 $\tilde{M}_j$         &         ${\bf 1} $         &            ${\bf 1} $       &             ${\bf 1}  $       &       0                      &               0              &  $(0,2+2j)$ \\
$\beta_{l,i}$ &      ${\bf 1} $         &            ${\bf 1} $       &             ${\bf 1}  $       &       0                      &               0            &   $(2,2-2i )$ \\
$\beta_{N-1,i}$ &     ${\bf 1} $         &            ${\bf 1} $       &             ${\bf 1}  $       &       0                      &               0            &   $(2,2-2i )$ \\ \hline
\end{tabular} 
\caption{The charges of fields in the Lagrangian description of the $(A_{N-1},A_{(k=mN)-1})$ theories. The gauge group and matter content with $l=N-1$  have been singled out, such that in the table $l$ runs over $1,\dots, N-2$. The columns before the double vertical lines are the gauge groups in the quiver, and the remaining three columns are the baryonic global symmetries and R-charges. The operators $\tilde{M}_j$ that remain coupled at the fixed point have $j=m+1,\dots,mN-1$. The singlets $\beta_{l,i}$ are flipping fields, where for $l=1$ $i$ runs over $2,\dots,m$, and for $l=2,\dots,N-1$ $i$ runs over $2,\dots,m+1$.  \label{tab:atab22} }
\end{table}
 
One can verify that the following properties of this Lagrangian SCFT match onto those of the $(A_{N-1},A_{(k=mN)-1})$ theories, as summarized in Table \ref{tab:tp1} (with $\ell=1$):

\begin{itemize}
 
\item Recall from around \eqref{eq:cbo} that the Argyres-Douglas theories under consideration have $\frac{1}{2}(N-1)(k-2)$ Coulomb branch operators $u_{i}$ with dimension
	\ba{
	\Delta(u_{l,i}) = \frac{i}{m+1}\ ,\qquad i = m+2,\dots,\ell m\ ,\qquad l = 2,\dots, N\ ,
	}
	(here we are also including an additional $l$ subscript on the $u_i$ to label the set of operators with degenerate dimensions for a given $i$), and $\CN=2$ R-charges
	$r(u_{l,i}) = \frac{2i}{m+1}$ and  $R(u_{l,i}) = 0$.
The mapping of the Coulomb branch operators to the fields in the quiver gauge theory description is given by \cite{Agarwal:2017roi,Benvenuti:2017bpg}
	\ba{
	u_{l,i} =  \left\{   \begin{array}{cl}  \tr \phi_l^i & i = m+2,\dots,l m\ ;\ l = 2,\dots, N-1 \\  M_{j=i-1} & i = m+2,\dots,N m\ ;\ l = N \end{array} \right.
	}

\item  One can also identify the superpartners $\CO'_{l,i}$ of the Coulomb branch operators that correspond to the level-two descendants of the $\CN=2$ chiral multiplet whose primary is $u_{\ell,i}$. Their dimensions satisfy $\Delta(\CO'_{l,i}) = \Delta(u_{l,i}) + 1$, and their $\CN=2$ R-charges are
	$r(\CO'_{l,i}) = r(u_{l,i}) - 2$, and  $R(\CO'_{l,i})=2$.
The authors of \cite{Benvenuti:2017bpg}  find that the $(N-1)(m-1)$ operators $\beta_{l,i}$ with $l =1,\dots,N-1$ and $i=2,\dots,m$ map to $\CO'$ operators, which are paired with Coulomb branch operators as 
\ba{
\left\{ \beta_{l -1,2m+2-i}    \right\} \leftrightarrow   \{  u_{l, i}  \}  \ ,\qquad   i=m+2,\dots,2m\ ,\qquad l = 2,\dots,N\ .
}
An additional $\frac{m}{2}(N-1)(N-2)$ baryonic  operators formed from traces of the product of two quarks $(Q_{l},\tilde{Q}_{l})$ with powers of vector multiplet scalars $\phi_{l}$ complete the set of $\CO'$ operators. 

\item The complex dimension of the conformal manifold is $N-2$ \cite{Benvenuti:2017bpg}. 

\item The moment map operators at the $\CN=2$ fixed point have dimension $\Delta =2$, and R-charges $r = 0$, $R= 2$,
and correspond to the set of $N-1$ operators
	\ba{
	\{ \tr \phi_{N-1}^{mN-m-1} q\tilde{q}\ ,\ \beta_{l,m+1} \}\ ,\qquad l=2,\dots,N-1\ .
	}
These are in correspondence with the $N-1$ mass deformations of the Argyres-Douglas SCFTs \cite{Benvenuti:2017bpg}.
	
An additional $2(2^{N-1}-1)$ Higgs branch operators correspond to baryons composed of gauge invariant products of $N-1$ quarks and adjoints, and have dimension \cite{Benvenuti:2017bpg}
	\ba{ 
	\Delta = k - \frac{k}{N}\ . \label{eq:hbop}
	}

\end{itemize}


\bibliographystyle{./ytphys}
\bibliography{./refs}

\end{document}